\documentclass[preprint,11pt]{article}
\usepackage{fullpage}
\ifdefined\usebigfont

\usepackage{times}
\usepackage[fontsize=13pt]{scrextend}
\usepackage[left=1.56in,right=1.56in,top=1.74in,bottom=1.74in]{geometry}
\usepackage{bigints}
\usepackage{csquotes}
\else
\fi

\usepackage{amssymb,amsfonts,amsmath,amsthm,amscd,dsfont,mathrsfs,bbm}
\usepackage{graphicx,float,psfrag,epsfig,amssymb}
\usepackage[usenames,dvipsnames,svgnames,table]{xcolor}
\definecolor{darkgreen}{rgb}{0.0,0,0.9}
\usepackage[pagebackref,letterpaper=true,colorlinks=true,pdfpagemode=none,citecolor=OliveGreen,linkcolor=BrickRed,urlcolor=BrickRed,pdfstartview=FitH]{hyperref}
\usepackage{wrapfig}
\usepackage{relsize}
\usepackage{color}
\usepackage{pict2e}
\usepackage{subcaption}
\usepackage{nameref}
\usepackage{makecell}
\usepackage[font={small}]{caption} 
\usepackage{mathtools}
\usepackage{enumitem}

\let\chapter\section
\usepackage[ruled,vlined]{algorithm2e}
\usepackage[noend]{algorithmic}
\usepackage{diagbox}

\DeclareMathAlphabet{\mathpzc}{OT1}{pzc}{m}{it}

\newtheorem{propo}{Proposition}[section]
\newtheorem{lemma}[propo]{Lemma}

\newtheorem{assumption}[propo]{Assumption}
\newtheorem{proposition}[propo]{Proposition}
\newtheorem{defi}[propo]{Definition}

\newtheorem{thm}[propo]{Theorem}
\newtheorem{rmk}[propo]{Remark}

\newtheorem{example}[propo]{Example}

\usepackage{hyperref}       
\usepackage{url}   

\addtocontents{toc}{\protect\setcounter{tocdepth}{2}}

\def\tX{\widetilde{X}}

\def\tX{\widetilde{X}}
\def\tx{\widetilde{x}}

\def\cZ{{\cal Z}}

\def\cG{{\cal G}}

\def\cD{{\cal D}}

\def\tp{{\tilde{p}}}

\def\cG{\mathcal{G}}

\def\hy{\widehat{y}}

\def\reals{{\mathbb R}}

\def\eps{{\varepsilon}}
\def\prob{{\mathbb P}}
\def\E{{\mathbb E}}

\def\var{{\rm Var}}

\def\L0{{L_i}}

\def\de{{\rm d}}
\def\<{\langle}
\def\>{\rangle}

\def\hth{\widehat{\theta}}

\def\F{{\sf F}}
\def\ind{{\mathbb I}}

\def\F{{\sf F}}
\def\normal{{\sf N}}

\def\tX{{\widetilde{X}}}

\def\sT{{\sf T}}

\def\tg{\widetilde{g}}

\def\v*{v_i}
\def\T*{T_i}

\def\u*{u_i}
\def\F*{F_i}

\def\tW{\widetilde{W}}

\def\cU{{\mathcal{U}}}

\def\tT{\widetilde{T}}

\def\tx{\tilde{x}}
\def\tT{\widetilde{T}}

\def\bw{{w}}

\def\cL{\mathcal{L}}

\def\th{\theta}
\def\hth{{\widehat{\theta}}}

\def\cX{\mathcal{X}}

\def\hp {\widehat{p}}

\def\l1u{W}

\newcommand{\ajcomment}[1]{}

\makeatletter
\newcommand{\labitem}[2]{%
\def\@itemlabel{\text{#1}}
\item
\def\@currentlabel{#1}\label{#2}}
\makeatother

\DeclareMathAlphabet{\mathpzc}{OT1}{pzc}{m}{it}

\def\cP{\mathcal{P}}

\def\heta{{\hat{\eta}}}

\def\tw{\widetilde{w}}

\def\hp{\widehat{p}}
\def\normal{\mathsf{N}}
\def\vphi{\varphi}
\def\tv{\mathsf{TV}}

\def\heta{\widehat{\eta}}
\def\bern{\mathsf{Bern}}
\def\kell{\mathsf{KL}}
\def\unif{\mathsf{Unif}}
\def\CE{\mathsf{CE}}

\newcommand{\kl}{\mathsf{KL}}

\def \multi{\mathsf{multi}}

\def \cN{\mathcal{N}}

\def \bx {\mathbf{x}}
\def \bv {\mathbf{v}}
\def \bA {\mathbf{A}}
\def \bw {\mathbf{w}}

\def \bV {\mathbf{V}}
\def \tp{\widetilde{p}}
\def \tq{\widetilde{q}}
\def \cG{\mathcal{G}}
\def \hy{\widehat{y}}
\def\hpi{\widehat{\pi}}

\def \asym{\mathsf{asym}}
\def\finite{\mathsf{finite}}
\def\gan{\mathsf{GAN}}
\def\agn{\mathsf{AGN}}
\def\grasp{GRASP\;}

\begin{document}
\title{\bf GRASP: A Goodness-of-Fit Test for Classification Learning}

\author{ 
Adel Javanmard\thanks{Data Sciences and Operations Department, University
of Southern California} \thanks{A.~Javanmard is partially supported by the Sloan Research Fellowship
in mathematics, an Adobe Data Science Faculty Research Award and the
NSF CAREER Awards DMS-1844481 and DMS-2311024.} \and 
Mohammad Mehrabi\footnotemark[1] \thanks{The names of the authors are in alphabetical order. }
}
\maketitle
\begin{abstract}
Performance of classifiers is often measured in terms of average accuracy on test data. Despite being a standard measure, average accuracy fails in characterizing the fit of the model to the underlying conditional law of labels given the features vector ($Y|X$), e.g. due to model misspecification, over fitting, and high-dimensionality. In this paper, we consider the fundamental problem of  assessing the goodness-of-fit for a general binary classifier. Our framework does not make any parametric assumption on the conditional law $Y|X$, and treats that as a black box oracle model which can be accessed only through queries.  We formulate the goodness-of-fit assessment problem as a tolerance hypothesis testing of the form
\[
H_0: \E\Big[D_f\Big(\bern(\eta(X))\|\bern(\heta(X))\Big)\Big]\leq \tau\,,
\]
where $D_f$ represents an $f$-divergence function, and $\eta(x)$, $\heta(x)$ respectively denote the true and an estimate likelihood for a feature vector $x$ admitting a positive label. We propose a novel test, called \grasp for testing $H_0$, which works in finite sample settings, no matter the features (distribution-free).
We also propose model-X \grasp designed for model-X settings where the joint distribution of the features vector is known. Model-X \grasp uses this distributional information to achieve better power. We evaluate the performance of our tests through extensive numerical experiments.
\end{abstract}
\section{Introduction}
In classification learning, one is given a set of training data $\{(x_i,y_i)\}_{i\le n}$ (with $x_i$  representing multi-dimensional feature vector and $y_i$ representing label variables), and aims to learn a model which can be used to predict the labels on new feature vectors. Classification algorithms are backbone of machine learning systems and undoubtedly one of the prominent statistical learning tools in data processing. There has been a plethora of classification methods proposed in the literature ranging from logistic regression and generalized linear models to more complex models such as boosting, random forests, and neural networks. In practice, the performance of these methods is often assessed in terms of accuracy on a test (hold-out) dataset, with the hope that it is a good indicator of the predictive performance of the model on unseen data points. Despite being used widely, the classification accuracy alone does not necessarily characterize the deviations of the learnt model from the underlying data generating process. Indeed relying solely on it as a measure of performance can be misleading due to model misspecification, and over/under fitting. This leads to the following fundamental question:\\

\textit{(*) How well does a classifier learn the ground truth data generating law between the feature vector and the label? }\\

%

Developing a statistical test for the above question has a myriad of applications. It can flag the inherent and systematic flaws of a model, and its poor generalization to unseen population. Further, it provides a more holistic and honest assessment of the model performance, which is of paramount importance with the rise of reproducibility issues in modern data analysis. In particular, with the practice of data sharing, many datasets are used routinely as benchmark to compare different models. However, perpetual use of public datasets, without proper mechanism to access them to ensure validity of inferences, causes spurious discoveries and overfitting; learning models which performs well on benchmark datasets but generalize poorly to unseen datasets~\cite{rosset2014novel,javanmard2018online,dwork2015preserving,dwork2017guilt}. Another application of such test is for models built by commercial ``machine learning as a service'' providers such as Google and Amazon. They provide platforms where one can upload a dataset and a data classification task and pay  to construct a model. Therefore, it is important to decide if the current model is  sufficiently good (in a statistically sound sense) or to continue training process, which incurs additional cost. 
{\color{black}{Among other applications, the proposed methodology can be used in $K$-fold framework to choose the optimal model with respect to its goodness-of-fit, rather than its empirical accuracy on the hold-out set. }}

In statistics, question (*) is often formulated as goodness-of-fit test. However, most of the existing literature focus on specific parametric models, such as logistic regression, and do not apply to more complicated models such as neural network or random forest. An exception is the very recent seminal work of \cite{zhang2021classification}. We refer to Section~\ref{sec:related} for a detailed discussion.

In this paper, we develop a goodness-of-fit test for a broad class of data generating rules (unknown) and black-box models, with only query access. We propose a novel methodology named \grasp (short for {\bf G}oodness-of-fit with {\bf Ra}ndomization and {\bf S}coring {\bf P}rocedure) for this end, which controls type I error in finite sample settings, no matter the features, and does not make any parametric assumption (distribution-free). We also propose a (less conservative) variant test which comes with asymptotic validity. 
Both variants show high power in identifying  deviations of the classification procedure from the true conditional law of the labels.
We also consider model-X settings where no knowledge of the conditional distribution labels ($Y|X$) is assumed, but we do assume the joint distribution of the features $X$ is known, e.g., by having access to abundant unlabeled data. This setting has been studied in several recent work on variable selection and conditional testing; see ~\cite{candes2018panning,barber2020robust,bates2021metropolized,javanmard2021pearson} for a non-exhaustive list. We propose model-X \grasp which leverages this information to obtain a better statistical power. 

Our focus will be on the binary classification setup. We denote the feature vectors  by $x\in \cX\subset\reals^d$ and binary labels by $y\in \{0,1\}$, and define the underlying conditional rule as $\eta(x)=\prob(Y=+1|X=x)$.  This simply reflects the likelihood of a feature vector $x$ admitting the positive label. We have access to $\eta(x)$ only through a set of queries
$\cD$ consisting of $n$ samples drawn i.i.d. from a common law $\cP$ over $\cX\times \{0,1\}$.  We are provided with an estimate model $\heta:\cX\rightarrow [0,1]$, and we would like to assess the performance of this model. We assume that this model has been trained on a dataset disjoint from $\cD$ ($\cD$ has no share in the training procedure.) The training mechanism is optional, e.g., can be a neural network or decision tree, among many others. 
As a first step towards answering (*), we impose the following hypothesis testing problem:
\begin{align}\label{eq:GoTtest}
H_0&: \E[D_f(\bern(\eta(X))\|\bern(\heta(X))]\leq \tau\,,
\end{align}
where expectation is with respect to the distribution of $X$.
The $f$-divergence $D$ is a given metric to measure the distance between $\heta(\cdot)$ and $\eta(\cdot)$ (See equation~\eqref{eq: bern-f} for formal definition). Different divergence functions $f$ lead to different distance measures. For instance, setting $f(t) =\frac{1}{2} |t-1|$ gives us the total variation distance $\E\left[|\heta(X)-\eta(X)|\right]$. By letting  $\tau$ be zero, one can test for the perfect alignment of the test model $\heta$ and the ground truth rule $\eta$. 


\subsection{Related works}\label{sec:related}
\noindent\textbf{Goodness-of-fit.} Once a model is fitted to data, it is important to assess the quality of the fit. Several methods have been developed for testing goodness-of-fit of generalized linear models under the low-dimensional setting ($d\ll n$) with  a focus on logistic regression and multinomial models \cite{hosmer1980goodness, le1991goodness, tsiatis1980note, su1991lack, lin2002model,hosmer2002goodness, osius1992normal, farrington1996assessing}. For the high-dimensional setting, \cite{shah2018goodness} proposed a framework for testing goodness-of-fit  of high-dimensional linear models by using parametric bootstrap to calibrate the estimate model for scaled residuals. \cite{jankova2020goodness} proposed the generalized residual prediction (GRP) test for goodness-of-fit testing of high-dimensional generalized linear models. The aforementioned works focus on a class of parametric models, particularly for linear and generalized linear models. For a non-parametric setting, \cite{zhang2021classification} proposed the binary adaptive goodness-of-fit test (BAGofT), where it tests for the perfect match of the test model and the ground-truth conditional probability, in an asymptotic regime. {

While the existing methods for testing goodness-of-fit mostly focus on contexts where the estimate model has asymptotic convergence to the ground truth law, we propose a method that is flexible for arbitrarily complex classifiers independent of their predictive performance. In addition, it can be used under the high-dimensional setting ($d\gg n$) as well and still generates a high resolution p-value for moderately large number of samples.  In addition, we consider the tolerance testing scheme which is more general than the perfect match testing (i.e. $\eta=\heta$, a.s.), making it more useful in practice. Our proposed method also allows to consider a variety of metrics, including the average absolute distance, and the excess cross entropy, among many others.  

{\color{black} We would like to highlight a distinction between our problem and the conventional setup for testing goodness-of-fit. In our setup, the model estimate $\heta$ is learned on the training data and is subsequently evaluated on an independent test data. In hypothesis \eqref{eq: null}, the model estimate $\heta$ is fixed and the randomness stems solely from the test data. This differs from other goodness-of-fit setups, such as those described in \cite{zhang2021classification, jankova2020goodness}, where the statistical inference takes into account the variability of the training set.  For example, the phrase `finite sample size' in those work refers to the size of the training data, while in our setting it refers to the size of the test data.\smallskip}

\noindent\textbf{Prediction error.} Estimating the prediction performance of a model is one of the core tasks in data-driven applications \cite{hastie2009elements}. In particular, it can serve as a benchmark for model selection. Covariance penalty (CP) and cross validation (CV) are two of widely used methods to estimate the prediction performance of a regression model \cite{akaike1998information, mallows2000some, efron1986biased, efron2004estimation}. However, It has been shown recently that such methods are not statistically accurate in measuring the out-of-the sample performance of models. In fact, \cite{rosset2019fixed, wager2020cross} showed that covariance penalty and AIC statistic \cite{akaike1998information} methods reflect the in-sample prediction error. In addition, \cite{bates2021cross} revealed potential inaccuracies of cross validation, and argued that it indeed measures \emph{average} prediction accuracy over many \emph{hypothetical} datasets. In this work, we take another perspective on model's prediction error. For a broad class of models (e.g., neural networks, decision trees, boosting algorithms, etc), and for the widely used negative likelihood loss (a.k.a. cross entropy), our framework allows to compare the model loss with the optimal oracle loss. Formally, for the estimate probabilistic model $\hp_{Y|X}(y|x)$, for specific choice of $f$-divergence function, we can test the null hypothesis testing:

\begin{align*}
\E_{(x,y)\sim\cP}[-\log \hp_{Y|X}(y|x)]- \min_{\tp_{Y|X}}\left\{\E_{(x,y)\sim\cP}[-\log \tp_{Y|X}(y|x)]\right\}\leq \tau\,.
\end{align*}
Note that the minimum for the second term is achieved at $\tilde{p}_{Y|X} = p_{Y|X}$.
\smallskip

\noindent\textbf{Calibration.} With the rise of deploying machine learning systems in real-life, the confidence of these systems in their predictions is of a great importance. Classification procedures often output a confidence value $\hpi\in [0,1]$ along with their predicted value $\hy$, which is supposed to indicate the model certainty about $\hy$ being equal to $y$ (correct prediction). Calibration of a model refers to aligning such certainty with its long-run accuracy. A rather surprising observation made recently is that many modern machine learning methods are not well calibrated~\cite{guo2017calibration,nixon2019measuring, vaicenavicius2019evaluating,widmann2019calibration,kumar2019verified}. Expected calibration error (ECE) \cite{naeini2015obtaining} is a well-known metric to measure the calibration of models, where it is formally defined as $\E_{\hpi}[|\prob(\hy=y|\hpi)-\hpi| ]$. 


In \cite{lee2022t}, authors proposed a method for testing for the perfect calibration of generic predictive models in multi-class classification problems. The initial intuition behind model calibration is to test for the closeness of classification models to the ground truth law, but mathematically has a different formulation for the goodness-of-fit testing problem considered in the current work. In an extreme instance, a model is well calibrated if $\E[\eta(X)|\heta(X) = \eta]=\eta$ (regression setting), but in \eqref{eq:GoTtest} with e.g, the average absolute difference metric, a perfect fit ($\tau$=0) corresponds to $\heta(X)=\eta(X)$, almost surely. 
 
 {\color{black}
 \smallskip
 
\noindent\textbf{Hypothesis testing for nonparametric regression functions.} In \cite{mukherjee2018optimal}, the authors construct confidence sets for the regression function in nonparametric binary regression with an unknown design density. It is assumed that the the regression function $\eta(.)$ and the marginal probability density function of the features  belong to a continuous class of Sobolev type spaces. Other than results on adaptive parameter estimation, this work provides a framework for testing the null hypothesis that the regression function is equal to 1/2 versus its alternative, while allowing the marginal density function to be a general function in a Sobolev type space. Furthermore, it is shown that 
the complexity of the null hypothesis does not affect the minimal rate of separation between the null and the
alternative.
More on this line of research, \cite{lepski1999minimax} considers the detection problem for a response function $f$ in a stochastic model $\de X(t)=f(t)\de t+\eps \de W(t)$ with $W(t)$ being the standard Wiener process. The detection problem corresponds to testing the null hypothesis $f=0$ against its alternative. In addition, \cite{ingster2009minimax} considers simple hypothesis testing of the form $f=f_0$ in a multivariate setup $y=f(x)+\eps$ with isotropic Gaussian noise in $[0,1]^d$. We refer to \cite{ingster2003nonparametric} for more details on nonparametric hypothesis testing on stochastic Gaussian models. }

\subsection{Summary of contributions and organization}
In this paper, we introduce a novel method for tolerance testing of average distance of classifiers to the underlying conditional law of labels. Our proposed method, called \grasp can be used for arbitrarily complex black-box models (with only query access), with no parametric assumption whatsoever, and no matter the features. \grasp can be used in high-dimensional setting where the features dimension can exceed the sample size. We propose two variants of \grasp, one controls the type I error in finite sample settings, and the other (less conservative) version controls the type I error in asymptotic setting.  Through extensive numerical experiments, we show that \grasp achieves high statistical power. 

For model-X settings, where the distribution of features can be well approximated, we propose model-X \grasp which is built upon similar ideas as in the distribution-free version but harnesses the knowledge of features distribution to improve statistical power. 

Here is an outline of the next sections:
\begin{itemize}
\item Section \ref{section: problem-formulation}: We first provide a brief review on $f$-divergence functions, and then formulate the goodness-of-fit test as a tolerance hypothesis testing problem. We end this section by a short review on some convex analysis definitions. 
\item Section \ref{section: df}: We start by focusing on the distribution-free regime, and propose high-level intuitions behind the \grasp algorithm. We next move to formally introduce \grasp procedure along with its test statistics and decision rules. Next we prove that the type I error of \grasp (size of the test) can be controlled in finite sample settings, for arbitrary data generative rules, and general classifiers.  We also introduce a less conservative version of \grasp that has asymptotic control on the type I error.  
The \grasp test uses a score function $T$ in forming the test statistics. It is worth noting that the size of the test is controlled for arbitrary score functions. The choice of score function though impacts the power of the test and we will discuss some choices in Section~\ref{sec:scoreF}. We conclude Section \ref{section: df} by characterizing one-sided confidence intervals and $p$-values for hypothesis \eqref{eq:GoTtest}.

\item Section \ref{section: model-X}: We move to the model-X setting, where abundant unlabeled data points are available. We propose model-X \grasp that uses this data to learn the features distribution and  utilize it for a statistically more powerful procedure. Similar to the distribution-free setting, we show that the size of the test is controlled under the pre-determined level, for finite number of data points. Further, a less conservative decision rule is introduced which has asymptotic control on the type I error. 

In Section~\ref{sec:scoreF} we discuss the role of score function $T$ on the power. We derive the form of the optimal score function, which depends on the data generative law $\eta(x)$.
We discuss two approaches: $(i)$ \emph{model-agnostic} which replaces $\eta(x)$ by $1/2$ (random guessing) in the optimal score function; $(ii)$ \emph{GAN-based} approach 
which uses generative adversarial networks (GANs) to estimate the required densities to use in the optimal score function. 

\item Section \ref{section: numerical}: We provide extensive numerical experiments to evaluate the performance of the distribution-free and model-X \grasp (type I control), the power of these tests and the advantage of model-X framework in achieving a higher power, as well as the impact of the score function $T$ on the power. 
\end{itemize}
 
\subsection{Notation} For an integer $k$ let $[k]$ stand for the set $\{1,2,\dots,k\}$. We denote the distribution of a Bernoulli random variable with success parameter $p$ with $\bern(p)$, and let $\multi(p_1,\dots,p_L)$ denote a multinoulli distribution with $L$ categories, where the probability of observing category $\ell\in [L]$ is $p_\ell$. For the case of $p_1=\dots=p_L=1/L$ we use the shorthand $\multi(L)$. In addition, for positive real values $a<b$,  let $\unif([a,b])$ indicate the uniform distribution on the interval $[a,b]$. For a random variable $X$, we write $\cL(X)$ to refer to the probability density function of $X$. We denote the density function of a chi-squared distribution with $L$ degrees of freedom by $\chi^2_L$, and let $\chi^2_{L}(\beta)$ represents the $\beta$-th quantile of an $\chi^2_L$ distribution. {\color{black}{We use upper case letters for random variables, and lower case letters to indicate deterministic values, e.g, realizations of a random variable.}} We drop the subscript under the expectation with showing the corresponding distribution, whenever it is clear from the context.

\section{Problem formulation}\label{section: problem-formulation}

Under a binary classification setting, for a given test model $\heta:\cX\to [0,1]$, we are interested in measuring its average distance to the oracle model $\eta(x)=\prob(y=+1|x)$. The test model $\heta(\cdot)$ can be any arbitrarily complex predictive model, such as a fitted logistic regression, or the last layer of a trained neural network. 
We focus on a class of distances between $\eta$ and $\heta$ which are inspired by $f$-divergence of distributions. We start by the definition of $f$-divergence of two density functions. 

\begin{defi}($f$-divergence)\label{def: f-div}
Consider a convex and continuous function $f:\reals \to \reals$. For two probability density functions $p, q$ that are defined with respect to the Lebesgue measure $\mu$ over $\cX\subset \reals^d$, define the $f$-divergence between them as the following
\[
D_f(p\|q)=\int q f\left(\frac{p}{q}\right) \de \mu\,.
\]
\end{defi}

Specializing this definition to Bernoulli distributions, with parameters $a,b$, we obtain the following definition:
%
\begin{equation}\label{eq: bern-f}
D_f(\bern(a)\|\bern(b))=bf\left(\frac{a}{b} \right)+(1-b)f\left(\frac{1-a}{1-b} \right)\,.
\end{equation}
For the rest of this paper, we will focus on the class of measures $D_f(\bern(\eta(x))\|\bern(\heta(x)))$ parametrized by function $f$ as distance between the test model $\heta$ and the ground truth model $\eta$. For some nonnegative value $\tau$, we consider the following hypothesis testing problem:
\begin{align}
H_0&: \E\left[D_f\Big(\bern(\eta(X))\|\bern(\heta(X))\Big)\right]\leq \tau\nonumber\\
H_A&:\E\left[D_f\Big(\bern(\eta(X))\|\bern(\heta(X))\Big)\right]> \tau\label{eq: null}\,.
\end{align}
with $H_0$ representing the null hypothesis and $H_A$ the alternative.



\begin{lemma} \label{lemma: f-functions}The followings hold:
\begin{enumerate}
\item By choosing $f(t)=\frac{1}{2}|t-1|$ (total variation distance) we get
\[
\E\left[D_{\tv}(\bern(\eta(X))\|\bern(\heta(X)))\right]=\E[|\heta(X)-\eta(X)|]\,.
\]
\item 
For the choice of $f(t)=t\log t$ (KL divergence), we get
\[
\E\left[D_{\mathsf{KL}}(\bern(\eta(X))\|\bern(\heta(X)))\right]=\CE(\heta)-\CE(\eta)\,,
\]
where for a model $\heta(x):\cX\to[0,1]$, its cross entropy loss  is given by
\[
\CE(\heta)=-\E[\eta(X)\log \heta(X)+(1-\eta(X))\log (1-\heta(X)) ]\,.
\]
The minimum cross-entropy loss is achieved when $\heta=\eta$.
\item For the choice of $f(t)=(\sqrt{t}-1)^2$ (Hellinger distance), we get
\begin{align*}
\E\left[D_{\mathsf{H}}(\bern(\eta(X))\|\bern(\heta(X)) )\right]=\E\left[\Big(\sqrt{\eta(X)}-\sqrt{\heta(X)}\Big)^2+\Big(\sqrt{1-\eta(X)}-\sqrt{1-\heta(X)}\Big)^2 \right]\,.
\end{align*}
\end{enumerate}

\end{lemma}
The Proof of Lemma \ref{lemma: f-functions} is given in Section \ref{proof: lemma: f-functions}. It is worth noting that by considering different $f$, our framework allows to  for a variety  distance measures between oracle model $\eta$ and the estimate model $\heta$. 
We conclude this section by  two definitions that will be used later in Section \ref{section: model-X}.

\begin{defi}(subdifferential)
For a convex function $f:\reals\to \reals$, the subdifferential $\partial f(t)$ at a point $t$ is given by the following set of real values
\[
\partial f(t)=\{u\in \reals: f(s)-f(t)\geq u(s-t)\,, \forall s\in \reals   \}\,.
\] 
In addition, for differentiable $f$, we have $\partial f(t)=f'(t)$.
\end{defi}

\begin{defi}(conjugate dual)\label{def: conjugate}
The \textit{conjugate dual} function of $f:\reals\to \reals$ is defined as 
\[
f^*(t)=\sup_{s\in \reals}\; (st-f(s))\,.
\]
In addition, for convex lower semi-continuous $f$, we can write $f$ in terms of its conjugate dual as
\[
f(s)=\sup_{t\in \reals}\; (st-f^*(t))\,.
\]
\end{defi}

\section{Distribution-free setting} \label{section: df}
In this section, we develop a new methodology for testing the null hypothesis in \eqref{eq: null} without imposing any structure on the conditional law $\eta(x)$,  covariates $x$ distribution, or predictive model $\heta(x)$. In order to provide a high level intuition behind the main idea of the procedure, we first introduce a sampling scheme which characterizes the $f$-divergence between the models $\eta$ and $\heta$ as the conditional $f$-divergence distance of a sampled value and $\unif[0,1]$ distribution. 
\begin{propo}\label{propo: f}
For $(x,y)\sim \cP$ let
\begin{equation}\label{eq: w}
 w=\begin{cases} 
u_1\sim\unif[0,\heta(x)],& \quad  y=1\,,\\
u_2\sim \unif[\heta(x),1],&\quad y=0\,.\\
\end{cases}
\end{equation}
Then we have
\[
\E_X\left[D_{f}(\cL_{W|X}\|\unif([0,1]))\right]=\E_X[D_f(\bern(\eta(X))\|\bern(\heta(X)))]\,.
\]
As a special case, if $\heta = \eta$ then $W|X\sim \unif([0,1])$.
\end{propo}
The proof of Proposition \ref{propo: f} is given in Section \ref{proof: propo: f}. Proposition  \ref{propo: f} implies that deviation of sampled $w$ from the uniform $[0,1]$ can be counted as evidence for large distance of $\eta$ and $\heta$.  In the next section we elaborate the formal process to construct a set of statistics that will be used later to define  the decision rules for testing null hypothesis \eqref{eq: null}.

\subsection{Test statistic} 
{
Algorithm \ref{algorithm: balls-bins} describes the procedure for constructing the test statistic $\bV_{n,L}$. The construction consists in two main steps:
\newline\noindent\textbf{Counterfeit sampling.} For each data point $(x,y)$, we follow the 
procedure described in \eqref{eq: w}  and construct the sampled value $w$. We then construct randomizations $\tw_1,\dots\tw_M$ from the uniform distribution $[0,1]$. 
\newline\noindent\textbf{Score and label.} In this step, by using a score function $T: \cX\times [0,1] \to \reals$, we try to score the original sample $T(x,w)$ and corresponding values $T(x,\tw_j)$. Then the original data point $(x,y)$ will be labeled based on the relative location of $T(x,w)$ among the counterfeit values $\{T(x,\tw_j)\}_{j=1:M}$. The final output statistic $\bV_{n,L}\in \reals^L$ denote the count of each label among the whole $n$ samples. 
}
 Note that the number of labels $L$ is an input parameter in Algorithm \ref{algorithm: balls-bins}. 
The construction of test statistic $\bV_{n,L}$ is inspired by the PCR test proposed by \cite{javanmard2021pearson}, where a similar structure (counterfeit sampling-scoring-labeling) is used but for a different task, namely to test conditional independence between variables. Also in  \cite{javanmard2021pearson}, the counterfeits are drawn from a conditional distribution (depending on variables under test), while here the counterfeits are drawn from $\unif([0,1])$, no matter the features or labels.

\begin{algorithmic}[t]
	\begin{algorithm}
	{
		\SetAlgoLined
		\REQUIRE $n$ data points $(x_j,y_j)\in \cX \times \{0,+1\}$, the model $\heta:\cX\to[0,1]$, a score function $T: \cX \times [0,1] \to \reals$,  and integers $M,L\geq1$ such that $M+1=KL$ for some integer $K$. 
		\ENSURE Test statistics $\bV_{n,L}=[V_1,...,V_L]\in \reals^{L}$. \\
			\For{$j\in \{ 1,2,...,n\}$}{
			\begin{itemize}
			\item  Let  \[
			          w_j=\begin{cases} 
                                  u_1\sim \mathsf{unif}[0,\heta(x_j)],& \quad y_j=1\,,\\
                                 u_2\sim \mathsf{unif}[\heta(x_j),1],&\quad y_j=0\,.\\
                                  \end{cases}
                                  \]

			\item  Draw $M$ i.i.d. samples $\tw_j^{(1)},..., \tw_j^{(M)}$ from $\unif[0,1]$.
                         \item Use score function $T$ to score the initial sample $(x_j,w_j)$ and its $M$ constructed counterfeits $\{(x_j,\tw_j^{(1)}),..., (x_j,\tw_j^{(M)})\}$:
                         \begin{align*}
                         T_j&=T(x_j,w_j)\\
                         \tT_j^{(i)}&=T(x_j,\tw_j^{(i)})\,, \quad \text{for } i\in [M]\,.
                         \end{align*}
 	    \item Let $R_j$ denote the rank of $T_j$ among $\{T_j,\tT_j^{(1)},..., \tT_j^{(M)}\} $:
	    \[
	  R_j=  1+\sum\limits_{i=1}^{M}\ind_{\left\{T_j \geq \tT_j^{(i)}\right \}}
	    \]
		\item Assign label $L_j\in \{ 1,2,...,L\}$ to sample $j$ if $ (L_j-1)K+1\leq   R_j \leq KL_j$.
		
				\end{itemize}
}

\For{$\ell\in \{1,2,...,L\}$}{
			\begin{itemize}
				\item Let $V_\ell$ be the number of samples with label $\ell$, i.e.	$V_\ell=\Big|\big \{j\in \{1,2,...,n\}: L_j=\ell \big\}  \Big|\,.$
				
				\end{itemize}
		}
		
				\caption{Construction of GRASP test statistic (Distribution-free setting)}\label{algorithm: balls-bins}
				}
	\end{algorithm}
\end{algorithmic}

\subsection{Decision rule} We introduce two decision rules based on the statistics $\bV_{n,L}=[V_1,\dots V_{L}]$ given by Algorithm \ref{algorithm: balls-bins}. For this end,  we first construct the following two test statistics that will be used later for the decision rules: 
%
%
%
 \begin{eqnarray}\label{eq: test-stats}
\begin{split}
U_{\tau}^{\mathsf{finite}}(\bV_{n,L})=\min_{p \in \reals^L} \quad & \frac{1}{n}\sum\limits_{\ell=1}^{L}\frac{\left(V_\ell-np_\ell\right)^2}{p_\ell+ \frac{1}{L}}\\
			\text{s.t} \quad &p_\ell\geq 0\,, \quad \forall \ell\in[L]\,,\quad
			\sum\limits_{\ell=1}^{L}p_\ell=1\,,\quad \frac{1}{L}\sum_{\ell=1}^L f(Lp_\ell)\leq \tau \,. \\  
			\\
{U}_{\tau}^{\mathsf{asym}}(\bV_{n,L})=\min_{p \in \reals^L} \quad & \frac{1}{n}\sum\limits_{\ell=1}^{L}\frac{\left(V_\ell-np_\ell\right)^2}{p_\ell}\\
			\text{s.t} \quad &p_\ell\geq 0\,, \quad \forall \ell\in[L]\,,\quad
			\sum\limits_{\ell=1}^{L}p_\ell=1\,,\quad \frac{1}{L}\sum_{\ell=1}^L f(Lp_\ell)\leq \tau \,. 
\end{split}
 \end{eqnarray}

Note that the choice of function $f$ and the value of $\tau$ in the test statistics ${U}_{\tau}^{\asym}(\bV_{n,L})$ and ${U}_{\tau}^{\finite}(\bV_{n,L})$ are determined by the null hypothesis \eqref{eq: null}. We next consider the following two decision rules based on \eqref{eq: test-stats}. At the significance level $\alpha$, the decision rules are given by
 \begin{equation}\label{eq: decision-asym}
 \Phi^{\asym}_{n,L,\alpha,\tau}=\begin{cases} 1 ~~(\text{reject } H_0)\,, & \text{if }  {U}_{\tau}^{\asym}(\bV_{n,L}) \geq \chi^2_{L-1}(1-\alpha) \,,\\
 0~~(\text{fail to reject } H_0)\,, & \text{otherwise }  \,.\\
 \end{cases}
 \end{equation}
 The next decision rule is given by
 \begin{equation}\label{eq: decision-finite}
 \Phi^{\finite}_{n,L,\alpha,\tau}=\begin{cases} 1 ~~(\text{reject } H_0)\,, &  \text{ if }{U}_{\tau}^{\finite}(\bV_{n,L}) \geq L+\sqrt{\frac{2L}{\alpha}}\,,\\
 0 ~~(\text{fail to reject } H_0)\,, & \text{otherwise } \,.\\
 \end{cases}
 \end{equation}
 
 The rationale behind the $\asym$ and $\finite$ names comes from the fact that we show  that later they have asymptotic and finite-sample control guarantees for type I error, respectively. 
 
As we show in the next lemma, the asymptotic rule is less conservative than the finite rule.
\begin{propo}\label{propo: compare}
For $L$ sufficiently large, the asymptotic decision rule \eqref{eq: decision-asym} rejects more than the finite decision rule \eqref{eq: decision-finite}. More precisely, for $L\geq 55$ and every $\alpha\in (0,1)$, if $\Phi_{n,L,\alpha,
\tau}^{\finite} =1$ then $\Phi_{n,L,\alpha,\tau}^{\asym} =1$. 
 \end{propo}
We refer to Section \ref{proof: propo: compare} for the proof of Proposition \ref{propo: compare}. 

We conclude this section by providing some insight behind the test statistics. Consider the case of perfect alignment ($\tau = 0$), for which we showed $w_j$ are uniform in $[0,1]$.
Therefore, under the null each sample is identically distributed as its counterfeits and its label, assigned in the last step of Algorithm~\ref{algorithm: balls-bins}, follows $\multi(L)$ distribution. The test statistics in~\eqref{eq: test-stats} reduces to the Pearson's chi-square test statistic. When $\tau>0$, we will show in Theorem~\ref{thm: discretize-distance} that the labels follow a multinoulli distribution $\multi(p_1,\dotsc, p_L)$ which should be close to the uniform multinoulli distribution $\multi(L)$ in the sense that $\frac{1}{L}\sum_{\ell=1}^{L}f(Lp_\ell)\le \tau$.
However, the nominal probabilities $p_\ell$ are unknown and so in our construction of test statistics~\eqref{eq: test-stats}, we form an optimization problem over $p_\ell$ to impose this constraint and then consider the minimum Pearson's chi-square test value over the feasible probability vectors $(p_1,\dotsc, p_{L})$.

\subsection{Size of the test}
We will show that the proposed decision rules control the size of the test under the pre-assigned level $\alpha$. We first state the following technical assumption which posits a density function for specific conditional law for $T(X,W)$ and $T(X,\tW)$. 
\begin{assumption} \label{assumption: cdfs} Consider the following conditional cumulative distribution functions:
\begin{align*}
F_o(t;x)&=\prob(T(X,W)\leq t|X=x)\,,\\
F_c(t;x)&=\prob(T(X,\tW)\leq t|X=x)\,.
\end{align*}

where $(X,W)$ is is given by~\eqref{eq: w}, and $\tW$ is drawn independently from $\unif[0,1]$. Assume that the density functions of these cdfs exist and indicate them by $f_o(t;x)$ and $f_c(t;x)$. In addition, a new measure on the interval $[0,1]$ can be considered with cdf $\omega(u;x):=F_o(F_c^{-1}(u;x);x)$. Suppose that this measure is absolutely continuous with respect to the Lebesgue measure, and denote its density function (Radon–Nikodym derivative) by $\psi(u;x)$. 
\end{assumption}

\begin{propo}[distribution-free]\label{propo: discretize-distance}
 Let $\bV_{n,L}=[V_1,...,V_L]$ be the output of Algorithm \ref{algorithm: balls-bins}, then $\bV_{n,L}$  has a multinomial distribution with $L$ outcomes and nominal probabilities $p_1,\dots,p_L$. In addition,  under Assumption \ref{assumption: cdfs} the following holds:
  \begin{align*}
\frac{1}{L}\sum\limits_{\ell=1}^{L}{f(Lp_\ell)} &\leq \E\left[D_f\big(\cL(T(X,W)|X)\|\cL(T(X,\tW)|X)\big)\right]\\
 &\leq \E\left[D_f\big(\cL_{W|X}\|\cL_{\tW|X}\big)\right]\\
 &=\E\left[D_f\big(\cL_{W|X}\|\unif([0,1])\big)\right]\,.
 \end{align*}

\end{propo}
The proof of this Proposition is given in Section \ref{proof: propo: discretize-distance}. It is easy to observe that the second inequality in Proposition \ref{propo: discretize-distance} can be achieved by using the score function $T(x,w)=w$. The next theorem is an immediate consequence of Proposition \ref{propo: f} and Proposition \ref{propo: discretize-distance}. 

\begin{thm}[distribution-free]\label{thm: discretize-distance}
 Let $\bV_{n,L}=[V_1,...,V_L]$ be the output of Algorithm \ref{algorithm: balls-bins} that has a multinomial distribution with nominal probabilities $p_1,\dots p_L$, as per Proposition \ref{propo: discretize-distance}. Then for a score function $T:\cX\times [0,1]\to \reals$ satisfying Assumption \ref{assumption: cdfs}, the following upper bound holds:
 \begin{align*}
\frac{1}{L}\sum_{\ell=1}^{L}{f(Lp_\ell)}   \leq \E\left[D_f\Big(\bern(\eta(X))\|\bern(\heta(X))\Big)\right]\,.
 \end{align*}
 \end{thm}

Note that by definition of $f$-divergence for discrete distributions we have 
\[
D_f(\multi(p_1,\dots,p_L)||\multi(L)) = \frac{1}{L}\sum_{\ell=1}^Lf(Lp_\ell)\,.
\]
Therefore, Theorem \ref{thm: discretize-distance} implies that under the null hypothesis \eqref{eq: null} the $f$-divergence between the multinoulli distribution $\multi(p_1,\dots,p_L)$ and the uniform multinoulli distribution $\multi(L)$ should be bounded by $\tau$. As can be observed from \eqref{eq: test-stats}, our decision rules are based on optimization problems over probability vectors $\{p_{\ell}\}$, which minimizes a chi-squared type statistic subject to the constraint $\frac{1}{L}\sum_{\ell=1}^{L}f(Lp_\ell)\le \tau$. 



In other words, by virtue of Theorem \ref{thm: discretize-distance}, we can reduce the null hypothesis regarding the quantity of interest $\E\left[D_f(\bern(\eta(X))\|\bern(\heta(X)))\right]$ to a hypothesis on the quantity $\frac{1}{L}\sum_{\ell=1}^{L}f(Lp_\ell)$ which is more amenable to test. 

Algorithm \ref{algorithm: simple-balls-bins} outlines a simple version of Algorithm \ref{algorithm: balls-bins} with the score function $T(x,w)=w$. In this case, scores of counterfeits will be uniform random variables on $[0,1]$, and we label the original scores simply by partitioning the $[0,1]$ interval into $L$ subintervals of equal size.  It is worth noting that the second inequality of Proposition \ref{propo: discretize-distance} becomes an equality for the choice of score function $T(x,w)=w$. At first glance, this is expected to always results in a higher power compared to other choices of $T$. This argument is not valid though, since the gap in the first inequality of \ref{thm: discretize-distance} is undetermined, and so we keep both versions (score function $T(x,w)=w$, and general score function) in the paper.

The next result provides control over the size of our test with the two decision rules \eqref{eq: decision-finite} and \eqref{eq: decision-asym}, in the finite sample and asymptotic settings.

\begin{algorithmic}[t]
	\begin{algorithm}
		\SetAlgoLined
		{
		\REQUIRE $n$ data points $(x_j,y_j)\in \cX \times \{0,+1\}$, the model $\heta:\cX\to[0,+1]$,  and an integer $L\geq1$. 
		\ENSURE Test statistic  $\bV_{n,L}=[V_1,...,V_L]\in \reals^L$. \\
			\For{$j\in \{ 1,2,...,n\}$}{
			\begin{itemize}
			\item  Let  \[
			          w_j=\begin{cases} 
                                  u_1\sim \mathsf{unif}[0,\heta(x_j)],& \quad y_j=1\,,\\
                                 u_2\sim \mathsf{unif}[\heta(x_j),1],&\quad y_j=0\,.\\
                                  \end{cases}
                                  \]

		\item Assign label $L_j\in \{ 1,2,...,L\}$ to sample $j$ if $ \frac{L_j-1}{L}\leq  w_j \leq \frac{L_j}{L}$.
		
				\end{itemize}
}

\For{$\ell\in \{1,2,...,L\}$}{
			\begin{itemize}
				\item Let $V_\ell$ be the number of samples with label $\ell$, i.e.	$V_\ell=\Big|\big \{j\in \{1,2,...,n\}: L_j=\ell \big\}  \Big|\,.$
				
				\end{itemize}
		}
		
				\caption{GRASP test statistic (distribution-free setting and with the score function $T(x,w) = w$)}\label{algorithm: simple-balls-bins}
				}
	\end{algorithm}
\end{algorithmic}

\begin{thm}\label{thm: size-DF}
 Let $\bV_{n,L}=[V_1,\dots,V_L]$ be the output of either Algorithms \ref{algorithm: balls-bins} or \ref{algorithm: simple-balls-bins}. Consider the decision rules $\Phi^{\asym}_{n,L,\alpha,\tau}$ and $\Phi^{\finite}_{n,L,\alpha,\tau}$, which are given respectively in \eqref{eq: decision-asym} and \eqref{eq: decision-finite}, with the test statistics $U_{\tau}^{\asym}(\bV_{n,L})$ and $U_{\tau}^{\finite}(\bV_{n,L})$ as per \eqref{eq: test-stats}. Under the null hypothesis \eqref{eq: null}, we have:
 \[
 \prob\left( \Phi^{\finite}_{n,L,\alpha,\tau}=1 \right) \leq \alpha\,,~~\lim\sup_{n\to \infty}\prob\left(  \Phi^{\asym}_{n,L,\alpha,\tau}=1\right) \leq \alpha\,. 
\]
 
 \end{thm}
We refer to Section \ref{proof: hm: size-DF} for the proof of Theorem \ref{thm: size-DF}.
This implies that deploying the decision rule $\Phi^{\finite}_{n,L,\alpha,\tau}$ controls the type I error at level $\alpha$, for every finite $n$. In addition,  the decision rule $\Phi^{\asym}_{n,L,\alpha,\tau}$ has an asymptotic control over the type I error at the significance level $\alpha$.  Let us stress that the result of Theorem \ref{thm: size-DF} is valid for every choice of parameters $K,L$ and score function $T$.

\subsection{P-values and one-sided confidence intervals}

Considering the definition of rules $\Phi^{\asym}_{n,L,\alpha,\tau}$ and $\Phi^{\finite}_{n,L,\alpha,\tau}$ and the results of Theorem \ref{thm: size-DF}, we construct the following p-values for the hypothesis testing problem in \eqref{eq: null}.
\begin{align*}
p_{n,L,\tau}^{\finite}=\begin{cases}
 1\,,  &U_\tau^{\finite}(\bV_{n,L})\leq L \,,\\
 1 \wedge  \frac{2L}{(U_\tau^{\finite}(\bV_{n,L})-L)^2}\,, &\text{ otherwise}\,.
\end{cases}
\end{align*}

\[
p_{n,L,\tau}^{\asym}=1-F_{L-1}(U_{\tau}^{\asym}(\bV_{n,L}))\,,
\]
where $F_{L-1}(t)$ denote the cdf of a Chi-squared random variable with $L-1$ degrees of freedom.  Super-uniformity of these $p$-values, under the null hypothesis, follows simply from Theorem \ref{thm: size-DF}. Formally, for every $t\in [0,1]$ we have
\begin{align*}
\prob\left(p_{n,L,\tau}^{\finite}\leq t \right) \leq t\,, \quad \forall n,L\geq 1\,.\\
\lim_{n\to \infty}\sup\prob\left(p_{n,L,\tau}^{\asym}\leq t \right) \leq t\,, \quad \forall L\geq1\,.
\end{align*}
We next use the duality between confidence intervals and hypothesis testing to construct a one-sided confidence interval for the quantity of interest $\E_x[D_f(\bern(\eta(x))\|\bern(\heta(x)))]$. The intuition behind this construction comes from the fact that for a fixed value $\bV_{n,L}$, the test statistics $U_{\tau}^{\asym}(\bV_{n,L})$ and $U_{\tau}^{\finite}(\bV_{n,L})$ are nonincreasing in $\tau$. 
\begin{propo}\label{propo: CI}
For $\alpha\in (0,1)$, let
\begin{align*}
\tau^{\finite}_{n,L,\alpha}&=\sup\left\{\tau\geq0 : U^{\finite}_{\tau}(\bV_{n,L})\geq L+\sqrt{2L/\alpha}\right\}\,,\\
\tau^{\asym}_{n,L,\alpha}&=\sup\left\{\tau\geq0 : U^{\asym}_{\tau}(\bV_{n,L})\geq \chi^2_{L-1}(1-\alpha)\right\}\,.
\end{align*}
Then the followings hold
\begin{align*}
\prob\Big(\E \Big[D_f\Big(\bern(\eta(X))\|\bern(\heta(X))\Big)\Big]\geq \tau^{\finite}_{n,L,\alpha}\Big) &\geq 1-\alpha\,,\\
\lim_{n\to\infty}\inf \prob\Big(\E\Big[D_f\Big(\bern(\eta(X))\|\bern(\heta(X))\Big)\Big]\geq \tau^{\asym}_{n,L,\alpha}\Big) &\geq 1-\alpha\,.
\end{align*}
Note that the probabilities in the above equations are with respect to the randomness in $\tau^{\finite}_{n,L,\alpha}$ and $\tau^{\asym}_{n,L,\alpha}$, which stem from the randomness in 
$\bV_{n,L}$.
\end{propo}
We refer to Section \ref{proof: propo: CI} for the Proof of Proposition \ref{propo: CI}.

{\color{black}
\subsection{Choice of the score function}

In this section, we provide some insights on the choice of score function $T$ in Algorithm \ref{algorithm: balls-bins}.  First, from Theorem \ref{thm: discretize-distance} we have
\[
 \frac{1}{L}\sum_{\ell=1}^{L}{f(Lp_\ell)} \leq  \E\left[D_f\Big(\bern(\eta(X))\|\bern(\heta(X))\Big)\right]\,.
\]
Note that in Algorithm \ref{algorithm: balls-bins} we test for $\frac{1}{L}\sum_{\ell=1}^{L}{f(Lp_\ell)}  \leq \tau$ (which holds under null), and therefore the statistical power of our method in the first place depends on the gap between the quantity of primary interest  $\E\left[D_f\Big(\bern(\eta(X))\|\bern(\heta(X))\Big)\right]$ and $ \frac{1}{L}\sum_{\ell=1}^{L}{f(Lp_\ell)} $. 


We next explore scenarios under which the gap in the chain of inequalities presented in Proposition \ref{propo: discretize-distance} can be tightened. Recall the chain of inequalities summarized below:
\begin{align}
\frac{1}{L}\sum\limits_{\ell=1}^{L}{f(Lp_\ell)} &\leq \E\left[D_f\big(\cL(T(X,W)|X)\|\cL(T(X,\tW)|X)\big)\right] \nonumber \\
 &\leq \E\left[D_f\big(\cL_{W|X}\|\cL_{\tW|X}\big)\right] \nonumber \\
 &=\E\left[D_f\big(\cL_{W|X}\|\unif([0,1])\big)\right]\nonumber\\
 &=\E_X[D_f(\bern(\eta(X))\|\bern(\heta(X)))]\,. \label{eq: chain-again-df}\,,
 \end{align}
where the last step is proved in Proposition~\ref{propo: f}.
Our goal is to develop a score function that narrows the gap between the left and right-hand sides in the above chain of inequalities. Note that the second inequality becomes tight when $T(x,w)=w$, a simple score function outlined in Algorithm \ref{algorithm: simple-balls-bins}. While it is a straightforward choice, it does not account for possible dependence between $x$ and $w$, and it is also not clear how this choice would impact the first inequality.

To examine the first inequality, we consider the limit case of $K, L\to\infty$  and see if the gap in the first inequality becomes tight in this asymptotic case. Our next result answers this question in the negative. In contrast, we later answer the similar question in the positive for the model-X setup. Further details can be found in Proposition \ref{propo: model-x}.

\begin{propo}\label{propo: K_L grows-df}
Recall the density function $\psi(u,X)$ from Assumption~\ref{assumption: cdfs}.
Consider the similar setup of Proposition \ref{propo: discretize-distance} with the following two additional assumptions:
i) For a positive value $C$ we have $\psi(u;X)\leq C$ a.s. over [0,1], and  ii) $\psi(u,X)$ is differentiable on $(0,1)$ and there exists a positive constant $B$ such that a.s. we have $|\frac{\partial }{\partial u} \psi(u,X)| \leq B$.  Then the following holds as $K,L$ grow to infinity: 
\begin{align}
 \lim_{L\to \infty}\lim_{K\to \infty}  \frac{1}{L}\sum_{\ell=1}^L f(Lp_\ell)&=\int_0^1 f(\E_X[\psi(u;X)])\de u\nonumber\\
 &\leq \E_X\left[ \int_0^1 f(\psi(u;X))\de u\right]\nonumber\\
 &=\E_X\left[D_f\big(\cL(T(X,W)|X)\|\cL(T(X,\tW)|X)\big)\right]\,.\label{eq:Jensen}
 \end{align}
 \end{propo}

We refer to Section \ref{proof: propo: K_L grows-df} for the proof of Proposition \ref{propo: K_L grows-df}.

The above Proposition highlights an important issue: even with $K,L\to\infty$, we can still have a gap in the first inequality in~\eqref{eq: chain-again-df} for any strictly convex  divergence function $f$, no matter the choice of score function $T$. One can try to minimize this gap for some $T$, but apart from being a challenging question it is not clear how this choice would impact the second inequality in~\eqref{eq: chain-again-df}.  

In Section~\ref{section: numerical}, Experiment~\ref{ex:w}, we evaluate the performance of our test using $T(x,w) = w$. We repeat the same numerical study in Experiment~\ref{ex:res} using another score function which also depends on $x$. Concretely, we regress $w$ on to $x$, and take $T(x,w) = |w - x^\sT \hth|$ the residuals.  As we see the former choice of $T(x,w)$ slightly outperforms the regression-type one in power, for different choices of $f$-divergence. These experiments indicate that the simple choice of $T(x,w) = w$ can be competitive and in general other more complicated score score which takes into account the dependence between $w$ and $x$ may achieve a lower power. 

\subsection{Testing perfect fit of the model}
By setting $\tau = 0$ in hypothesis~\eqref{eq:GoTtest}, our framework allows to test whether we have perfect fit for the model at hand $\heta$. We next provide a modification of our test tailored for this special case, which has demonstrated higher power in our numerical experiments. The main adjustment to the framework is that the score functions are now defined at \emph{dataset level}, rather than \emph{sample level}.  Let $\bx,\bw$ respectively stand for $\{x_i\}_{i=1:n}$, $\{w_i\}_{i=1:n}$ with sampling process for $w_i$ described in \eqref{eq: w}. In addition, for $j\in [M]$ we define $\bw_j$ to be a vector of size $n$ with iid samples drawn from $\unif([0,1])$. By Proposition~\ref{propo: f}, under the null, we have $W|X\sim \unif([0,1])$ and therefore, $T(\bx,\bw), T(\bx,\bw_1), \dotsc, T(\bx,\bw_M)$ are exchangeable. Using this observation, we construct the following $p$-value for the null:
\begin{equation}\label{eq: pval-crt}
p=\frac{1+\sum\limits_{j=1}^{M}\ind\left(T(\bx,\bw)\geq T(\bx,\bw_j) \right)}{M+1}\,,
\end{equation}
Using the exchangeability property, in the next proposition we show that the above $p$-value is super uniform under null and hence we have control on the size of the test.

The above construction  of $p$-value is inspired by the Conditional Randomization Test (CRT) \cite{candes2018panning}. Although the choice of score function $T$ is optional, it is recommended to use a score function that captures the variation between $w$ and $x$, and hence $T(\bx,\bw)$ is smaller than most of its counterfeits $T(\bx,\bw_j)$, resulting in a small $p$-value under the alternative hypothesis. One simple choice is to use the residual (e.g., mean squared error) of a linear regression model when regressing variable $w$ on $x$. Alternatively, more complex predictive methods such as Lasso, random forest, or neural networks can be used to fit a model, and the residual of the fitted model can be reported as the score value. 


Our next proposition shows the super-uniformity of p-value~\eqref{eq: pval-crt} under the null, and its proof is deferred to Section \ref{proof: propo: pval-crt}.
\begin{propo}\label{propo: pval-crt}
Under the null hypothesis \eqref{eq: null} with $\tau=0$, the p-value \eqref{eq: pval-crt} is super uniform, i.e. for every $t\in [0,1]$ we have $\prob(p\leq t)\leq t$. 
\end{propo}

In Example~\eqref{ex:misfit}, we evaluate the performance of our test in a setting where the size of the test data is significantly smaller than the training size, and is comparable to the feature dimension. The modification made in defining the test score at the data-set level  allows us to achieve non-trivial power even for small size test data. For the score function, we  regress $w$ on $\bx$ using a three-layer neural network and define the score function as the mean-squared-error of this model across the dataset. As we discuss, although the test error is not an indicative measure of goodness-of-fit, our test returns a significantly smaller $p$-value for the model that is closer to the underlying conditional law. We refer to Example~\eqref{ex:misfit} for further details.



}

\section{Model-X setting}\label{section: model-X}
In many applications, we may have access to a large amount of unsupervised data (covariate data without corresponding labels) in addition to a limited number of labeled observations.
Motivated by this trend,~\cite{candes2018panning} proposed model-X setup where it assumes no knowledge of the conditional distribution of labels $Y|X$, but assumes that the joint distribution of the covariates is known, or can be well approximated. Model-X setup is also relevant in experimental design, where we control the covariate distribution and so it is known to us, e.g., in randomized controlled trials, as well as sensitivity analysis of quantitive models~\cite{saltelli2008global}. This modeling assumption has paved the way to address several statistical problems which are open or notoriously hard otherwise. For example,~\cite{candes2018panning} used this setup to extend the knockoff framework of~\cite{barber2015controlling} to high-dimensional regime and provided a methodology for variable selection with controlled false discovery rate in this regime. Also, it develops a conditional randomization test using the knowledge of covariates distribution (see ~\cite{tansey2022holdout,berrett2020conditional,javanmard2021pearson} for other related work on this topic.)

Using similar ideas as in \grasp, in this section we propose model-X \grasp, which uses the covariates distribution to improve the power in flagging the differences between the estimate model $\heta$ and the oracle model $\eta$.

Our next proposition is analogous to Proposition~\ref{propo: f} and extends it to the model-X setup.
\begin{propo}\label{propo: f-model-x}
For $(x,y)\sim \cP$ let
\begin{equation}\label{eq: w-model-x}
 w=\begin{cases} 
u_1\sim\unif[0,\heta(x)],& \quad y=1\,,\\
u_2\sim \unif[\heta(x),1],&\quad y=0\,.\\
\end{cases}
\end{equation}

Then the following holds:
\[
D_{f}\left(\cL_{X,W}\|\cP_X\times \unif([0,1])\right)=\E\left[D_f\Big(\bern(\eta(X))\|\bern(\heta(X))\Big)\right]\,.
\]
\end{propo}

The proof of Proposition \ref{propo: f-model-x} is given in Section \ref{proof: propo: f-model-x}.  Proposition \ref{propo: f-model-x} implies that the distance of interest  $\E[D_f(\bern(\eta(X))\|\bern(\heta(X)))]$  can be seen as the $f$-divergence between multivariate random variables  $(X,W)$ and $(\tX,\tW)$ with $X,\tX\sim \cP_X$, independently and $\tW\sim \unif([0,1])$. 

\subsection{Test statistic and decision rule}
Algorithm \ref{algorithm: balls-bins-model-x} describes the procedure for constructing statistics $\bV_{n,L}$ under model-X setup for testing the null hypothesis in \eqref{eq: null}. The overall procedure is similar to the distribution-free counterpart proposed in Algorithm \ref{algorithm: balls-bins}. The major difference is that here the counterfeits are of the form $(\tilde{x},\tw)$, with the covariate component $\tx$ drawn from $\cP_X$ and $\tw$ drawn from $\unif[0,1]$. In contrast,  in the distribution-free version (cf. Algorithm \ref{algorithm: balls-bins}) the covariate $x$ was fixed for an original sample and its counterfeits.

\begin{algorithmic}[t]
	\begin{algorithm}
		\SetAlgoLined
		\REQUIRE $n$ data points $(x_j,y_j)\in \cX \times \{0,+1\}$, the model $\heta:\cX\to[0,+1]$, a score function $T: \cX \times [0,1] \to \reals$,  and integers $M,L\geq1$ such that $M+1=KL$ for some integer $K$. 
		\ENSURE Test statistics $\bV_{n,L}=[V_1,...,V_L]\in \reals^L$. \\
			\For{$j\in \{ 1,2,...,n\}$}{
			\begin{itemize}
			\item  Let  \[
			          w_j=\begin{cases} 
                                  u_1\sim \mathsf{unif}[0,\heta(x_j)],& \quad y_j=1\,,\\
                                 u_2\sim \mathsf{unif}[\heta(x_j),1],&\quad y_j=0\,.\\
                                  \end{cases}
                                  \]

			\item  Draw $M$ i.i.d. samples $\tw_j^{(1)},\dots, \tw_j^{(M)}$ from $\unif[0,1]$ and {$\tx_j^{(1)},\dots,\tx_j^{(M)}$ from $\cP_x$}. 
                         \item Use score function $T$ to score the initial sample $(x_j,w_j)$ and its $M$ constructed counterfeits $\{(\tx_j^{(1)},\tw_j^{(1)}),..., (\tx_j^{(M)},\tw_j^{(M)})\}$:
                         \begin{align*}
                         T_j&=T(x_j,w_j)\\
                         \tT_j^{(i)}&=T(\tx_j^{(i)},\tw_j^{(i)})\,, \quad \text{for } i\in [M]\,.
                         \end{align*}
 	    \item Let $R_j$ denote the rank of $T_j$ among $\{T_j,\tT_j^{(1)},..., \tT_j^{(M)}\} $:
	    \[
	  R_j=  1+\sum\limits_{i=1}^{M}\ind_{\left\{T_j \geq \tT_j^{(i)}\right \}}
	    \]
		\item Assign label $L_j\in \{ 1,2,...,L\}$ to sample $j$ if $ (L_j-1)K+1\leq   R_j \leq KL_j$.
		
				\end{itemize}
}

\For{$\ell\in \{1,2,...,L\}$}{
			\begin{itemize}
				\item Let $V_\ell$ be the number of samples with label $\ell$, i.e.	$V_\ell=\Big|\big \{j\in \{1,2,...,n\}: L_j=\ell \big\}  \Big|\,.$
				
				\end{itemize}
		}
		
				\caption{Construction of GRASP test statistic (Model-X setting)}\label{algorithm: balls-bins-model-x}
	\end{algorithm}
\end{algorithmic}
\medskip

\noindent{\bf Decision rule.}
Let $\bV_{n,L}$ be the statistic returned by Algorithm \ref{algorithm: balls-bins-model-x}. We construct the test statistics $U^{\asym}_{\tau}(\bV_{n,L})$ and $U^{\finite}_{\tau}(\bV_{n,L})$ following the same formulation as in \eqref{eq: test-stats}. The obtained test statistics are then used in \eqref{eq: decision-finite} and \eqref{eq: decision-asym} to give the decision rules $\Phi_{n,L,\alpha,\tau}^{\finite}$  and
$\Phi_{n,L,\alpha,\tau}^{\asym}$. In the next section, we show that these rules control the size of our test ($\Phi_{n,L,\alpha,\tau}^{\finite}$ in finite sample settings and $\Phi_{n,L,\alpha,\tau}^{\asym}$ in asymptotic regimes).  

\subsection{Size of the model-X \grasp test}
We first start with the following assumption, which assumes a density function for the random variable $T(X,W)$. This is the unconditional version of Assumption~\ref{assumption: cdfs}, and is used in our analysis of the test in a model-X setting. 

\begin{assumption}\label{assumption: cdfs-model-x}  Consider the following cumulative distribution functions:
\begin{align*}
F_o(t)&=\prob(T(X,W)\leq t)\,,\\
F_c(t)&=\prob(T(\tX,\tW)\leq t)\,,
\end{align*}

where $x, \tx\sim \cP_X$ independently, $w$ is given by \eqref{eq: w-model-x} and $\tw \sim \unif[0,1]$. Assume that the density functions of cdfs $F_o$ and $F_c$ exist and show them by $f_o(t)$ and $f_c(t)$, respectively. In addition, consider a measure on the interval $[0,1]$ defined with the cdf $\omega(u):=F_o(F_c^{-1}(u))$. Suppose that this measure is absolutely continuous with respect to the Lebesgue measure, and denote its density function (Radon–Nikodym derivative) by $\psi(u)$. 
\end{assumption}

In our next result we show that the $f$-divergence between the multinomial distribution of $\bV_{n,L}$ (output of algorithm \ref{algorithm: balls-bins-model-x}) and the uniform multinomial distribution is bounded by the distance of the random variables $T(x,w)$ and $T(\tx,\tw)$.
\begin{propo}\label{propo: model-x}
Let $\bV_{n,L}=[V_1,...,V_L]$ be outputs of Algorithm \ref{algorithm: balls-bins-model-x},  then $\bV_{n,L}$ has a multinomial distribution with $L$ outcomes and nominal probabilities $p_1,\dots,p_L$. Under the setting of Assumption \ref{assumption: cdfs-model-x}, for a score function $T:\cX \times [0,1]\to \reals$, the following holds
 \begin{align*}
 \frac{1}{L}\sum_{\ell=1}^L f(Lp_\ell)&\leq D_f(\cL(T(X,W))\|\cL(T(\tX,\tW)))\,.
 \end{align*}
 In addition, if the function $\psi(u)$ from Assumption \ref{assumption: cdfs-model-x} is continuous, the above inequality becomes equality as $K,L$ grow to infinity: 
\[
 \lim_{L\to \infty}\lim_{K\to \infty}  \frac{1}{L}\sum_{\ell=1}^L f(Lp_\ell)=D_f(\cL(T(X,W)),\cL(T(\tX,\tW)))\,.
 \]
\end{propo}
The proof of this proposition is given in Section \ref{proof: propo: model-x}. The next Theorem follows from the combination of Propositions \ref{propo: f-model-x} and \ref{propo: model-x}, along with the data processing inequality for $f$-divergence.
\begin{thm}[Model-X]\label{thm: discretize-distance:model-x}
Let $\bV_{n,L}=[V_1,...,V_L]$ be the output of Algorithm \ref{algorithm: balls-bins-model-x} that has a multinomial distribution with nominal probabilities $p_1,\dots p_L$ as per Proposition \ref{propo: model-x}. Then for a score function $T:\cX\times [0,1]\to \reals$ satisfying Assumption \ref{assumption: cdfs-model-x} the following holds 
 \begin{align*}
\frac{1}{L}\sum_{\ell=1}^L f(Lp_\ell)&\leq 
\E\left[D_f(\bern(\eta(X))\|\bern(\heta(X)))\right]\,.
 \end{align*}
\end{thm}
The proof of this theorem is given in Section \ref{proofs: thm: discretize-distance:model-x}.
The result of this theorem allows us to perform statistical inference on the complex expression $\E\left[D_f(\bern(\eta(x))||\bern(\heta(x)))\right]$ by focusing on the deviation of $\bV_{n,L}$ from the uniform multinomial distribution, similar to the distribution-free version. 

We are now ready to show that the two decision rules \eqref{eq: decision-finite} and \eqref{eq: decision-asym} control the size of our testing in the model-X setup, respectively in the finite sample and asymptotic regimes.

\begin{thm}
 Let $\bV_{n,L}=[V_1,\dots,V_L]$ be the output of Algorithm \ref{algorithm: balls-bins-model-x}. Consider the decision rules $ \Phi^{\asym}_{n,L,\alpha,\tau}$ and $\Phi^{\finite}_{n,L,\alpha,\tau}$ which are given respectively by \eqref{eq: decision-asym} and \eqref{eq: decision-finite}, with test statistics $U_{\tau}^{\asym}(\bV_{n,L})$ and $U_{\tau}^{\finite}(\bV_{n,L})$ in \eqref{eq: test-stats}. Then under the null hypothesis \eqref{eq: null}, this holds:
 \[
 \prob\left( \Phi^{\finite}_{n,L,\alpha,\tau}=1 \right) \leq \alpha\,,~~\lim\sup_{n\to \infty}\prob\left(  \Phi^{\asym}_{n,L,\alpha,\tau}=1\right) \leq \alpha\,. 
\]
 \end{thm}
 
 The proof of this theorem follows along the same lines as in the proof of Theorem \ref{thm: size-DF} .

\subsection{Choice of the score function}\label{sec:scoreF}
%
In summary, our analysis in the previous section showed the following chain of inequalities: 
\begin{align}
\frac{1}{L}\sum_{\ell=1}^L f(Lp_\ell)&\leq D_f(\cL(T(X,W))\|\cL(T(\tX,\tW)))\nonumber\\
& \leq D_f(\cL(X,W)\|\cL(\tX,\tW))\nonumber\\
&=\E\left[D_f(\bern(\eta(X))\|\bern(\heta(X)))\right]\label{eq: series}\,,
\end{align}
where the first inequality follows from Proposition \ref{propo: model-x}, the second one is an application of the data processing inequality, and the third line is the claim proved in Proposition \ref{propo: f-model-x}.   


In order to increase the power of the test (make it less conservative), our guide is to choose score functions for which the gap between the leftmost side and the rightmost side in the chain of inequality \label{eq: series} is minimum.


For the first inequality, as shown in Proposition \ref{propo: model-x} (second part), the gap vanishes for large values of $L$ and $K$. Focusing on the second inequality, we seek score functions $T$ such that 
\begin{equation}\label{eq: choice-T}
D_{f}(\cL(T(X,W))\|\cL(T(\tX,\tW)))=D_{f}(\cL(X,W)\|\cL(\tX,\tW))\,.
\end{equation}
A trivial choice is $T(x,w) = (x,w)$, however, we would like to have scores with one-dimensional range, since we need to compare the scores values of the original samples and the counterfeits.

We proceed our discussion with providing a brief background on the variational representation of $f$-divergences, with the following statement borrowed from \cite[Lemma 1]{nguyen2010estimating}.
\begin{lemma}\label{lemma: variational}
For two probability density functions $q,p$ on $\cX\subset \reals^d$ and a set of measurable functions $\cG$, we have
\begin{equation}\label{eq: variational}
D_f(p\|q)\geq \sup_{\vphi\in \cG} \int (\vphi p- f^*(\vphi)q)\de \mu
  \,,
\end{equation}
where $f^*$ is the conjugate dual of $f$. In addition, the equality is achieved if the subdifferential $\partial f(p/q)$ contains an element of $\cG$.
\end{lemma}
For the reader’s convenience, we provide the proof of Lemma \ref{lemma: variational} in Section \ref{proof: variational}.

\begin{proposition}\label{propo: achievablity}
For two probability density functions $p,q$ over $\cX\subset \reals^d$ and function $f$ as per Definition \ref{def: f-div}, let $\sigma\in \partial f(p/q)$. Then, for $X_p\sim p$ and $X_q\sim q$ the $f$-divergence between distributions of $\sigma(X_p)$ and $\sigma(X_q)$ is equal to the $f$-divergence of distributions $p$ and $q$. Formally, 
\[
D_f(\cL(\sigma(X_p))\|\cL(\sigma(X_q)))=D_f(p\|q)\,.
\]
\end{proposition}
The proof of this proposition is given in \ref{proof: propo: achievablity}.

Going  back to our discussion on the choice of score function $T(x,w)$, observe that by an application of Proposition \ref{propo: achievablity}, condition \eqref{eq: choice-T} is satisfied for $T(x,w)\in \partial f\left(p_{X,W}(x,w)/p_{\tX,\tW}(x,w)\right)$. Since $\tW\sim\unif[0,1]$, independent from $\tX\sim\cP_X$,  we have 
\[
\frac{p_{X,W}(x,w)}{p_{\tX,\tW}(x,w)}=p_{W|X}(w|x)\,.
\]
In addition, from the construction of $w$, cf. Algorithm \ref{algorithm: balls-bins-model-x}, we have
\begin{align}
\prob(w|x)&=\prob(y=1|x)\prob(w|x,y=1)+\prob(y=0|x)\prob(w|x,y=0)\nonumber\\
&=\frac{\eta(x)}{\heta(x)}\ind(w\leq \heta(x))+\frac{1-\eta(x)}{1-\heta(x)}\ind(\heta(x)\leq w)\label{eq: model-X-w|x}\,.
\end{align}

Putting things together, the optimal score function is given by

{\color{black} 
\[
T(x,w) \in \partial f\left(\frac{\eta(x)}{\heta(x)}\ind(w\leq \heta(x))+\frac{1-\eta(x)}{1-\heta(x)}\ind(\heta(x)\leq w)\right)\,.
\]
}
We next note that our test statistics in Algorithm \ref{algorithm: balls-bins-model-x} is based on the \emph{relative ranking} of the score values, not the absolute values.
For an increasing function $g$, the rank values obtained by the score functions $T$ and $g\circ T$ are the same. In particular, since $f$ is convex its derivative is a non-decreasing function, and so we can simply consider the ratio of densities. {\color{black}{When $f$ is not strictly-convex, we may have ties in the ordering with respect to the above derivation. We use the following rule which also breaks ties. }} 
\begin{equation}\label{eq: T-opt}
T^{\mathsf{opt}}(x,w)=\frac{\eta(x)}{\heta(x)}\ind(w\leq \heta(x))+\frac{1-\eta(x)}{1-\heta(x)}\ind(\heta(x)\leq w)\,.
\end{equation}
As can be seen $T^{\mathsf{opt}}(x,w)$ involves the true model $\eta(x)$ which is unknown. To cope with this issue, we consider two routes:
\begin{itemize}
\item {\bf Model-agnostic approach:} We use $T^{\mathsf{opt}}(x,w)$ with replacing $\eta(x)= \frac{1}{2}$.
\item{\bf GAN-based approach:} We propose an optimization inspired by generative adversarial networks (GANs) to approximate  the ratio of densities $p_{X,W}(x,w)/p_{\tX,\tW}(x,w)$.
\end{itemize}

Our next subsection gives a brief background overview on GANs and the details of our GAN-based approach.

\subsubsection{Generative Adversarial Nets (GANs)}
The GANs framework was introduced by \cite{goodfellow2014generative} to generate samples from the population of given data samples $x\in \cX\subset\reals^d$. This framework can be perceived as a game between a generator that tries to learn data distribution $p_X$ and produce fake samples that are statistically close to the original data, and a discriminator that tries to discriminate samples of the generator from the original data samples. 

It is a repeated game where each player aims to improve her model with respect to her objective. Starting from noise sample $z\in \cZ\subset \reals^k$ with density function $p_Z$, the generator produce fake samples  in the data space $\cX$ via a mapping $G(z;\th_g)$, where $G:\cZ\to\cX$ is a differentiable neural network with parameters $\th_g$. The discriminator, on the other side, has access to another neural network $D(x;\th_d)$ that represents the likelihood of data $x$ coming from the original data distribution $p_X$. This game can be cast as the following min-max optimization problem
\begin{equation}\label{eq: GAN}
\min_{\th_g}\max_{\th_d} \left\{ \E_{X} \Big[\log D(X;\th_d) \Big] +\E_{Z}\Big[ \log\big(1-D(G(Z;\th_g);\th_d)\big) \Big] \right\}\,.
\end{equation}
As evident from the above formulation, the discriminator aims to increase the likelihood of the original samples coming from $p_X$, while simultaneously lower the likelihood of fake sample $G(z;\th_g)$ coming from the same distribution. The generator, on the other hand, aims to find a model $G(z;\th_g)$ which generates samples with high likelihood of coming from $p_X$.  As shown in \cite{goodfellow2014generative}, the above optimization problem has a global optimum at $p_g=p_X$, where $p_g$ stands for the distribution of samples $G(z;\th_g)$ with $z\sim p_Z$.

 We follow the GANs framework by considering $\cL(X,W)$ as the  original data distribution. In addition, we assume that the generator at each round produces the randomizations $(\tx,\tw)$, and update the discriminator to discriminate $(x,w)$ from $(\tx,\tw)$. The major distinction with the original GAN setup is dropping a separate network for the generator as the generated samples are always coming from $\cL(\tX,\tW)$. Optimization problem~\eqref{eq: GAN} then reduces to:
 \begin{equation}\label{eq: D}
\max_{\th_d} \left\{ \E_{(X,W)} \Big[\log D((X,W);\th_d) \Big] +\E_{(\tX,\tW)}\Big[ \log\big(1-D((\tX,\tW);\th_d)\big) \Big] \right\}\,.
\end{equation}
The next result connects the optimal discriminator $D(x)$ (solution of \eqref{eq: D}) and the likelihood ratio function. This proposition is borrowed from \cite[Proposition 1]{goodfellow2014generative}. 

\begin{propo}\label{propo: D-p}
Let $D^*(x,w)$ be the maximizer of optimization problem \eqref{eq: D}. Then we have
\[
D^*(x,w)=\frac{p_{X,W}(x,w)}{p_{X,W}(x,w)+p_{\tX,\tW}(x,w)}\,.
\]
\end{propo}

 Since we only have one network, we only update the parameters of the discriminative network $D$. For this end, we adopt the same optimization procedure used in \cite{goodfellow2014generative, nowozin2016f}, where at each step given a mini-batch of samples $\left\{(x^{1},w^{1}),\dots,(x^{m},w^{m})\right\}$ and $\left\{(\tx^{1},\tw^{1}),\dots,(\tx^{m},\tw^{m})\right\} $ we use the following stochastic gradient ascent:


 \[
 \th_d^{(t+1)}=\th_d^{(t)}+ \frac{\beta_t}{m}\sum_{i=1}^{m} \nabla_{\th_d}\left[\log D\Big((x^{i},w^{i});\th_d^{(t)}\Big)+  \log\Big(1- D\big((\tx^{i},\tw^{i});\th_d^{(t)}\big)\Big)\right]\,, 
 \] 
 with step size $\beta_t$.
 The gradients of the above optimization problem can be efficiently computed using back-propagation algorithms.   Let $\hth_d$ be the resulting estimate after convergence. Considering Proposition~\ref{propo: D-p} and our discussion below Proposition~\ref{propo: achievablity}, we consider the following score function 
 \[
 T(x,w)=\frac{D((x,w),\hth_d)}{1-D((x,w),\hth_d)}\,.
 \]


\section{Numerical Experiments}\label{section: numerical}
In this section, we evaluate the performance of our proposed methodology on various synthetic datasets. Consider a binary classification problem under the logistic regression setting. Throughout the experiments, we let the feature vectors have an isotropic Gaussian distribution and the conditional probability be given by 
\begin{equation}\label{eq: logistic}
\eta(x)=\frac{1}{1+\exp(-x^\sT \th_0)}\,,~ x\sim \normal(0,I_d)\,.
\end{equation}
We construct $\th_0$ by drawing one time realization from the  $\normal(0,\sigma^2I_d)$  distribution and this $\th_0$ value is fixed for the rest of the experiments.  We set the feature dimension $d=200$ and standard deviation $\sigma=0.25$.  Let $\heta(x)$ denote the model that we want to perform the goodness-of-fit testing. We consider $\heta(x)$ to be another logistic model with parameter $\th_1$. Concretely, $\heta(x)={1/[1+\exp(-x^\sT \th_1)]}$. Obviously $\heta(x)=\eta(x)$ if $\th_1=\th_0$.
\smallskip

\begin{example} {\bf( Size of the test)} \emph{We begin with studying the size of our proposed test statistics. We focus on the  $\th_1=\th_0$ setting, which implies that $\E[D_f(\bern(\eta(X))||\bern(\heta(X))]=0$ for every divergence function $f$. We set $\tau=0$ in the null hypothesis \eqref{eq: null} and consider three significance levels $\alpha=0.05,0.1$ and $0.15$. The  rejection rate of $H_0$ is reported as the average over $200$ independent experiments. Since $\tau=0$ the optimization problem in \eqref{eq: test-stats} is independent of the choice of $f$. Its only feasible solution is given by $p_1=\dots=p_L=1/L$, and so the test statistics amount to}
\[
U_{0}^{\asym}(\bV_{n,L})=\frac{L}{n}\sum_{\ell=1}^L \left(\bV_\ell-\frac{n}{L}\right)^2\,,\quad   U_{0}^{\finite}(\bV_{n,L})=\frac{2L}{n}\sum_{\ell=1}^L \left(\bV_\ell-\frac{n}{L}\right)^2\,. 
\]

\emph{ Table \ref{table: size-DF} summarizes the rejection rates for the distribution-free Algorithm \ref{algorithm: simple-balls-bins} and for different values of $(n,L)$. Each cell of the table has an ordered pair, where the first and second entry respectively corresponds to the asymptotic decision rule \eqref{eq: decision-asym} and the finite decision rule \eqref{eq: decision-finite}.} 

\emph{We next consider model-X setting for a similar experiment. We consider the model-agnostic approach in choosing the score function $T$ discussed at the end of Section~\ref{sec:scoreF}, namely,}
\begin{equation}\label{eq: score-nominal-model-x}
T(x,w)=\ind(w\leq \heta(x))~\frac{1}{2\heta(x)}+\ind(\heta(x)\leq w)~\frac{1}{2(1-\heta(x))}\,.
\end{equation}
\emph{Table \ref{table: size-model-x} presents the rejection rate of the model-X Algorithm \ref{algorithm: balls-bins-model-x} for three significance levels $\alpha=0.05, 0.1, 0.15$ for different number of samples ($n$), number of labels ($L$) and randomizations per label $(K)$. Each cell of Tables \ref{table: size-model-x} consists of an ordered pair, where the first and second numbers respectively correspond to the asymptotic decision rule \eqref{eq: decision-asym} and the finite decision rule \eqref{eq: decision-finite}. Reported numbers are averaged over $200$ independent experiments. }
 	\begin{table}[]
\begin{center}
			\scalebox{0.9}{
	\begin{tabular}{|c|c|c||c|c||c|c| }\hline
&\multicolumn{2}{c||}{$\alpha=0.05$}&\multicolumn{2}{c||}{$\alpha=0.1$}& \multicolumn{2}{c|}{$\alpha=0.15$}\\\hline
 \diagbox{$n$}{$L$}&\makebox{$50$}&\makebox{$100$}&\makebox{$50$}&\makebox{$100$}&\makebox{$50$}&\makebox{$100$}\\ \hline
 5000&  (0.055,0) &(0.03,0)     &(0.105,0) &(0.075,0)   &(0.16,0)&(0.1,0) \\ \hline
 20000&  (0.07,0) &(0.05,0)      &(0.1,0) &(0.1,0)       &(0.185,0)&(0.16,0) \\\hline
 50000&  (0.06,0) &(0.05,0)       &(0.105,0) &(0.105,0)   &(0.15,0)&(0.145,0) \\\hline
\end{tabular}
}
\end{center}
\caption{Size of the distribution-free GRASP, cf. Algorithm \ref{algorithm: simple-balls-bins}. In this experiment, $\heta(x)=\eta(x)$ is considered with $\eta(x)$ being a logistic model \eqref{eq: logistic}.  We consider both decision rules (asymptotic~\eqref{eq: decision-asym}) and (finite~\eqref{eq: decision-finite}). The first and the second entry of each pair in the table represent the rejection rate of these decision rules, respectively. The numbers are averaged over $200$ independent experiments for three significance levels $\alpha=0.05,0.1, 0.15$ and for different sample sizes ($n$) and number of labels ($L$). } \label{table: size-DF}
\end{table}
	\begin{table}[!h]
\begin{center}
			\scalebox{0.9}{
	\begin{tabular}{|c|c|c||c|c||c|c| }\hline
&\multicolumn{2}{c||}{$\alpha=0.05$}&\multicolumn{2}{c||}{$\alpha=0.1$}& \multicolumn{2}{c|}{$\alpha=0.15$}\\\hline
 \diagbox{$(n,K)$}{$L$}&\makebox{$50$}&\makebox{$100$}&\makebox{$50$}&\makebox{$100$}&\makebox{$50$}&\makebox{$100$}\\ \hline
 (5000,1)&  (0.055,0)& (0.04,0)               &(0.115,0) & (0.08,0)       &(0.165,0)&(0.11,0) \\ \hline
 (5000,5)&  (0.025,0)&(0.07,0)   &(0.12,0)&(0.12,0)   &(0.185,0)&(0.125,0) \\\hline
 (10000,1)& (0.085,0)  &(0.035,0)   &(0.125,0)&(0.095,0)   &(0.18,0)&(0.105,0) \\\hline
  (10000,5)& (0.065,0)  &(0.06,0)   &(0.07,0)&(0.095,0)   &(0.17,0)&(0.125,0) \\\hline
\end{tabular}
}
\end{center}
\caption{Size of the model-X GRASP,  cf. Algorithm \ref{algorithm: balls-bins-model-x}. In this experiment, $\heta(x)=\eta(x)$ is considered with $\eta(x)$ being a logistic model \eqref{eq: logistic}.  Algorithm \ref{algorithm: balls-bins-model-x} is run with the score function $T(x,w)$ given in \eqref{eq: score-nominal-model-x}.  We consider both decision rules (asymptotic~\eqref{eq: decision-asym}) and (finite~\eqref{eq: decision-finite}). The first and the second entry of each pair in the table represent the rejection rate of these decision rules, respectively. The numbers are averaged over $200$ independent experiments for three significance levels $\alpha=0.05,0.1, 0.15$ and for different sample sizes ($n$) and number of labels ($L$), and number of randomizations per label ($K$). } \label{table: size-model-x}
\end{table}
\end{example}

\begin{example}\label{ex:w}{\bf (Power of the test)} \emph{In this experiment, we consider the null hypothesis  \eqref{eq: null} for a logistic model $\heta(x)$ with parameter $\th_1=-\th_0$. Formally, we have} 
 \begin{equation}\label{eq: logistic-test}
 \heta(x)=\frac{1}{1+\exp(-\th_1^\sT x)}\,, \text{ with } \th_1=-\th_0\,.
 \end{equation}
 
  \emph{In this experiment, we consider three divergence functions (Kl, TV, and Hellinger given in Lemma \ref{lemma: f-functions}) between the true model $\eta(\cdot)$ and the test model $\heta(\cdot)$. The computed nominal values of these divergences are $\tau_0^{\kl}=2.7819,\tau_0^{\tv}=0.7330,$ and $\tau_0^{\mathsf{H}}=0.9576$. More precisely, we have}
  \begin{align}\label{eq: tau_0}
  \tau_0^{\kl}&=\E\left[D_{\kl}\Big(\bern(\eta(X))\|\bern(\heta(X))\Big)\right]\,,\\
   \tau_0^{\tv}&=\E\left[D_{\tv}\Big(\bern(\eta(X))\|\bern(\heta(X))\Big)\right]\,,\\
   \tau_0^{\mathsf{H}}&=\E\left[D_{\mathsf{H}}\Big(\bern(\eta(X))\|\bern(\heta(X))\Big)\right]\,.
  \end{align}
  
   \emph{In order to analyze the power of proposed methods, we choose $\tau$ that are smaller than $\tau_0$ values, consequently we expect that the null hypothesis \eqref{eq: null} must be rejected. For each divergence function, we consider four $\tau$ values and report the average number of times (out of $50$ independent experiments) that the proposed method rejects the null hypothesis $H_0$. Table \ref{table: power-simple} reports the results for distribution-free GRASP, outlined in Algorithm \ref{algorithm: simple-balls-bins}, for a variety of choices of $(n,L)$.
Each cell of Table \ref{table: power-simple} consists of an ordered pair where the first and second coordinate respectively stands for the asymptotic decision rule \eqref{eq: decision-asym} and the finite decision rule \eqref{eq: decision-finite} at significance level $\alpha=0.1$.  }
   
 \emph{To analyze the statistical power of model-X GRASP, as outlined in Algorithm \ref{algorithm: balls-bins-model-x}, we adopt a similar setup with the test model $\heta(x)$ given by \eqref{eq: logistic-test}, and run the model-X GRASP with the score function \eqref{eq: score-nominal-model-x}. Table \ref{table: power-model-x} presents the  results  for different values of the sample size ($n$), number of labels ($L$), and randomizations per label $(K)$, at the significance level $\alpha=0.1$. Each cell of Table \ref{table: power-model-x} has an ordered pair with the first numbers indicates the rejection rate with the asymptotic decision rule \eqref{eq: decision-asym} and the second number indicates the rejection rate with the finite decision rule  \eqref{eq: decision-finite}. Reported numbers are averaged over $50$ independent experiments. }

\emph{Since the problem setting for the last two power analysis experiments (distribution-free \grasp and model-X \grasp) are the same, by a simple comparison between Table \ref{table: power-simple} and Table \ref{table: power-model-x}, it can be observed that for each of the $f$-divergence functions, the model-X \grasp algorithm achieves a higher statistical power than the distribution-free \grasp procedure. This is expected as the model-X \grasp utilizes the covariate distribution to obtain higher statistical power. }
	\begin{table}[]
\begin{center}
			\scalebox{0.82}{
	\begin{tabular}{|c|c|c|c|c||c|c|c|c||c|c|c|c| }\hline
&\multicolumn{4}{c||}{$\mathsf{KL}$}&\multicolumn{4}{c||}{$\mathsf{TV}$}&\multicolumn{4}{c|}{$\mathsf{Hellinger}$}  \\\hline
&\multicolumn{4}{c||}{$\tau_0^{\kl}=2.7819$}&\multicolumn{4}{c||}{$\tau_0^{\tv}=0.7330$}&\multicolumn{4}{c|}{$\mathsf{\tau_0^{\mathsf{H}}}=0.9576$} \\ \hline
 \diagbox{$(n,L)$}{$\tau$}&\makebox{$0.72$}&\makebox{$0.82$}&\makebox{$0.96$}&\makebox{$1.02$}&\makebox{$0.4$}&\makebox{$0.44$}&\makebox{$0.48$}&\makebox{$0.52$}&\makebox{$0.28$}&\makebox{$0.32$}&\makebox{$0.36$}&\makebox{$0.4$}          	\\ \hline
	$(5000,50)$&(1,1) &(1,0.02)&(0,0) & (0,0)&(1,1) &(1,0.52) &(0.18,0) &(0,0)& (1,1) &(1,0.08) &(0.08,0) &(0,0) \\\hline
	$(20000,50)$&(1,1)&(1,1)&(0,0) &(0,0) & (1,1)&(1,1) &(1,0.86) &(0,0) &(1,1)&(1,1) &(1,0.04) &(0,0)\\\hline
	$(50000,50)$&(1,1)  &(1,1)  &(0.54,0)  &(0,0) &(1,1)  & (1,1)& (1,1)&(0,0)  & (1,1)& (1,1)&(1,0.04)  &(0,0) \\\hline
	$(5000,100)$& (1,1) & (1,0) &(0,0)  &(0,0) &(1,1)  &(1,0) &(0.02,0)  & (0,0)   & (1,0.86)& (1,0) &(0.06,0)  &(0,0) \\\hline
	$(20000,100)$& (1,1) & (1,1) &(1,0)  &(0,0) &(1,1)  &(1,1) &(1,0)  & (0,0)   & (1,1)& (1,1) &(1,0)  &(0,0) \\\hline
	$(50000,100)$& (1,1) & (1,1) &(1,1)  &(0.44,0) &(1,1)  &(1,1) &(1,1)  & (0,0)   & (1,1)& (1,1) &(1,1)  &(0.06,0) \\\hline
\end{tabular}
}
\end{center}

\caption{Statistical power of distribution-free GRASP, as outlined in Algorithm \ref{algorithm: simple-balls-bins},  with the true model $\eta(x)$ given by \eqref{eq: logistic} and the test model $\heta(x)$ in \eqref{eq: logistic-test}. In this experiment,  we consider three $f-$divergences (KL, TV, and Hellinger as per Lemma \ref{lemma: f-functions}) for different choices of sample size ($n$) and number of labels ($L$). Each cell has two numbers, where the first and second number respectively refer to the rejection rates under the asymptotic decision rule \eqref{eq: decision-asym} and the finite decision rule \eqref{eq: decision-finite}, at significance level $\alpha=0.1$. The reported numbers are averaged over $50$ independent experiments } \label{table: power-simple}
\end{table}


\begin{table}[!h]
			\scalebox{0.78}{
	\begin{tabular}{|c|c|c|c|c||c|c|c|c||c|c|c|c| }\hline
&\multicolumn{4}{c||}{$\mathsf{KL}$}&\multicolumn{4}{c||}{$\mathsf{TV}$}&\multicolumn{4}{c|}{$\mathsf{Hellinger}$}  \\\hline
&\multicolumn{4}{c||}{$\tau_0=2.7819$}&\multicolumn{4}{c||}{$\tau_0=0.7330$}&\multicolumn{4}{c|}{$\mathsf{\tau_0=0.9576}$} \\ \hline
 \diagbox{$(n,K,L)$}{$\tau$}&\makebox{$1.5$}&\makebox{$1.7$}&\makebox{$1.9$}&\makebox{$2$}&\makebox{$0.6$}&\makebox{$0.63$}&\makebox{$0.66$}&\makebox{$0.7$}&\makebox{$0.58$}&\makebox{$0.62$}&\makebox{$0.70$}&\makebox{$0.8$}          	\\ \hline
 	$(5000,5,50)$&(1,1)&(1,0.12)&(0.82,0)&(0,0)&(1,1) &(1,1)&(1,0)&(0.04,0)&(1,1)& (1,0.98)&(1,0)&(0.5,0)\\\hline
	$(5000,5,100)$&(1,1)&(1,0.38)&(1,0)&(0.28,0)&(1,1)&(1,0) &(1,0)&(0,0)&(1,1)&(1,0)& (1,0)&(0.02,0) \\\hline
	$(10000,5,50)$&(1,1)&(1,1)&(1,0)&(0.1,0)&(1,1) &(1,1)&(1,1)&(0.84,0)&(1,1)&(1,1) &(1,0)&(1,0) \\\hline
	$(10000,5,100)$&(1,1)&(1,1)&(1,0)&(1,0)&(1,1) &(1,1)&(1,0)&(0,0)&(1,1)& (1,1)&(1,0)&(1,0)\\\hline
	$(5000,1,50)$&(1,1)&(1,0)&(0,0)&(0,0)& (1,1)&(1,0.78)&(1,0)&(0,0)&(1,1)&(1,0) &(1,0)&(0,0)\\\hline
	$(5000,1,100)$&(1,1)&(1,0)&(0.38,0)&(0,0)&(1,0.96) &(1,0)&(0.96,0)&(0,0)&(1,0.44)&(1,0) &(1,0)&(0,0)\\\hline
	$(10000,1,50)$&(1,1)&(1,0.1)&(0.02,0)&(0,0)&(1,1) &(1,1)&(1,0.38)&(0.04,0)&(1,1)& (1,1)&(1,0)&(0.74,0)\\\hline
	$(10000,1,100)$&(1,1)&(1,1)&(1,0)&(0.06,0)&(1,1) &(1,1)&(1,0)&(0,0)&(1,1)&(1,1) &(1,0)&(0.76,0)\\\hline
	\end{tabular}
}

\caption{Statistical power of model-X GRASP,  as outlined in Algorithm \ref{algorithm: balls-bins-model-x}, with the true model $\eta(x)$ given by \eqref{eq: logistic} and the test model $\heta(x)$ in \eqref{eq: logistic-test}. We consider three $f-$divergences (KL, TV, and Hellinger as per Lemma \ref{lemma: f-functions}) for different choices of sample size ($n$) and number of labels ($L$), and number of randomizations per label ($K$). The model-X algorithm is run with the score function $T(x,w)$ given in \eqref{eq: score-nominal-model-x}. Each cell contains an ordered pair, where the first and second number respectively refer to the rejection rates under the asymptotic decision rule \eqref{eq: decision-asym} and the finite decision rule \eqref{eq: decision-finite}, at significance level $\alpha=0.1$. The reported numbers are averaged over $50$ independent experiments.} \label{table: power-model-x}
\end{table}
\end{example}

{\color{black}\begin{example}\label{ex:res}({\bf Score function for distribution-free GRASP)} \emph{In this experiment, we evaluate the performance of distribution-free GRASP with other choices of score function.  So far, we have used the score function $T(x,w)=w$. However, we want to examine whether one can achieve higher statistical power by selecting a score function that accounts for variability between $x$ and $w$. To this end, we perform a similar experiment as the one reported in Table \ref{table: power-simple} for $(n,L)=(5000,50)$, but with a new score function. We fit a linear model with response value $w$ and feature vectors $x$ on an auxiliary dataset of size $4000$. We denote the fitted linear model by $\hat{\theta}$ and consider the score function $T(x,w)=|w-x^\sT \hth|$. We report the results for three f-divergence functions (TV, KL-divergence, and Hellinger distance) for both asymptotic and finite-sample decision rules in Figure \ref{plot: T-df}. We averaged the reported numbers over $50$ experiments at a significance level of $\alpha=0.1$. By comparing the rejection rates with the reported numbers in the first row of Table \ref{table: power-simple}, we find that the previous score function $T(x,w)=w$ performs slightly better than the fitted model. For instance, for KL divergence, using $T(x,w)=w$ with both asymptotic and finite decision rules achieves full power (rejection rate 1.00) at $\tau=0.72$, while using the fitted score function already achieves trivial power (rejection rate $0$) at $\tau=0.7$. Similar observations can be made for other divergence functions as well. Likewise, for TV and Hellinger divergence, we also observe that score function $T(x,w)=w$ outperforms the regression-based score function slightly.}
\begin{figure}[t]
\centering
\includegraphics[scale=0.25]{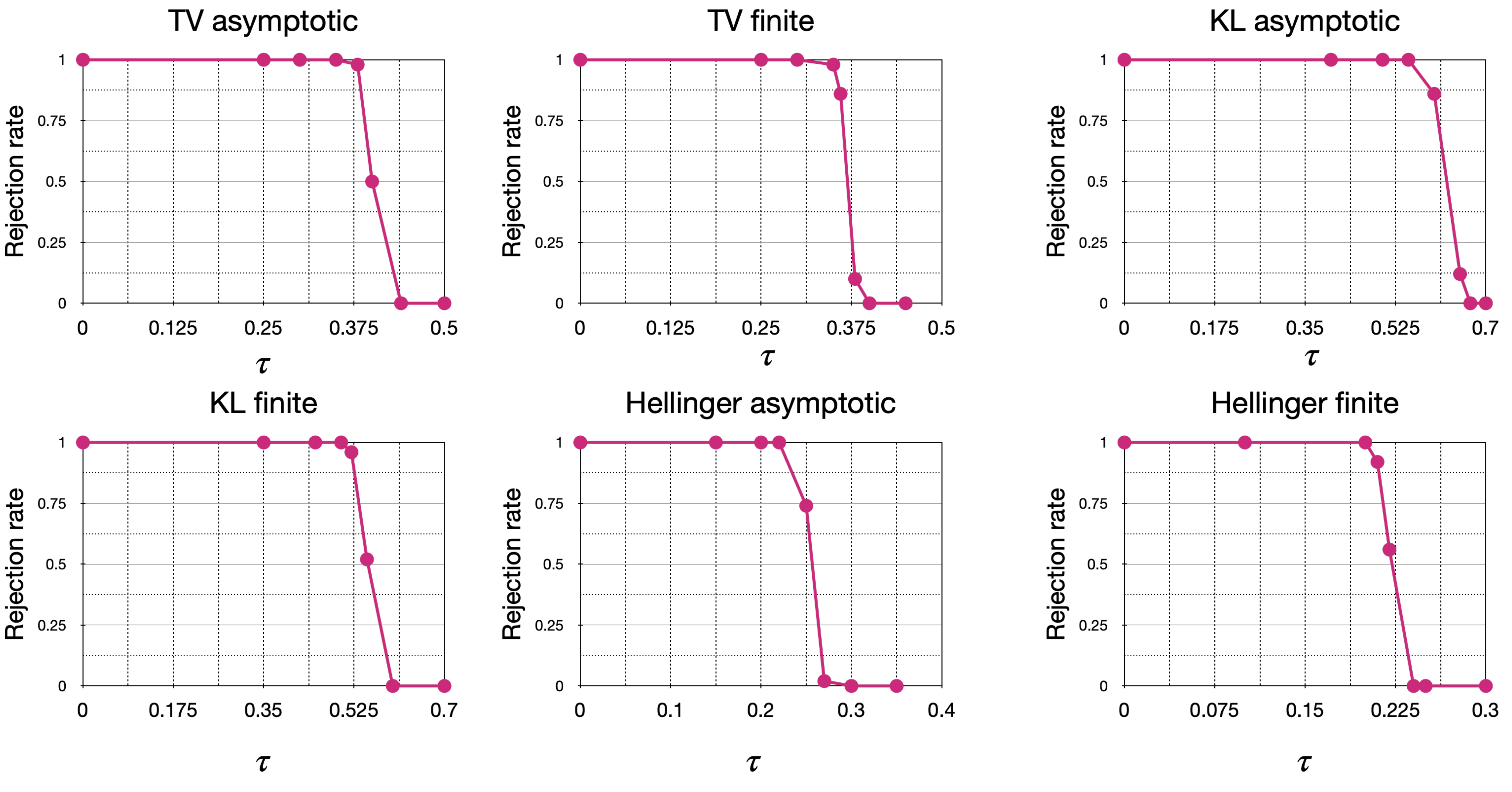}
\caption{Statistical power of  distribution free \grasp for trained score function $T(x,w)=|w-x^\sT \hth|$. The linear model $\hth$ is trained on an auxiliary dataset of size $4000$.  The true model $\eta(x)$ is given in \eqref{eq: logistic} and the test model $\heta(x)$ is given in \eqref{eq: logistic-test}. In this case, the nominal values are  $\tau_0^{\kl}=2.7819$, $\tau_0^{\tv}=0.7330,$ and $\tau_0^{\mathsf{H}}=0.9576$ (see \eqref{eq: tau_0}). For three $f-$divergence (KL, TV, Hellinger) we run distribution free \grasp with two decision rules $\Phi_{n,L,\alpha,\tau}^{\asym}$ and $\Phi_{n,L,\alpha,\tau}^{\finite}$ given in \eqref{eq: decision-asym} and \eqref{eq: decision-finite}. For each setting, we consider the null-hypothesis \eqref{eq: null} with different $\tau$ values. Here, sample size is $n=5000$ with number of labels $L=50$, and the significance level is $\alpha=0.1$.  Reported rates are averaged over $50$ experiments. \vspace{0.4cm}}\label{plot: T-df}
\end{figure}
\end{example}
}

\begin{example}{\bf (The choice of number of labels $L$)} \emph{In this experiment, we analyze the statistical power of the proposed methods for goodness-of-fit testing, while allowing a wide range of values for the inside parameter $L$ ( number of labels). We consider the previous setting with the true model $\eta(x)$ as in \eqref{eq: logistic} and the test model $\heta(x)$ as in \eqref{eq: logistic-test}. We consider the null hypothesis \eqref{eq: null} for three $f$ divergence functions $\kl,\tv$ and $\mathsf{H}$. In addition, the two decision rules \eqref{eq: decision-asym} and \eqref{eq: decision-finite} are being deployed. In this setting, the nominal values of divergence functions are $\tau_0^{\kl}=2.7819$, $\tau_0^{\tv}=0.7330,$ and $\tau_0^{\mathsf{H}}=0.9576$, where $\tau_0$ values are given in \eqref{eq: tau_0}. }

\emph{We consider the same range of $L$ values for different $f$-divergence (KL, TV, Hellinger) and decision rules ($\Phi_{n,L,\tau}^{\asym}$ and $\Phi_{n,L,\tau}^{\finite}$). However, we 
choose different values of $\tau$ in the hypothesis \eqref{eq: null}, depending on the setting to better capture the effect of $L$.
For distribution-free \grasp we consider the null hypothesis \eqref{eq: null} with $\tau_{\asym}^{\kl}=0.84, \tau_{\finite}^{\kl}=0.73, \tau_{\asym}^{\tv}=0.46,\tau_{\finite}^{\tv}=0.42,\tau_{\asym}^{\mathsf{H}}=0.34,\tau_{\finite}^{\mathsf{H}}=0.30$. Figure \ref{plot: df} depicts the statistical power curves under these settings versus the number of labels $L$. The sample size is set to $n=5000$ and the significance level to $\alpha=0.1$. The obtained numbers are averaged over $50$ independent experiments.}  

\emph{We repeat the same experiment for model-X \grasp with the score function \eqref{eq: score-nominal-model-x}. Here, we consider $\tau_{\asym}^{\kl}=1.87, \tau_{\finite}^{\kl}=1.6, \tau_{\asym}^{\tv}=0.67,\tau_{\finite}^{\tv}=0.62,\tau_{\asym}^{\mathsf{H}}=0.75$, and $\tau_{\finite}^{\mathsf{H}}=0.6$. The sample size is set to $n=5000$ with randomizations per label $K=1$. The significance level is set to $\alpha=0.1$ and the reported numbers are averaged over $50$ independent experiments. The behavior of the statistical power with respect to the change in the number of labels ($L$) can be seen in Figure \ref{plot: model-x}. As observed from the curves in Figures~\ref{plot: df} and \ref{plot: model-x}, the power favors a middle range of $L$ values before and after which the power starts to decline. }

\begin{figure}
\centering
\includegraphics[scale=0.25]{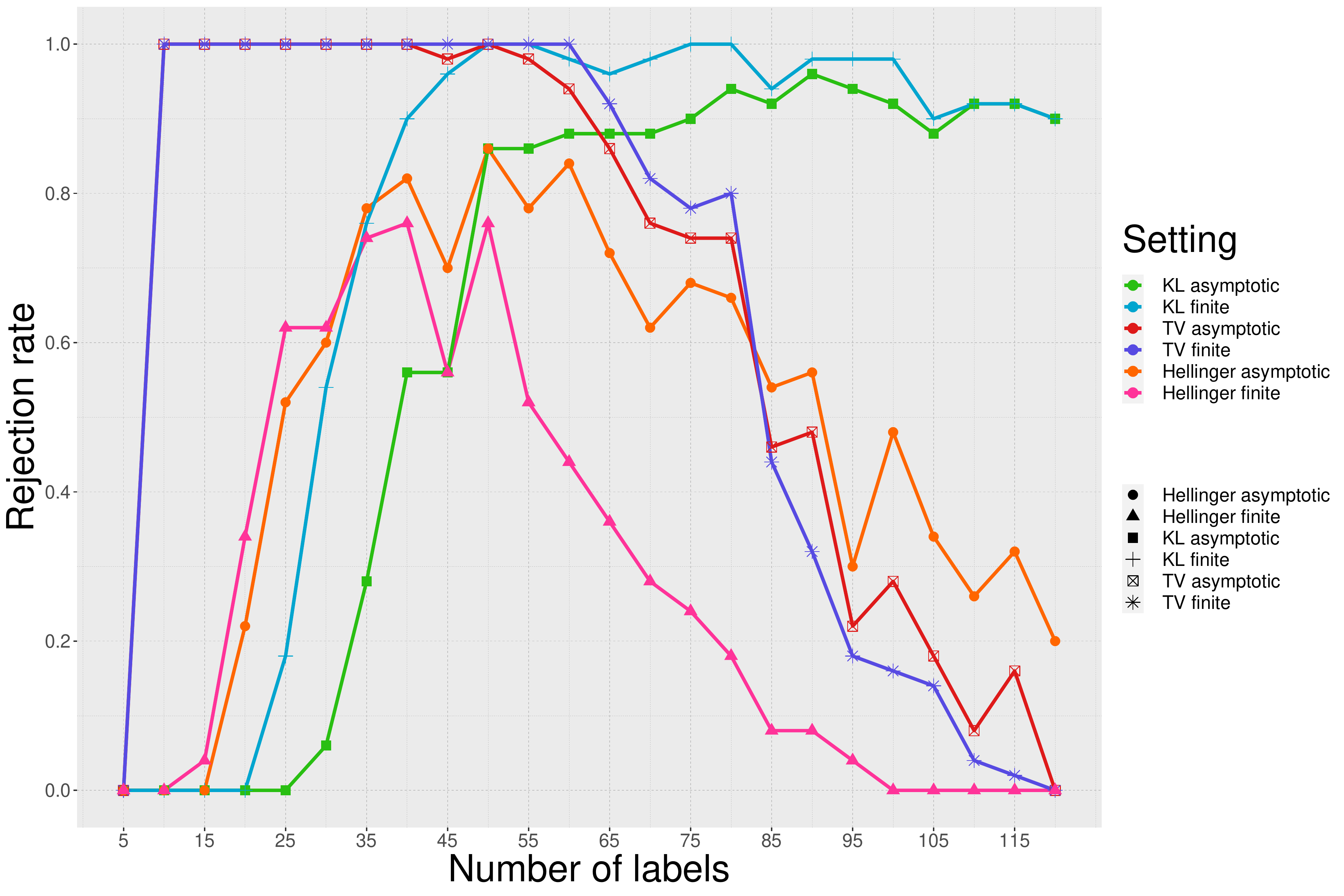}
\caption{Statistical power of  distribution free \grasp versus the number of labels $L$. The true model $\eta(x)$ is given in \eqref{eq: logistic} and the test model $\heta(x)$ is given in \eqref{eq: logistic-test}. In this case, the nominal values are  $\tau_0^{\kl}=2.7819$, $\tau_0^{\tv}=0.7330,$ and $\tau_0^{\mathsf{H}}=0.9576$ (see \eqref{eq: tau_0}). For three $f-$divergence (KL, TV, Hellinger) we run distribution free \grasp with two decision rules $\Phi_{n,L,\alpha,\tau}^{\asym}$ and $\Phi_{n,L,\alpha,\tau}^{\finite}$ given in \eqref{eq: decision-asym} and \eqref{eq: decision-finite}. For each setting, we consider the null-hypothesis \eqref{eq: null} with the following $\tau$ values: $\tau_{\asym}^{\kl}=0.84, \tau_{\finite}^{\kl}=0.73, \tau_{\asym}^{\tv}=0.46,\tau_{\finite}^{\tv}=0.42,\tau_{\asym}^{\mathsf{H}}=0.34$, and $\tau_{\finite}^{\mathsf{H}}=0.30$. Here, sample size is $n=5000$, the significance level is $\alpha=0.1$ and the reported rates are averaged over $50$ experiments. \vspace{0.4cm}}\label{plot: df}
\end{figure}

\begin{figure}
\centering
\includegraphics[scale=0.25]{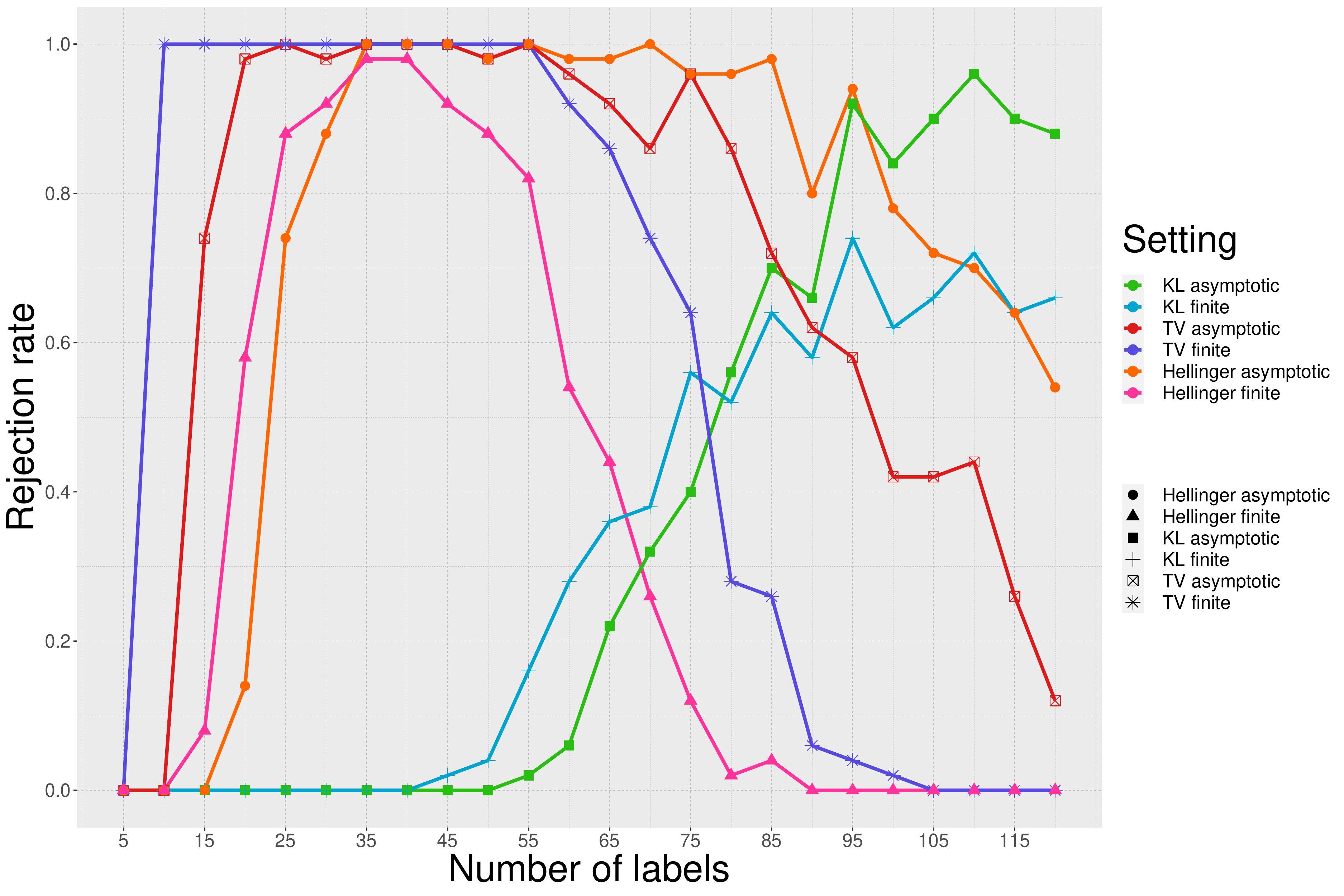}
\caption{Statistical power of  model-X \grasp versus the number of labels $L$. The true model $\eta(x)$ is given in \eqref{eq: logistic} and the test model $\heta(x)$ is given in \eqref{eq: logistic-test}. In this case, the nominal values are  $\tau_0^{\kl}=2.7819$, $\tau_0^{\tv}=0.7330,$ and $\tau_0^{\mathsf{H}}=0.9576$ (see \eqref{eq: tau_0}). For three $f-$divergence (KL, TV, Hellinger) we run  model-X \grasp with the score function $T(x,w)$ as per \eqref{eq: score-nominal-model-x} and the two decision rules $\Phi_{n,L,\tau}^{\asym}$ and $\Phi_{n,L,\tau}^{\finite}$ given in \eqref{eq: decision-asym} and \eqref{eq: decision-finite}. For each setting, we consider the null-hypothesis \eqref{eq: null} with the following $\tau$ values: $\tau_{\asym}^{\kl}=1.87, \tau_{\finite}^{\kl}=1.6, \tau_{\asym}^{\tv}=0.67,\tau_{\finite}^{\tv}=0.62,\tau_{\asym}^{\mathsf{H}}=0.75$, and $\tau_{\finite}^{\mathsf{H}}=0.6$. Here, sample size is $n=5000$, with $K=1$ randomizations per sample, and the significance level $\alpha=0.1$. The reported rates are averaged over $50$ experiments. \vspace{0.4cm}}\label{plot: model-x}
\end{figure}
\end{example}

\begin{example}\label{ex:misfit}{\bf (Statistical evidence for misfitted models)} 
\emph{In this experiment, we test for perfect fit of models to the underlying conditional law, by considering hypothesis testing problem \eqref{eq: null} with $\tau=0$. We consider feature vectors drawn independently from $\normal(0,I_d)$, where $d=300$. We use a planted model setup, where the data generating law is formulated by a two-layer neural network. Specifically, we consider a fully-connected neural network with $N=400$ ReLU neurons in the first layer and a sigmoid activation function as the output unit. The conditional distribution $\cL(Y|X)$ is given by:}
\begin{equation}  \label{eq: example-perfect-fit}
Y|X\sim \bern\left( {\mathsf{Sigmoid}}\big(  \th_0^\sT {\mathsf{ReLU}}(W_0 X)\big)\right)\,,
\end{equation}
\emph{where $W_0\in \reals^{N\times d}$ is the weight matrix, $\th_0\in \reals ^N$ is the weight vector of the second layer, and ${\mathsf{ReLU}}$ is the ReLU activation function applied entry-wise. The entries of the planted parameters $(W_0,\th_0)$ are drawn independently  from $\normal(0,1)$.}



\emph{We compare two neural networks, $\mathcal{N}_1$ and $\mathcal{N}_2$, trained on a similar dataset $\cD^{\mathsf{train}}=\{(x_i,y_i)\}_{i=1:T}$, but with different weight initialization schemes. Specifically, we initialize the weights of $\mathcal{N}_1$ using the Kaiming initialization method, while the weights of $\mathcal{N}_2$ are initialized by adding independent noise from a standard normal distribution with mean 0 and standard deviation $10^{-2}$ to each entry of $W_0$ and $\th_0$, We train both models on a training dataset of size $T=2^{16}$ (approximately 65K samples), and evaluate them on a test dataset of size 128.}

\emph{We first compare the predictive performance of $\cN_1$ and $\cN_2$ on test datasets. For this end, we consider $500$ independent draws of evaluation sets of size $128$ and report the average test accuracy of models on the drawn datasets.  Figure \ref{subfig:fig1} shows the boxplots of computed test accuracies for both models $\cN_1$ and $\cN_2$ over all $500$ experiments. The boxplots show that $\cN_1$ and $\cN_2$ have very similar accuracies, despite the fact that the $\ell_2$-distance between the model parameters of $\mathcal{N}_1$,  and $\mathcal{N}_2$ from the ground-truth model \eqref{eq: example-perfect-fit} are significantly different (362.12 for $\mathcal{N}_1$ and 80.87 for $\mathcal{N}_2$). This indicates that the perturbed initialization of $\mathcal{N}_2$ has resulted in a better local minimum in the optimization landscape compared to $\mathcal{N}_1$, despite their similar empirical accuracies, and therefore $\mathcal{N}_2$ is a better fit to $\cL(Y|X)$ than $\cN_1$. This further highlights the fact that solely focusing on test accuracy is not sufficient to differentiate between the models. }

\emph{We evaluate the performance of GRASP to determine the significance of evidence against the assumption that $\cN_1,\cN_2$ are perfectly fitted to $\cL(Y|X)$. We construct the $p$-values for the two models according to~\eqref{eq: pval-crt}, with $M = 500$. For the score function, we fit a three-layer neural network to regress $w$ on $\bx$, namely $\bv^\sT \mathsf{ReLU}(\bA_2 \mathsf{ReLU}(\bA_1 \bx))$, with $\bv\in\reals^{100\times 1}$, $\bA_2\in\reals^{100\times 300}$, $\bA_1\in\reals^{300\times 300}$. We define score function $T(\bx,\bw)$ as the MSE of this fitted model over the evaluation data set of size 128. For the $500$ independent draws of the evaluation sets of size $128$, we use the same datasets used earlier for test accuracies, and compute the GRASP p-values for $\cN_1$, $\cN_2$, as well as the nominal model ($\cL(Y|X)$). Figures \ref{subfig:fig2}, \ref{subfig:fig3}, \ref{subfig:fig4} depict the Q-Q plot of GRASP p-values (across the 500 experiments) for models $\cN_1$, $\cN_2$, and the nominal model $\cL(Y|X)$, respectively. It can be observed that the GRASP p-values for the nominal model follow a uniform distribution, while for the other two models the Q-Q plots deviate from the uniform distribution. In addition, the $\cN_2$ p-values have higher deviation from uniform distribution than the $\cN_1$ p-values, and this aligns with our expectation given that $\cN_2$ is closer to  the ground-truth model in $\ell_2$ distance than $\cN_1$. By hypothesis testing \eqref{eq: null} (for $\tau=0$) at the significance level $\alpha=0.1$, the nominal model has rejection rate $9.8\%$, and models $\cN_1,\cN_2$ have average rejection rates of $49.2\%, 21.2\%$, respectively. }

\begin{figure}[t]
    \centering
    
    \begin{subfigure}{0.4\textwidth}
        \includegraphics[width=\linewidth]{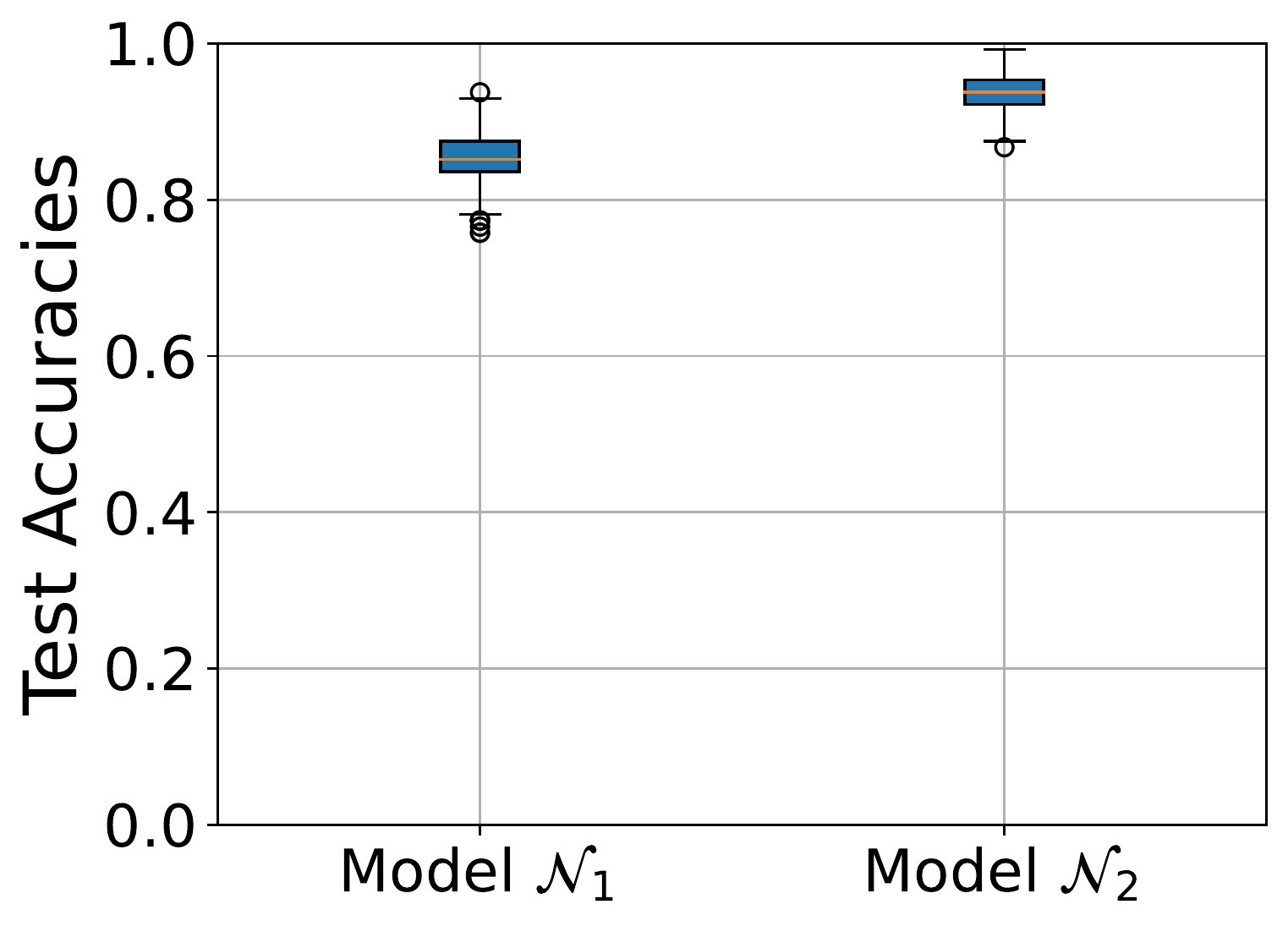}
        \caption{Boxplot for test accuracies of $\cN_1$ and $\cN_2$}
        \label{subfig:fig1}
    \end{subfigure}
    \hspace{3em}
    \begin{subfigure}{0.37\textwidth}
        \includegraphics[width=\linewidth]{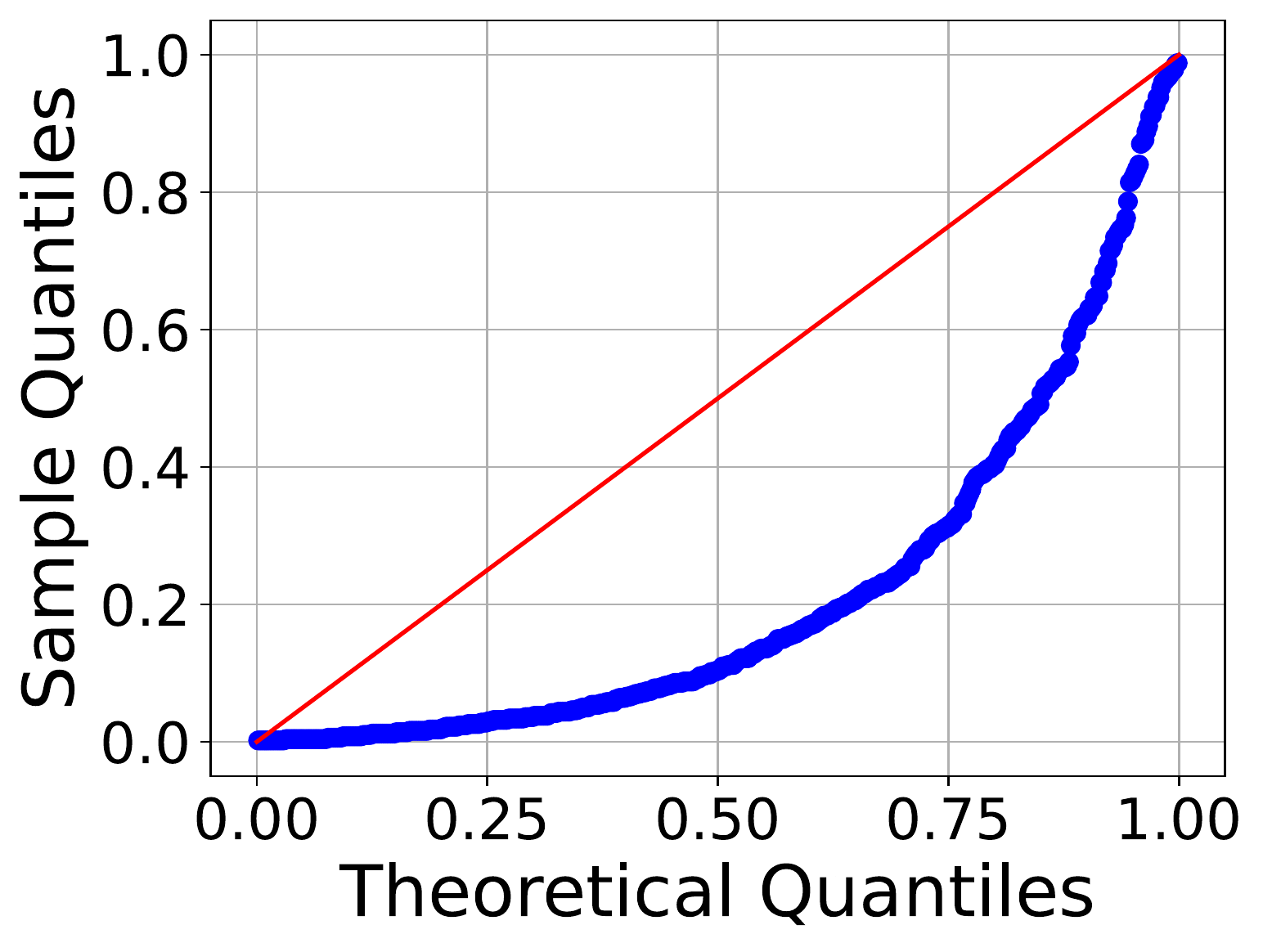}
        \caption{QQ plot of GRASP p-values for $\cN_1$}
        \label{subfig:fig2}
    \end{subfigure}
    %
    %
    \begin{subfigure}{0.37\textwidth}
        \includegraphics[width=\linewidth]{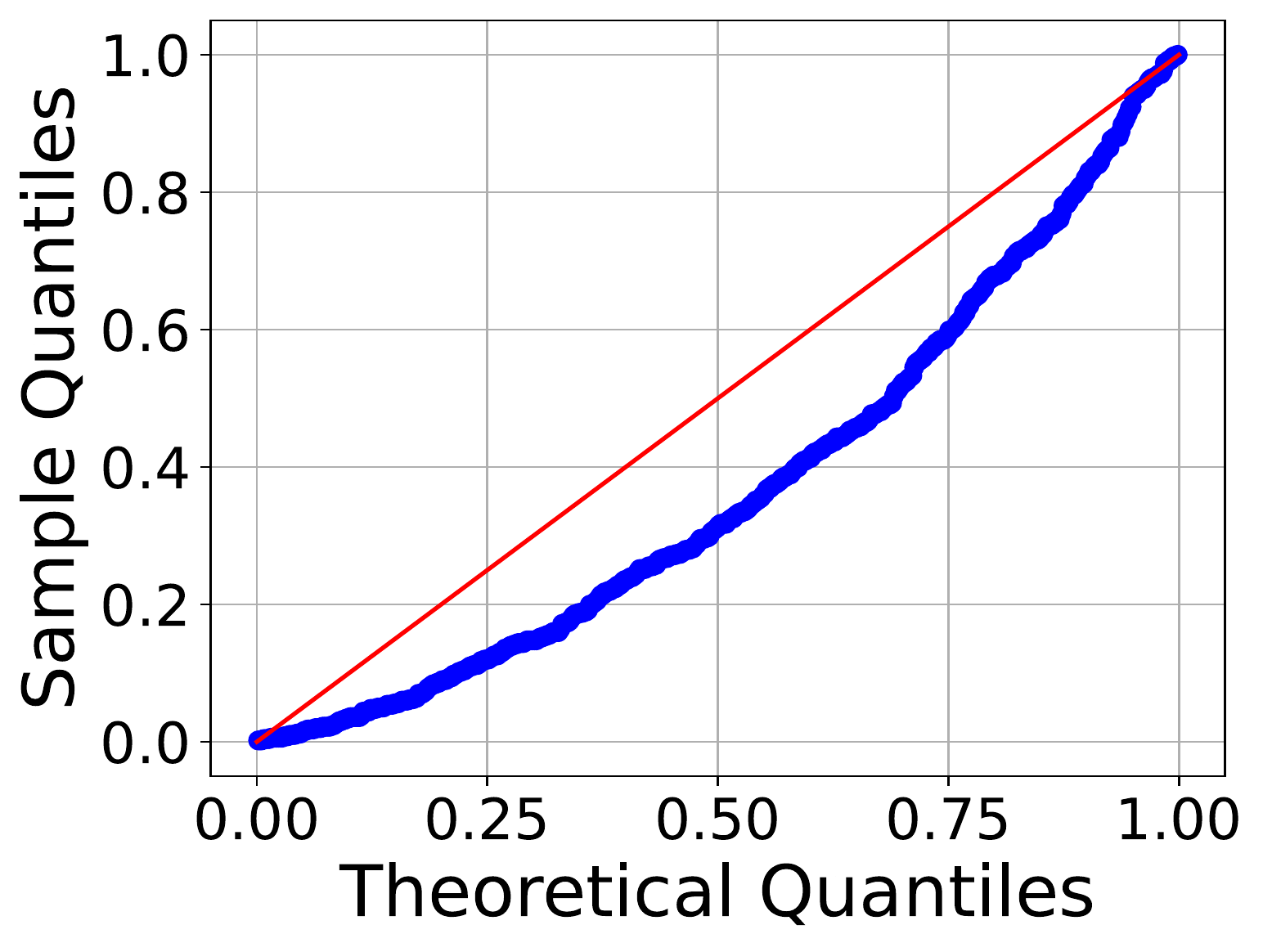}
        \caption{QQ plot of GRASP p-values for $\cN_2$}
        \label{subfig:fig3}
    \end{subfigure}
    \hspace{3em}
    \begin{subfigure}{0.37\textwidth}
        \includegraphics[width=\linewidth]{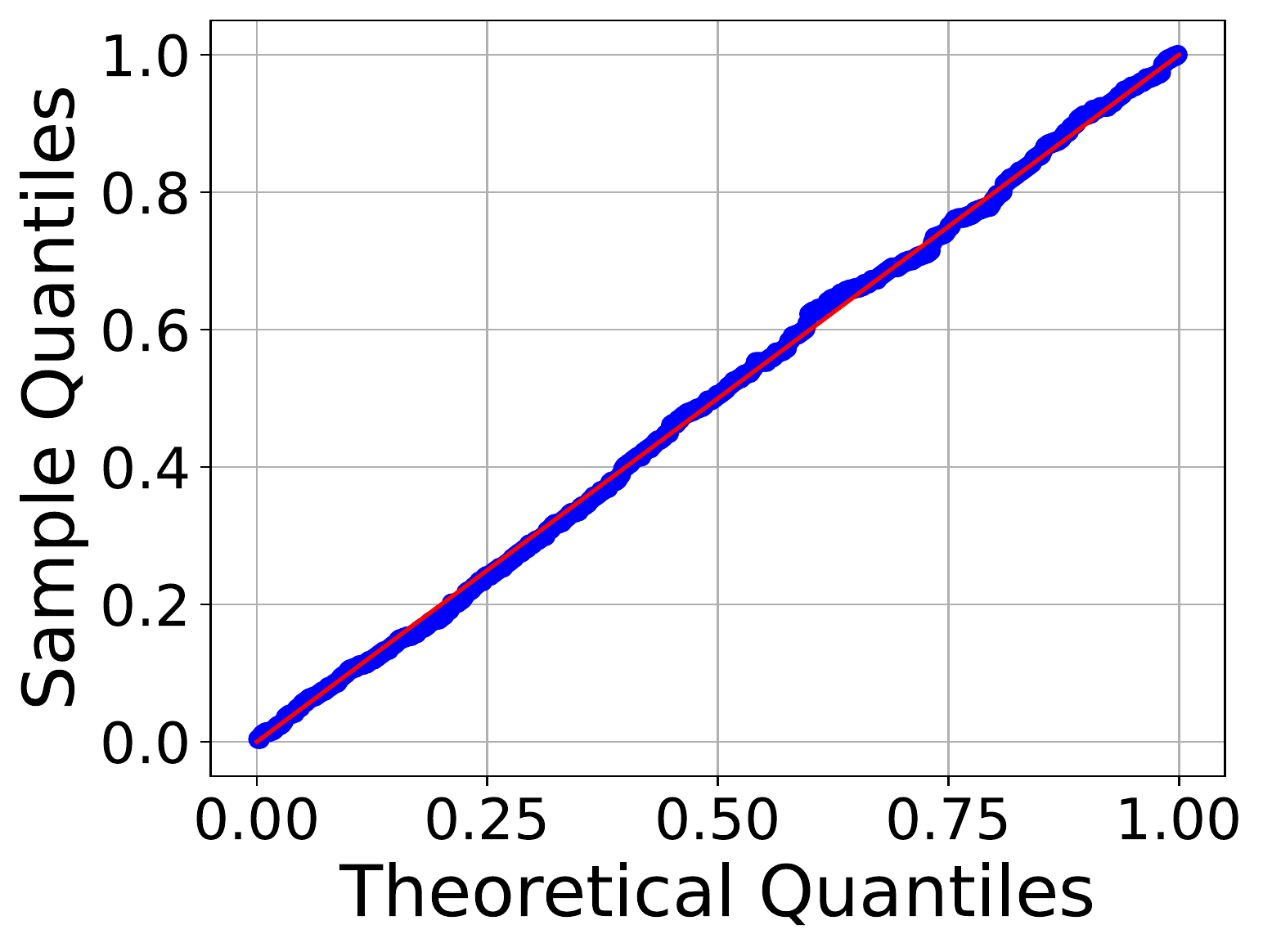}
        \caption{ {\scriptsize QQ plot of GRASP p-values for nominal model \eqref{eq: example-perfect-fit}} }
        \label{subfig:fig4}
    \end{subfigure}
    
    \caption{ Test accuracies and GRASP $p$-values for the two models $\cN_1$ and $\cN_2$ over $500$ draws of evaluation datasets of size $128$ (under the set-up of Example~\ref{ex:misfit}). While both models have very similar test accuracies, we observe that the GRASP $p$-values of $\cN_1$ show a significanlty higher deviation from the uniform distribution, compared to the  GRASP $p$-values of $\cN_2$. For testing the perfect fit ($\tau=0$), at the significance level $\alpha=0.1$, the nominal model has rejection rate $9.8\%$, and models $\cN_1,\cN_2$ have average rejection rates of $49.2\%, 21.2\%$, respectively.  }
    \label{fig:2x2_grid}
\end{figure}
%
%
%
%
\end{example}

\begin{example}{\bf (GAN-based versus model-agnostic score functions)} \emph{In this experiment, we compare the performance of model-X \grasp algorithm for two choices of score functions. Formally, the considered score functions are GAN-based approach and the model-agnostic given in \eqref{eq: score-nominal-model-x}. We consider the hypothesis testing problem \eqref{eq: null} for three different $f$-divergence functions TV, KL and the Hellinger distance, and two decision rules, asymptotic  \eqref{eq: decision-asym} and finite \eqref{eq: decision-finite}. For each setting (6 overall), we run the experiment with one of the score functions (12 experiments in total). The true model $\eta(x)$ is given by \eqref{eq: logistic} and the test model $\heta(x)$ is given by \eqref{eq: logistic-test}, with $\th_0$ a draw of $\normal(0,I_d)$, the feature dimension $d=200$ and $\th_1=-2\th_0$.   We let the number of samples be $n=5000$, number of labels   $L=50$, and $K=1$ randomizations per label. Figure \ref{fig: gan-agn} depict the performance of model-X \grasp for GAN based score function and agnostic (AGN) score function for three $f$-divergence functions, and  the two decision rules $\asym$ and $\finite$. The reported rates are averaged over $50$ instances. It is observed that in each setting,  the GAN-based score function achieves a higher statistical power for larger  $\tau$ values than its counterpart used with an agnostic score function, for both the asymptotic and the finite decision rules. }

\begin{figure}
	\centering
	\begin{subfigure}[b]{0.32\textwidth}
		\centering
		\includegraphics[scale=0.24]{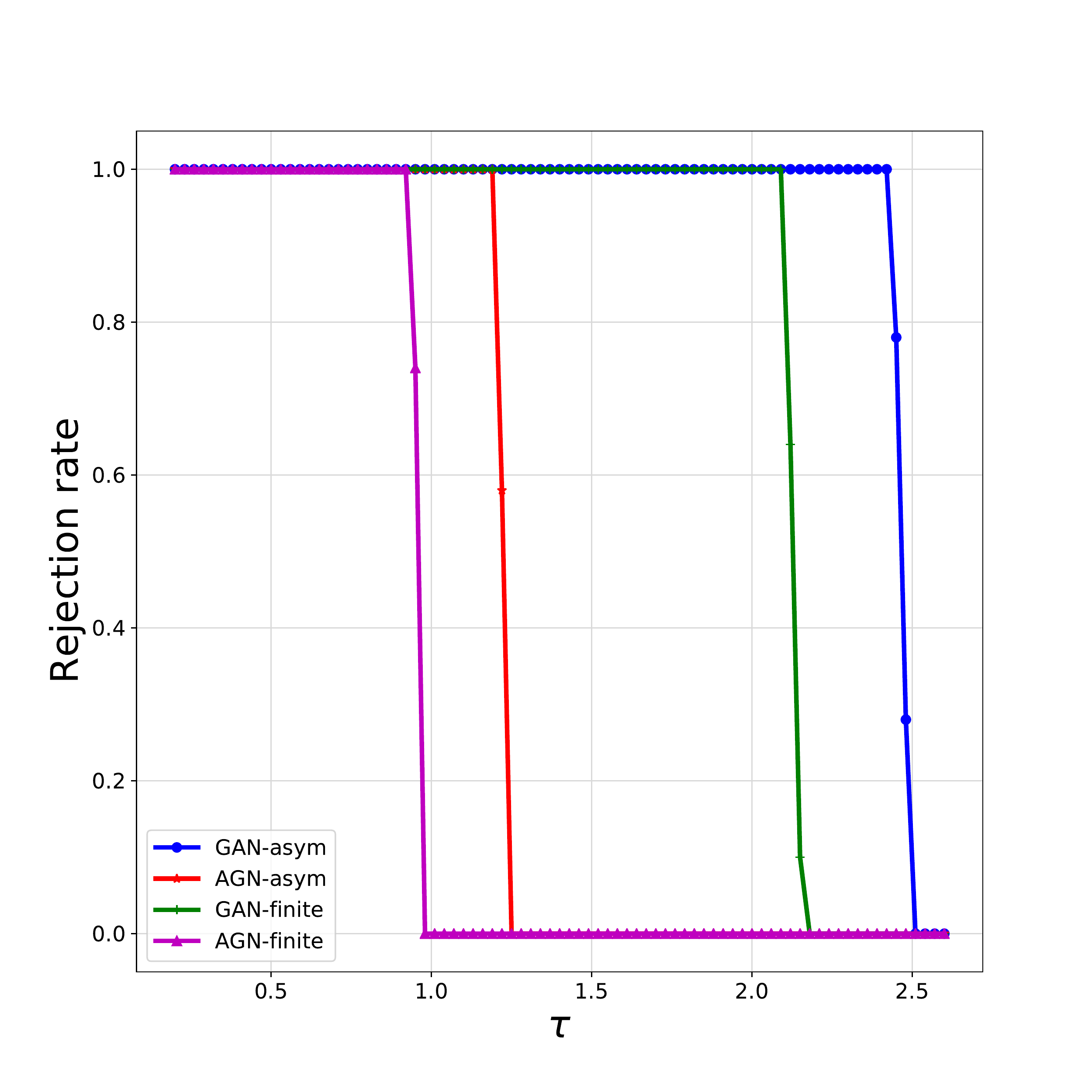}
		\caption{ KL}
		\label{fig:binary:q}
	\end{subfigure}
	\hfill
	\begin{subfigure}[b]{0.32\textwidth}
		\centering
		\includegraphics[scale=0.24]{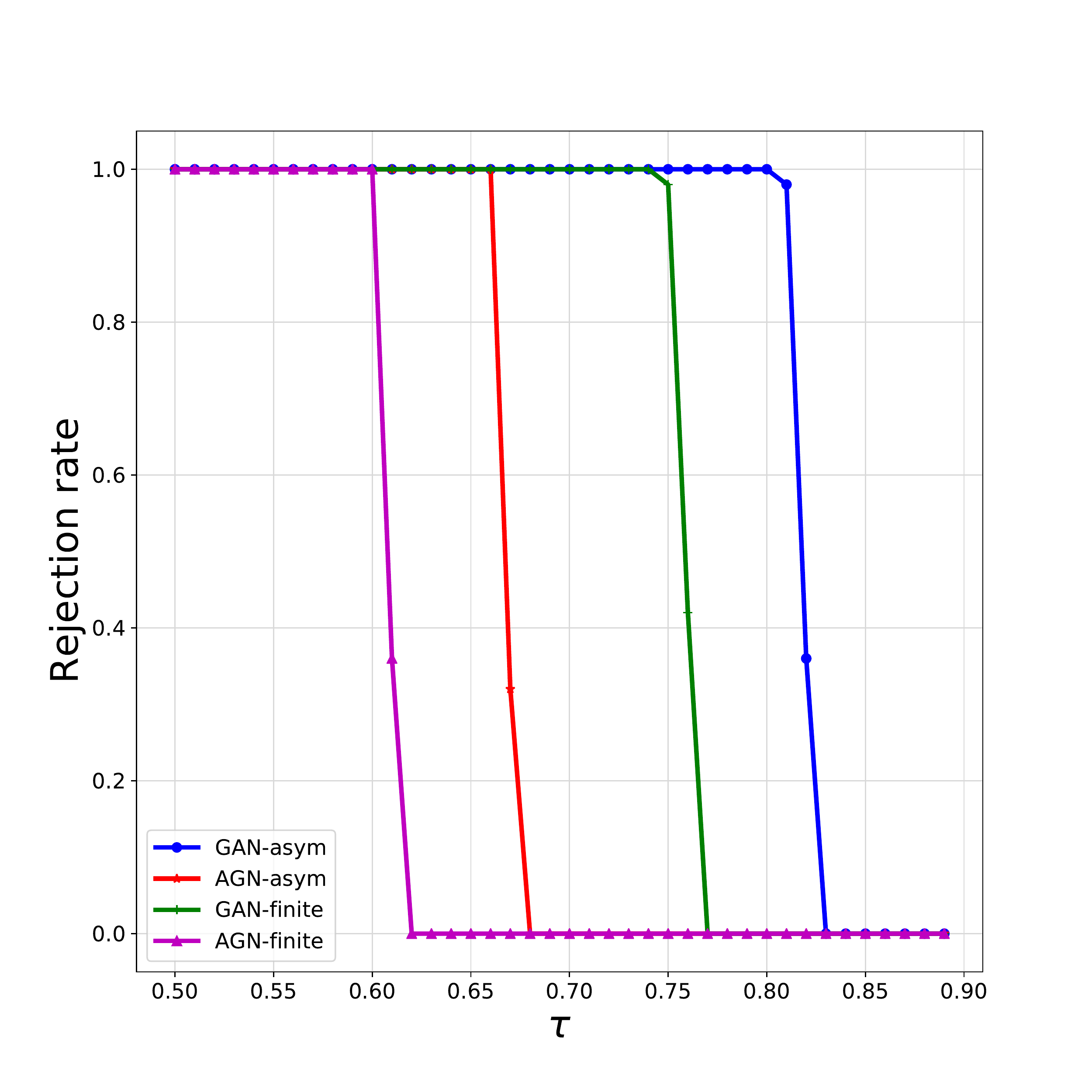}
		\caption{TV }
		\label{fig:binary:rho}
	\end{subfigure}
	\hfill
	\begin{subfigure}[b]{0.32\textwidth}
		\includegraphics[scale=0.24]{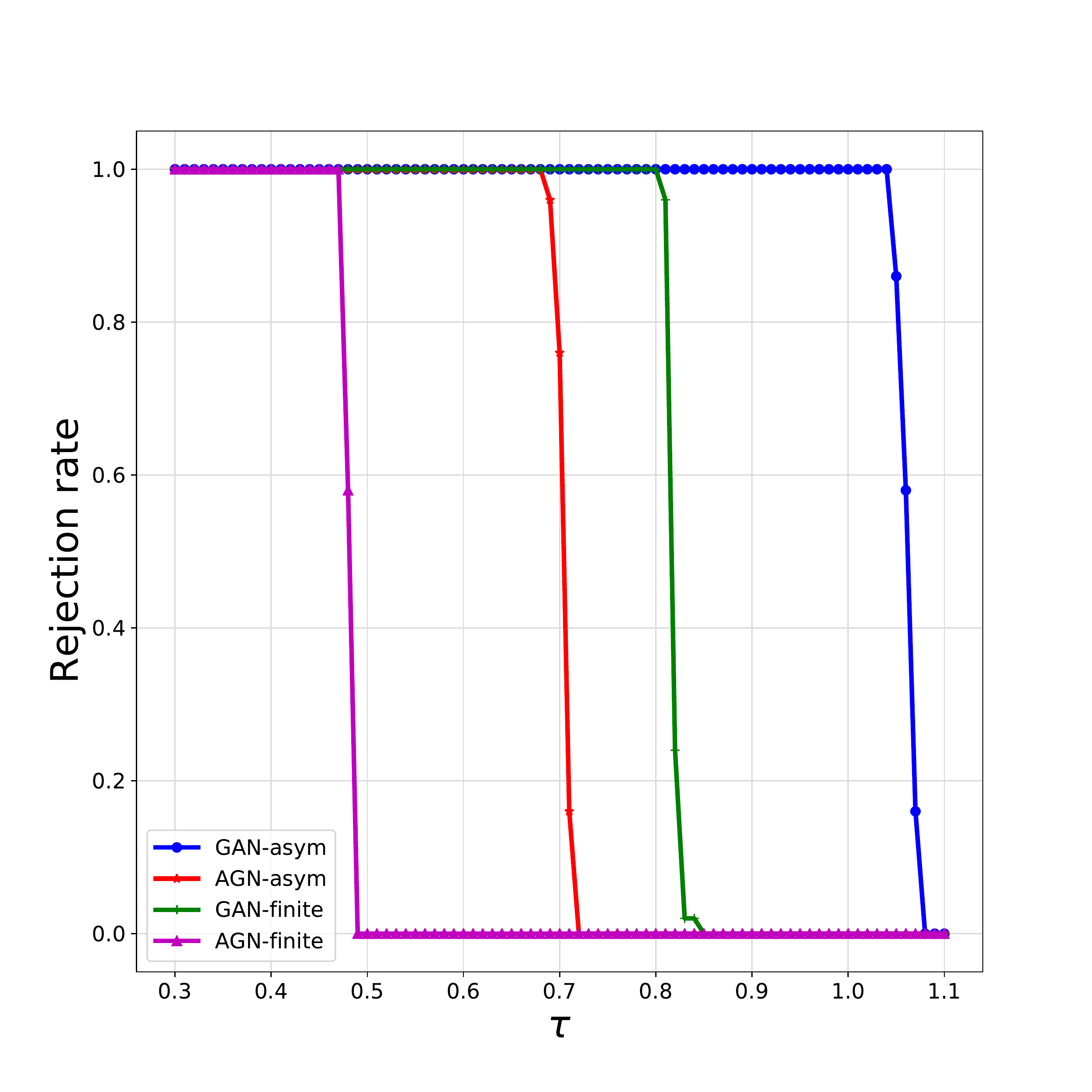}
		\caption{Hellinger}	
		\label{fig:binary:eps}	
	\end{subfigure}
	\caption{Comparison between the statistical power of model-X \grasp with a GAN-based score function and the agnostic score function \eqref{eq: score-nominal-model-x}. In this experiment, the test size is $n=5000$, number of labels $L=50$, and randomizations per label $K=1$. Hypothesis testing problem \eqref{eq: null} is considered for a variety of $\tau$ values,  three $f$-divergence functions, and two asymptotic and finite decision rules \eqref{eq: decision-asym} and \eqref{eq: decision-finite}. In each setting, it can be observed that the GAN-based score function achieves a higher statistical power than the agnostic score function. The reported numbers are average over $50$ experiments.}
	\label{fig: gan-agn}
\end{figure}


\emph{For the discriminator of the GAN-based score, we use a neural network with input dimension $201$ (as inputs are $(x,w)$ with $x\in \reals^{200}$ and $w\in \reals$). The first hidden layer consists of $256$ neurons with ReLU activation function. The second and third  hidden layers have respectively 128 and 64 neurons with ReLU activations.  The output is a single neuron with sigmoid activation function.  The cross-entropy loss is considered for the training process, and the network is trained on $64$k inputs. In order to prevent overfitting, the dropout probability 0.3 is considered for neural connections. }
\end{example}

{\color{black} 

\section{Solving the optimization problem for GRASP decision rules }

The optimization problems \eqref{eq: test-stats} have convex objectives and constraints ($f$-divergence ball), and therefore we have a convex optimization problem.  We use an iterative optimization procedure to find the optimal solution.  It is worth highlighting that vanilla projected gradient descent requires projection onto the $f$-divergence ball, which can be computationally complex for large $L$ or when selecting certain $f$ divergence functions. To circumvent this issue, we explore an alternative approach by leveraging conditional gradient methods (Frank-Wolfe \cite{frank1956algorithm}) that avoid the need for such projections. This adoption of conditional gradient method is also motivated by the observation that minimizing a linear objective over an $f$-divergence ball has a simpler to analyze dual formulation \cite{ben2013robust,namkoong2016stochastic}. We first provide a brief overview of conditional gradient methods.

 We consider a convex differentiable objective function $g:\reals^L\to \reals$ and a compact convex set $\cP\subset \reals^L$. Given the optimization problem of minimizing $g$ over $\cP$, the following iterative rule is considered by conditional gradient methods:
 
\begin{equation}\label{eq: conditional-gradient}
p_{t+1}=\gamma_t q_t+ (1-\gamma_t)p_t\,, \quad q_t= \arg \min_{q\in \cP} q^\sT \nabla g(p_t)\,,
\end{equation}
where $\gamma_t$ is the step size and can beset as $\gamma_t=\frac{1}{t+2}$.  

Getting back to initial optimization problems \eqref{eq: test-stats}, we have

\begin{equation}\label{eq: main-opts}
U^{\finite}_{\tau}(\bV_{n,L})=\min_{p \in \mathcal{U}_\tau} \frac{1}{n}\sum_{\ell=1}^L \frac{(V_\ell-np_\ell)^2}{p_\ell+1/L}\,,\quad  U^{\asym}_{\tau}(\bV_{n,L})=\min_{p \in \mathcal{U}_\tau} \frac{1}{n} \sum_{\ell=1}^L \frac{(V_\ell-np_\ell)^2}{p_\ell}\,,
\end{equation}
where the constraint set $\mathcal{U}_\tau$ is given by

\[
\cU_\tau=\left\{p\in \reals^L \text{ s.t. } \quad  p\geq 0,\quad  \sum_{\ell=1}^L p_\ell=1, \quad  \sum_{\ell=1}^L f(Lp_\ell)\leq L \tau   \right\}\,.
\]

We next focus on solving the linearization of objective functions  \eqref{eq: conditional-gradient} over the $f$-divergence ball, which is needed in the update rule  \eqref{eq: conditional-gradient}.  This problem has been studied before by \cite{ben2013robust,shapiro2017distributionally}  for general continuous distributions. For the reader's convenience, we state the result for the specific case of the problem over discrete distributions. 
\begin{propo}\label{propo: linear-f}
For  $x\in \reals^L$ and $\tau\geq 0$, let
\begin{align*}
(\lambda^*,\eta^*)\in \arg\min_{\lambda\geq 0, \eta\in \reals} \left[ \lambda \tau+\eta +\frac{\lambda}{L} \sum_{\ell=1}^L f^*\Big(\frac{-\eta-x_\ell}{\lambda} \Big)\right]\,.
\end{align*}
If $q^*\in \reals ^L$ is such that 
\[
\frac{-x_\ell-\eta^*}{L\lambda^*}\in \partial f(q^*_\ell)\,,\quad \forall \ell \in [L]\,,
\]
then $q^*$ is the minimizer of $q^\sT x$ over $\cU_\tau$.
\end{propo}

It is easy to observe that the objective function in Proposition \ref{propo: linear-f} is over a half-space in a two-dimensional space with decoupled constraints, and can be solved efficiently. Combining the result in Proposition \ref{propo: linear-f} with \eqref{eq: conditional-gradient} completes the iterative procedure to solve optimization problems \eqref{eq: main-opts}. Since objective functions \eqref{eq: main-opts} are different for asymptotic and finite decision rules, therefore we have distinct values for gradients ($x$ in $q^\sT x$ as per Proposition \ref{propo: linear-f}). Namely, we have 

\[
\nabla g^{\asym}(p)= n-\frac{\bV_{n,L}(\ell)}{np_\ell^2} \,,  \quad \nabla g^{\finite} (p)=\frac{(np_\ell-\bV_{n,L}(\ell))\left(np_\ell+\bV_{n,L}(\ell)+\frac{2n}{L}\right)}{n(p_\ell+\frac{1}{L})}  \,,
\]
where $g^{\asym}$ and $g^{\finite}$ are associate objective functions in \eqref{eq: main-opts}. Algorithms \ref{alg: opt-finite} and \ref{alg: opt-asym}  summarizes the iterative procedures for solving finite and asymptotic optimization problems \eqref{eq: main-opts}, respectively.

\begin{algorithm}[H]
\begin{algorithmic}[1]
\STATE (initialization): $p_\ell(0) \leftarrow 1/L\,,$ for all $\ell\in [L]$. 
\FOR{iteration $h = 1,\dotsc, H$}
\STATE $x_\ell(h)=\frac{(np_\ell(h)-\bV_{\ell})\left(np_\ell(h)+\bV_{\ell}+\frac{2n}{L}\right)}{n(p_\ell(h)+\frac{1}{L})} $
\STATE  $(\lambda(h),\eta(h)) = \arg \min\limits_{\lambda\geq 0, \eta\in \reals}^{} 
\left \{ \lambda \tau +\eta+\frac{\lambda}{L} \sum\limits_{\ell=1}^L f^*\Big(\frac{-\eta-x_\ell(h)}{\lambda} \Big)\right\}$ 
\STATE  let $q_\ell(h)$ be such that $ \frac{-\eta(h)-x_\ell(h)}{L\lambda(h)}\in \partial f(q_\ell(h))\,,$ for all $\ell\in [L]$
\vspace{1mm}
\STATE $p(h+1)= \gamma(h)q(h)+ (1-\gamma(h))p(h)$
\vspace{1mm}
\ENDFOR
\RETURN $U_\tau^{\finite}(\bV_{n,L})\leftarrow \frac{1}{n}\sum\limits_{\ell=1}^L \frac{(\bV_\ell-np_\ell(H))^2}{(p_\ell(H)+\frac{1}{L})^2}$
\end{algorithmic} \caption{Iterative procedure to solve optimization problem of $U^{\finite}_{\tau}(\bV_{n,L})$}\label{alg: opt-finite}
\end{algorithm}
\vspace{0.5cm}

\begin{algorithm}[H]
\begin{algorithmic}[1]
\STATE (initialization): $p_\ell(0) \leftarrow 1/L\,,$ for all $\ell\in [L]$. 
\FOR{iteration $h = 1,\dotsc, H$}
\STATE $x_\ell(h)=n-\frac{\bV_\ell^2}{np_\ell(h)^2}\,,$ for all $\ell\in [L]$
\STATE  $(\lambda(h),\eta(h))= \arg \min\limits_{\lambda\geq 0, \eta\in \reals}^{} 
\left \{ \lambda \tau +\eta+\frac{\lambda}{L} \sum\limits_{\ell=1}^L f^*\Big(\frac{-\eta-x_\ell(h)}{\lambda} \Big)\right\}$ 
\STATE  let $q_\ell(h)$ be such that $ \frac{-\eta(h)-x_\ell(h)}{L\lambda(h)}\in \partial f(q_\ell(h))\,,$ for all $\ell\in [L]$
\vspace{1mm}
\STATE $p(h+1)= \gamma(h)q(h)+ (1-\gamma(h))p(h)$
\vspace{1mm}
\ENDFOR
\RETURN $U_\tau^{\asym}(\bV_{n,L})\leftarrow \frac{1}{n}\sum\limits_{\ell=1}^L \frac{(\bV_\ell-np_\ell(H))^2}{p_\ell(H)^2}$
\end{algorithmic} \caption{Iterative procedure to solve optimization problem of $U^{\asym}_{\tau}(\bV_{n,L})$}\label{alg: opt-asym}
\end{algorithm}


}
%
%
%

\bibliographystyle{alpha}
\bibliography{mybib}
\clearpage
\setcounter{page}{1}
\begin{center}
{\LARGE Supplementary material for ``GRASP: A Goodness-of-Fit Test for Classification Learning''}
\vspace{1cm}

{\Large Adel Javanmard and Mohammad Mehrabi}
\end{center}
\appendix
\section{Proofs of theorems and lemmas}

\subsection{Proof of Lemma \ref{lemma: f-functions}}\label{proof: lemma: f-functions} 
Using \eqref{eq: bern-f} for $f(t)=\frac{1}{2}|t-1|$ we get
 \begin{align*}
\E\left[D_{\tv}(\bern(\eta(X))\|\bern(\heta(X)))\right]&=\E\left[\frac{\heta(X)}{2}\Big|\frac{\eta(X)}{\heta(X)}-1 \Big| + 
\frac{1-\heta(X)}{2}\Big|\frac{1-\eta(X)}{1-\heta(X)}-1 \Big| \right]\\
&=\E\left[ |\heta(X)-\eta(X)|\right]\,.
\end{align*}
By a similar argument for $f(t)=t\log(t)$ we get
\begin{align*}
\E\left[ D_{\kell}(\bern(\eta(X)\|\bern(\heta(X)))\right]
&=\E\left[ \heta(X)\frac{\eta(X)}{\heta(X)}\log \frac{\eta(X)}{\heta(X)} + 
(1-\heta(X))\frac{1-\eta(X)}{1-\heta(X)}\log \frac{1-\eta(X)}{1-\heta(X)}  \right]\\
&=\E\left[ \eta(X)\log \frac{\eta(X)}{\heta(X)} + 
(1-\eta(X))\log \frac{1-\eta(X)}{1-\heta(X)}  \right]\,.
\end{align*}
We then expand the expression inside logarithm to get 
\begin{align*}
\E\left[ D_{\kell}(\bern(\eta(X)\|\bern(\heta(X)))\right]
&=-\E\left[ \eta(X)\log{\heta(X)}+
(1-\eta(X))\log{(1-\heta(X))}  \right]\\
&\;+ \E\left[ \eta(X)\log{\eta(X)}+(1-\eta(X))\log{(1-\eta(X))}  \right]\\
&=\CE(\heta)-\CE(\eta)\,.
\end{align*}
Next, we prove the result for the Hellinger distance. By using \eqref{eq: bern-f} for $f(t)=(\sqrt{t}-1)^2$ we get
\begin{align*}
\E\left[ D_{\mathsf{H}}(\bern(\eta(X)||\bern(\heta(X)))\right]
&=\E\left[ \heta(X) \left(\sqrt\frac{\eta(X)}{\heta(X)}-1\right)^2+(1-\heta(X)) \left(\sqrt\frac{1-\eta(X)}{1-\heta(X)}-1\right)^2\right]\\
&=\E\left[\Big(\sqrt{\eta(X)}-\sqrt{\heta(X)}\Big)^2+ \Big(\sqrt{1-\eta(X)}-\sqrt{1-\heta(X)}\Big)^2\right]\,.
\end{align*}

\subsection{Proof of Proposition \ref{propo: f}}\label{proof: propo: f}

In oder to compute $\E[D_f(\cL(W|X)\|\unif([0,1]))]$, we first need to characterize the conditional distribution $\cL(W|X)$.  From the mechanism to construct $w$, it is easy to obtain
\begin{align}
p_{W|X}(w|x)&= \prob(y=+1|x)p_{W|Y,X}(w|y=+1,x) +\prob(y=0|x)p_{W|Y,X}(w|y=0,x)\nonumber\\
&=\frac{\eta(x)}{\heta(x)}\ind(w\leq \heta(x))+\frac{1-\eta(x)}{1-\heta(x)}\ind(w\geq \heta(x))\,. \label{eq: tw|x}
\end{align}
Using \eqref{eq: tw|x} in $f$-divergence definition \ref{def: f-div} yields 
\begin{align*}
\E\left[D_f(\cL(W|X)\|\unif([0,1]))\right]&=\E\left[\int_{0}^{1}f(p_{W|X}(w|X))\de w  \right]\\
&=\E\left[\int \limits_{0}^{\heta(X)} f\left(\frac{\eta(X)}{\heta(X)}\right) \de t+
\int \limits_{\heta(X)}^{1} f\left(\frac{1-\eta(X)}{1-\heta(X)}\right)\de t
\right]\,.\\
&=\E\left[\heta(X)f\left(\frac{\eta(X)}{\heta(X)}\right)+(1-\heta(X))f\left(\frac{1-\eta(X)}{1-\heta(X)}\right) \right]\\
&=\E\left[ D_f(\bern(\eta(X)\|\bern(\heta(X)))\right]\,.
\end{align*}
 \subsection{Proof of Proposition \ref{propo: compare}}\label{proof: propo: compare}
 From the optimization problems used in the definition of statistics  $U_\tau^{\finite}(\bV_{n,L})$ and $U_\tau^{\asym}(\bV_{n,L})$ in \eqref{eq: test-stats}, it is easy to get 
 $U_\tau^{\finite}(\bV_{n,L})\leq U_\tau^{\asym}(\bV_{n,L})$. In fact, this is due to the extra $1/L$ term in the denominator of  the test statistic  $ U_\tau^{\finite}(\bV_{n,L})$. We also need to compare the rejection thresholds. For this end, we first provide the following upper bound on the quantiles of a chi-squared distribution with $L-1$ degrees of freedom.  By using  (\cite{birge2001alternative}, Lemma 8.1),
  \[
  \chi^2_{L-1}(1-\alpha)\leq L-1+2\sqrt{(L-1)\log\frac{1}{\alpha}}+2\log\frac{1}{\alpha}\,.
  \]
Also it is easy to get that for $L\geq 60$ the following holds (we prove this later):
  \begin{equation}\label{eq: compare: tmp1}
  L-1+2\sqrt{(L-1)\log\frac{1}{\alpha}}+2\log\frac{1}{\alpha}\leq L+\sqrt{\frac{2L}{\alpha}}\,.
  \end{equation}
 Putting all together and by recalling the definition of $\Phi_{n,L,\alpha,\tau}^{\asym}$ we get
\begin{align*}
\Phi_{n,L,\alpha,\tau}^{\asym}&=\ind(U_\tau^{\asym}(\bV_{n,L})\geq \chi^2_{L-1}(1-\alpha))\\
&\geq \ind(U_\tau^{\finite}(\bV_{n,L})\geq \chi^2_{L-1}(1-\alpha))\\
&\geq  \ind\left(U_\tau^{\finite}(\bV_{n,L})\geq L+\sqrt{\frac{2L}{\alpha}} \right)=\Phi_{n,L,\alpha,\tau}^{\asym}\,.
\end{align*}
  
We only need to prove \eqref{eq: compare: tmp1}. By upper bounding $L-1$ with $L$ and multiplying both sides of \eqref{eq: compare: tmp1} by $0.5\sqrt{\alpha/L}$ the relation reduces to showing
  \[
  \sqrt{\alpha\log \frac{1}{\alpha}}+\sqrt{\frac{\alpha}{L}}\log\frac{1}{\alpha}\leq \frac{1}{\sqrt{2}}\,.
  \]
By simple algebraic calculation, it is easy to see that $\alpha\log{1/\alpha}\leq 1/e$ and $\sqrt{\alpha}\log{1/\alpha}\leq 2/e$ and so 
\[
 \sqrt{\alpha\log \frac{1}{\alpha}}+\sqrt{\frac{\alpha}{L}}\log\frac{1}{\alpha}\leq \frac{1}{\sqrt{e}}+\frac{2}{e\sqrt{L}}\,.
\]
Next note that  for $L\geq 55$ we have $ \frac{1}{\sqrt{e}}+\frac{2}{e\sqrt{L}}
 \leq \frac{1}{\sqrt{2}}$, which completes the proof.

\subsection{Proof of Proposition \ref{propo: discretize-distance}}\label{proof: propo: discretize-distance}
It is easy to observe that the joint distribution of $(V_1,\dots,V_L)$ is a multinomial distribution with probabilities $(p_1,\dots,p_{L})$ where $p_\ell$ denotes the probability of sample $(X,W)$ admitting label $\ell$ for $\ell\in [L] $.  As per Algorithm \ref{algorithm: balls-bins}, let $\tW_1,\dots,\tW_M$ be iid random variables with uniform distribution $\unif[0,1]$. To lighten the notation, we use the shorthands $T=T(X,W)$, and $T_j=T(X,\tW_j)$ for $j\in[M]$. In this case, the rank value is given by
\[
R={1+\sum_{j=1}^M \ind\big(T\geq T_j\big)}\,.
\]
 We get
\begin{align*}
\prob( (X,W) \text{ admits label } \ell)&=\sum_{j=K(\ell-1)}^{K\ell-1}\E[\prob(R=j+1|X)]\\
&=\sum_{j=K(\ell-1)}^{K\ell-1}\E\left[\prob(T \text{ is exactly larger than } j \text{ of } T_{j\in [M]}|X )  \right]\,.
\end{align*} 
We know that conditioned on $X$, random variables $T$, $T_{j\in [M]}$ are independent. This gives us 
\begin{align}
\prob( (X,W) \text{ admits label } \ell)&=\sum_{j=K(\ell-1)}^{K\ell-1}\E\left[\int\prob( \text{ exactly } j \text{ of } T_{j\in [M]} \text{ are smaller than } t|X ) \de P_{T|X}(t|X)   \right]\nonumber \\
&=\sum_{j=K(\ell-1)}^{K\ell-1}\E\left[\int \binom{M}{j} F_c(t;X)^{j}(1-F_c(t;X))^{M-j}   \frac{\partial F_o(t;X)}{\partial t} \de t  \right]\label{eq: tmp-prob} \,.
\end{align} 
 The last relation follows the iid property of random variables $T_{j\in [M]}$ conditioned on $X$ along with the definitions of $F_c(t;x), F_o(t;x)$ as per Assumption \ref{assumption: cdfs}.  By adopting the change of variable $u=F_c(t;X)$ in the integral \eqref{eq: tmp-prob} we arrive at
 \[
 \prob( (X,W) \text{ admits label } \ell) =\sum_{j=K(\ell-1)}^{K\ell-1}\E\left[\int_{0}^1 \binom{M}{j} u^{j}(1-u)^{M-j}  \frac{\partial F_o(F_c^{-1}(u;X);X)}{\partial u} \de u  \right]\,.
 \]

We next recall the function $\psi(u;X)=\frac{\partial F_o(F_c^{-1}(u;X);X)}{\partial u}$ from Assumption \ref{assumption: cdfs}, which allows us to write the above probability as

\begin{equation}\label{eq: p_ell}
p_\ell=\sum_{j=K(\ell-1)}^{K\ell-1} \binom{M}{j} \E\left[\int_{0}^{1}u^{j}(1-u)^{M-j}   \psi(u;X) \de u\right] \,. \quad \forall \ell\in [L]\,.
\end{equation}
For $\ell \in [L]$, we define the polynomial $\beta_\ell:[0,1]\to\reals$ as
\[
\beta_\ell(u)=\sum_{j=K(\ell-1)}^{K\ell-1} \binom{M}{j} u^{j}(1-u)^{M-j}\,.
\]
By using the above  in \eqref{eq: p_ell}, we obtain
\begin{equation}\label{eq: p_ell psi-above}
p_\ell= \E\left[\int_{0}^{1}\beta_\ell(u)   \psi(u;X) \de u\right] \,. \quad \forall \ell\in [L]\,,
\end{equation}
and hence
\begin{equation}\label{eq: D_f V_ell psi}
\sum_{\ell=1}^{L}\frac{f(p_\ell L)}{L}=  \frac{1}{L}\sum_{\ell=1}^{L}f\left(L \E\Big[\int_0^1 \beta_\ell(u)\psi(u;X) \de u\Big]\right)\,.
\end{equation}
We next upper bound the right-hand side in the above equation. Consider the following probability density function over interval $[0,1]$:
\begin{equation}\label{eq: phi}
\vphi_\ell(u)=\frac{\beta_\ell(u)}{\int_{0}^1\beta_\ell(u)\de u}\,,\quad \forall u\in[0,1]\,.
\end{equation}
By an application of Jensen's inequality and the fact that $f$ is a convex function, we have
\begin{equation}\label{eq: jensen-phi}
f\left(\E_{X}\Big[\E_{u\sim \vphi_\ell}[\psi(u;X)]\Big]\right)\leq \E_X\left[ \E_{u\sim \vphi_\ell}\Big[f (\psi(u;X))\Big]\right]\,.
\end{equation}
Then, by rewriting the expectation of $u$ in terms of density function $\vphi_\ell$  we get
\begin{equation}\label{eq: jenson-f}
f\left(\E_X\left[\int_0^1 \vphi_\ell(u) \psi(u;X) \de u\right] \right) \leq \E_X\left[ \int_0^1 \vphi_\ell(u)  f(\psi(u;X))\de u\right]\,.
\end{equation}
Plugging \eqref{eq: phi} into \eqref{eq: jenson-f} yields
\begin{equation}\label{eq: jenson-beta}
f\left(\E\left[ \frac{\int_0^1 \beta_\ell(u)\psi(u;X)\de u}{\int_{0}^1\beta_\ell(u)\de u} \right] \right) \leq 
\E\left[\frac{\int_0^1\beta_\ell(u) f(\psi(u;X))\de u} {\int_{0}^1\beta_\ell(u)\de u} \right]  \,.
\end{equation}
In addition, from the definition of $\beta_\ell(u)$ we have
\begin{align}
\int_0^1 \beta_\ell(u)\de u&= \sum_{j=K(\ell-1)}^{K\ell-1} \binom{M}{j}\int_0^1 u^{j}(1-u)^{M-j}\de u\nonumber\\
&=\sum_{j=K(\ell-1)}^{K\ell-1} \binom{M}{j} B(j+1,M-j+1)\nonumber\\
&=\sum_{j=K(\ell-1)}^{K\ell-1} \binom{M}{j} \frac{j!(M-j)!}{(M+1)!}=\frac{1}{L} \,,\label{eq: sum-bernstein} 
\end{align}
where $B(a,b)$ is the beta function. Using \eqref{eq: sum-bernstein} in \eqref{eq: jenson-beta} brings us to

\begin{equation} \label{eq: f-main-inequality}
f\left(L\E\Big[\int_0^1 \beta_\ell(u)\psi(u;X)\de u\Big] \right) \leq 
L\E\left[\int_0^1\beta_\ell(u) f(\psi(u;X))\de u \right]
\end{equation}

In the next step, combining \eqref{eq: f-main-inequality} with \eqref{eq: D_f V_ell psi} gives us

\begin{align}
 \sum_{\ell=1}^{L}\frac{f(p_\ell L)}{L}&\leq  \sum_{\ell=1}^{L} \E\left[\int_0^1 \beta_\ell(u)f(\psi(u;X)) \de u\right]\nonumber \\
&= \E\left[\int_0^1 f(\psi(u;X))\de u \right]\label{eq: discrete-f-upper-bound} \,,
\end{align}
where in the last relation we used the identity $\sum_{\ell=1}^L \beta_\ell(u)=1$, because
\begin{align*}
\sum_{\ell=1}^L \beta_\ell(u)&=\sum_{\ell=1}^{L}\sum_{j=K(\ell-1)}^{K\ell-1}\binom{M}{j}u^j(1-u)^{M-j}\\
&=\sum_{j=0}^{M}\binom{M}{j}u^j(1-u)^{M-j}=(u+1-u)^M=1\,.
\end{align*} 

We next use definition of $\psi(u;x)$ in \eqref{eq: discrete-f-upper-bound} to obtain
\begin{align}
\E\left[\int_0^1 f(\psi(u;X))\de u\right]&=   \E\left[\int_0^1 f\Big(\frac{\partial F_o(F_c^{-1}(u;X);X)}{\partial u}\Big)\de u\right]  
\label{eq: f-psi-o-c upper bound}\,.
\end{align}
In the next step, by using density functions $f_o(t;x),f_c(t;x)$ as per Assumption \ref{assumption: cdfs} in \eqref{eq: f-psi-o-c upper bound} we arrive at
\begin{align}
\E\left[\int_0^1 f(\psi(u;X))\de u\right]&=\E\left[ \int_0^1  f\Big( \frac{f_o(F_c^{-1}(u;X);X)}{f_c(F_c^{-1}(u;X);X)}  \Big) \de u\right] \nonumber\\
&=  \E\left[ \int_0^1 f\Big( \frac{f_o(t;X)}{f_c(t;X)}  \Big) f_c(t;X) \de t\right]\nonumber\\
&=\E\left[D_f\Big(\cL(T(X,W)|X)\|\cL(T(X,\tW)|X)\Big)\right]\label{eq: tmp-upper-f}\,,
\end{align}
where the last relation comes from Assumption \ref{assumption: cdfs} and the definition of $f$-divergence function. In the next step, by combining \eqref{eq: tmp-upper-f} and \eqref{eq: discrete-f-upper-bound} we get
\begin{align}\label{eq: tmp-upper-f-3} 
\sum_{\ell=1}^{L}\frac{f(p_\ell L)}{L} \leq  \E\Big[ D_f\left(\cL(T(X,W)|X)\|\cL(T(X,\tW)|X)\right)\Big]\,.
\end{align}
Further, by an application of the data processing inequality we have
 \begin{equation}\label{eq: tmp-upper-f-2}
 D_f\left(\cL(T(X,W)|X)\|\cL(T(X,\tW)|X)\right)\leq D_f\left(\cL(W|X)\|\cL(\tW|X)\right)\,.
 \end{equation}
 Finally, combining \eqref{eq: tmp-upper-f-2} and \eqref{eq: tmp-upper-f-3}  gives us
\begin{align*}
\sum_{\ell=1}^{L}\frac{f(p_\ell L)}{L} &\leq  \E\Big[D_f\left(\cL(T(X,W)|X)\|\cL(T(X,\tW)|X)\right)\Big]  \\
&\leq \E\left[D_f(\cL(W|X)\|\cL(\tW|X))\right]\,.
\end{align*}
\subsection{Proof of Theorem \ref{thm: size-DF}}\label{proof: hm: size-DF}
We know that $\bV_{n,L}$ has a multinomial distribution. Let $q_\ell$ denote the probability of occurrence for category $\ell\in [L]$. 
Invoking the result of \cite[Theorem 2]{balakrishnan2019hypothesis} for multinomial hypothesis testing with truncated chi-squared test statistics, we get

\begin{align}
	\alpha&\geq \prob\left(\sum_{\ell=1}^{L} \frac{(V_\ell-nq_\ell)^2-V_\ell}{\max\{q_\ell,\frac{1}{L} \}} \geq n\sqrt{ \frac{2}{\alpha} \sum\limits_{\ell=1}^{L} \left( \frac{q_\ell}{\max\{q_\ell,1/L \}} \right)^2 }    \right)\,.
\end{align}
In the next step, we use $\max(q_\ell,1/L)\geq 1/L$ along with $\sum_{\ell=1}^L q_\ell=1$ to get

\begin{align}
	\alpha&\geq \prob\left(\sum_{\ell=1}^L \frac{(V_\ell-nq_\ell)^2-V_\ell}{\max\{q_\ell,\frac{1}{L} \}} \geq n\sqrt{ \frac{2}{\alpha}L}     \right)\nonumber\\
	&= \prob \left(\sum_{\ell=1}^L \frac{(V_\ell-nq_\ell)^2}{\max\{q_\ell,\frac{1}{L} \}} \geq \sum\limits_{\ell=1}^{L}\frac{V_\ell}{\max\{q_\ell,\frac{1}{L} \}} + n\sqrt{ \frac{2}{\alpha}L}     \right)\nonumber\\\
	&\geq \prob \left(  \sum_{\ell=1}^{L}\frac{(V_\ell-nq_\ell)^2}{\max\{q_\ell,\frac{1}{L} \}} \geq L\sum\limits_{\ell=1}^{L}{V_\ell} +n\sqrt{ \frac{2}{\alpha}L}     \right)\nonumber\\
	&=\prob \left(\sum_{\ell=1}^L \frac{(V_\ell-nq_\ell)^2}{\max\{q_\ell,\frac{1}{L} \}} \geq n L+n\sqrt{ \frac{2}{\alpha}L}     \right) \nonumber\\
	&\geq \prob \left(\frac{1}{n}\sum_{\ell=1}^L \frac{(V_\ell-nq_\ell)^2}{q_\ell+\frac{1}{L} } \geq L+\sqrt{ \frac{2}{\alpha}L}     \right) \label{eq: V-prob-bound}\,.
\end{align}
By deploying Theorem \ref{thm: discretize-distance}, under the null  hypothesis \eqref{eq: null} we have
\begin{equation}\label{eq: size-tmp2}
\frac{1}{L}\sum_{\ell=1}^L f(Lq_\ell)\leq \E\left[D_f(\bern(\eta(X))\|\bern(\heta(X)))\right]\le \tau \,.
\end{equation}
This implies that  $(q_1,\dots,q_L)$ is a feasible point for the optimization problem in the definition of  $U_\tau^{\finite}(\bV_{n,L})$ in \eqref{eq: test-stats}, and so
\begin{equation}\label{eq: U_tau_upper}
\frac{1}{n}\sum_{\ell=1}^L \frac{(V_\ell-nq_\ell)^2}{q_\ell+\frac{1}{L} } \geq U_\tau^{\finite}(\bV_{n,L})\,.
\end{equation}
Plugging \eqref{eq: U_tau_upper} into \eqref{eq: V-prob-bound} yields
\[
 \prob \left( U_\tau^{\finite}(\bV_{n,L}) \geq L+\sqrt{ \frac{2}{\alpha}L}     \right) \leq \alpha\,.
\]
This completes the proof by the definition of decision rule $\Phi^{\finite}_{n,L,\tau}$.
  
We next proceed to the proof of the asymptotic result. 
Following a similar argument shows that $(q_{1}, \dotsc, q_{\ell})$ is also a feasible solution for the optimization problem in the definition of $U^{\asym}_{\tau}(\bV_{n,L})$, which implies that 
\begin{equation}\label{eq: tU upper bound}
\sum_{\ell=1}^{L}\frac{(V_\ell-nq_\ell)^2}{q_\ell}\geq U^{\asym}_{\tau}(\bV_{n,L})\,.
\end{equation}

In the next step, we use the following asymptotic result on the Pearson’s chi-squared test statistic (see e.g., \cite{lann1959testing} Theorem 14.3.1):

\begin{equation}\label{eq: tU-converge}
\lim_{n\to \infty}\sum\limits_{\ell=1}^{L}\frac{(V_\ell-nq_\ell)^2}{nq_\ell} \overset{(d)}{\to} \chi^2_{L-1}\,,
\end{equation}
where $\chi^2_{L-1}$ denotes the chi-squared distribution with $L-1$ degrees of freedom. As a direct result of \eqref{eq: tU-converge} we have
\[
\lim_{n\to \infty}\sup\prob\left(\sum\limits_{\ell=1}^{L}\frac{(V_\ell-nq_\ell)^2}{q_\ell} \geq \chi^2_{L-1}(1-\alpha)\right) =\alpha\,.
\]
Combining \eqref{eq: tU upper bound} with the above relation yields
\[
\lim_{n\to \infty}\sup\prob\left(  U^{\asym}_{\tau}(\bV_{n,L}) \geq \chi^2_{L-1}(1-\alpha)\right) \leq \alpha\,.
\]
The claim then follows simply from the definition of decision rule $\Phi^{\asym}_{n,L,\tau}$. 

\subsection{Proof of Proposition \ref{propo: f-model-x}}\label{proof: propo: f-model-x}
Let $\tX\sim \cP_X$ and $\tW\sim\unif([0,1])$, then we have
\begin{align*}
D_{f}\left(\cL(X,W)\|\cP_X\times \unif([0,1])\right)&=D_{f}\left(\cL(X,W)\|\cL(\tX,\tW)\right)\\
&=\int p_{\tX,\tW}(x,w)f\left(\frac{p_{X,W}(x,w)}{p_{\tX,\tW}(x,w)}\right)\de x \de w\\
&=\int p_{\tX}(x)p_{\tW|\tX}(w|x)f\left(\frac{p_X(x)p_{W|X}(x)}{p_{\tX}(x)p_{\tW|\tX}(w|x)}\right)\de x \de w\,.
\end{align*}
Since $p_{\tX}=p_X$ and $p_{\tW|\tX}(w|x)=1$(uniform distribution) we get
\begin{align}\label{eq: tmp-joint}
D_{f}\left(\cL(X,W)\|\cL(\tX,\tW)\right)&=\E\left[\int_0^1f\left({p_{W|X}(w|X)}\right)\de w\right]\,.
\end{align}
Also from the construction of $w$, it is easy to see that
\begin{align}\label{eq: tmp-conditional}
p_{W|X}(w|x)&= \prob(y=+1|x)p_{W|Y,X}(w|y=+1,x) +\prob(y=0|x)p_{W|Y,X}(w|y=0,x)\nonumber\\
&=\frac{\eta(x)}{\heta(x)}\ind(w\leq \heta(x))+\frac{1-\eta(x)}{1-\heta(x)}\ind(w\geq \heta(x))\,. \end{align}
Plugging \eqref{eq: tmp-conditional} into \eqref{eq: tmp-joint} brings us to
\begin{align*}
D_{f}\left(\cL(X,W)\|\cL(\tX,\tW)\right)&=\E\left[\int_{0}^{\heta(X)}f\left(\frac{\eta(X)}{\heta(X)}\right)\de w +  \int_{\heta(X)}^{1}f\left(\frac{1-\eta(X)}{1-\heta(X)}\right)\de w \right]\\
&=\E\left[\heta(X)f\left(\frac{\eta(X)}{\heta(X)}\right) +  (1-\heta(X))f\left(\frac{1-\eta(X)}{1-\heta(X)}\right)\right]\\
&=\E\left[D_f\left(\bern(\eta(X))\|\bern(\heta(X))\right) \right]\,,
\end{align*}
which completes the proof. 
\subsection{Proof of Proposition \ref{propo: CI}}\label{proof: propo: CI}
We adopt the shorthand $\tau_1$ for $\tau^{\finite}_{n,L,\alpha}$ and $\tau_0$ for $\E[D_f(\bern(\eta(X))\|\bern(\heta(X)))]$. From the definitions, $\tau_1$ is a function of $\bV_{n,L}$ and is random.  We only prove the result for the finite test statistics. The claim for the asymptotic statistics follows from the same argument. 

First, from the optimization problem used for $U_{\tau}^{\finite}(\bV_{n,L})$ in \eqref{eq: test-stats} it is easy to observe that for fixed values of $\bV_{n,L}$ statistics $U_{\tau}^{\finite}(\bV_{n,L})$ is non-increasing in $\tau$.  Therefore,
\begin{align*}
\prob(\E[D_f(\bern(\eta(x))\|\bern(\heta(x)))]\geq \tau_1)&=\prob(\tau_0\geq \tau_1 )\\
& = 1- \prob(\tau_1\geq \tau_0)\\
&\ge1-\prob(U^{\finite}_{\tau_0}(\bV_{n,L})\geq U^{\finite}_{\tau_1}(\bV_{n,L}) )\,.
\end{align*}
Second, from the definition of $\tau_1$ we have $U^{\finite}_{\tau_1}\geq L+\sqrt{2L/\alpha}$. Plugging this into the above relation yields
\[
\prob(\E_x[D_f(\bern(\eta(x))\|\bern(\heta(x)))]\geq \tau_1)\geq 1-\prob(U^{\finite}_{\tau_0}(\bV_{n,L})\geq  L+\sqrt{2L/\alpha} )\,.
\]
We next invoke the result of  Theorem \ref{thm: size-DF}, on the size of the test statistics $U^{\finite}_{\tau}(\bV_{n,L})$. When $\tau\geq \tau_0$ the null hypothesis holds and so we have 
\[
\prob(U^{\finite}_{\tau}(\bV_{n,L}) \geq L+\sqrt{2L/\alpha})\leq \alpha\,.
\]
This completes the proof. 

{\color{black}

\subsection{Proof of Proposition \ref{propo: K_L grows-df}} \label{proof: propo: K_L grows-df}
The first inequality in Proposition \ref{propo: K_L grows-df} is a direct result of Jenson's inequality applied on the convex function $f$. The last equation is also shown in the proof of Proposition \ref{propo: discretize-distance}, for the complete proof we refer to the chain of relations started in \eqref{eq: f-psi-o-c upper bound}. In this section, we focus on proving the first equation, in which we need to show that



\begin{align}\label{eq:claim-2-df}
 \lim_{L\to \infty}\lim_{K\to \infty}  \frac{1}{L}\sum_{\ell=1}^L f(Lp_\ell) = \int_0^1 f(\E[\psi(u;X)] )\de u\,.
\end{align}
Let $b_j(M,u)$ denote the $j$-th Bernstein polynomial of degree $M$, which is given by
\[
b_j(M,u)=\binom{M}{j}u^{j}(1-u)^{M-j} \,.
\]
Then the probability $p_\ell$ given in \eqref{eq: p_ell} can be written as
\begin{align}\label{p_ell-tmp-dff}
p_\ell=\sum_{j=K(\ell-1)}^{K\ell-1}  \int_0^1 b_{j}(M,u) \E[\psi(u;X)]  \de u \,. \quad \forall \ell\in [L]\,
\end{align}
In the next step, from the first assumption we have a.s.  $|\psi(u;X)|\leq C$, and by an application of dominated convergence theorem we can change the order of derivative and expectation and get
\[ 
 \frac{\de}{\de u}\E[\omega(u;X)]= \E[\psi(u;X)]\,.
\]
Next, by using this in \eqref{p_ell-tmp-dff} and then by partial integration we arrive at
\begin{equation}\label{eq: p_ell_bernstein-df}
p_\ell=-\sum_{j=K(\ell-1)}^{K\ell-1}  \int_0^1 \frac{\partial b_{j}(M,u)}{\partial u}  \E[\omega(u;X)] \de u \,. \quad \forall \ell\in [L]\,.
\end{equation}

On the other hand, by simple algebraic calculation, it is easy to get the following identity for Bernstein polynomials:
\begin{equation}\label{eq: bernstein-identity-df}
\frac{\partial b_j(M,u)}{\partial u}=M(b_{j-1}(M-1,u)-b_{j}(M-1,u))\,,
\end{equation}
where we set the convention $\binom{n}{k}=0$ for $k>n$ and $k<0$.  
Using \eqref{eq: bernstein-identity-df} in \eqref{eq: p_ell_bernstein-df} gives us
\[
p_\ell=M\sum_{j=K(\ell-1)}^{K\ell-1}\int_{0}^1 \E[\omega(u;X)](b_j(M-1,u)-b_{j-1}(M-1,u)) \de u \,. \quad \forall \ell\in [L]\,
\]
Moving the summation inside the integration we obtain

\begin{equation}\label{eq: p_ell-bernstein-simple-df}
p_\ell=M\int_0^1 \E[\omega(u;X)](b_{K\ell-1}(M-1,u)-b_{K(\ell-1)-1}(M-1,u)) \de u \,. \quad \forall \ell\in [L]\,
\end{equation}
On the other hand, it is easy to check that
\begin{align}\label{eq: bernstein-identity2-df}
\int_0^1 b_{K\ell-1}(M-1,u)\de u&=\int_0^1 b_{K(\ell-1)-1}(M-1,u)\de u=\frac{1}{M}\,.
\end{align}
Using the identities \eqref{eq: bernstein-identity2-df} and \eqref{eq: p_ell-bernstein-simple-df}, we write
\begin{align*}
\E\left[\omega\Big(\frac{\ell}{L};X\Big)\right]-\E\left[\omega\Big(\frac{\ell-1}{L};X\Big)\right]-{p_\ell} &= M\int_0^1\bigg(\E\left[\omega\Big( \frac{\ell}{L};X\Big)\right]- \E\left[\omega(u;X)\right] \bigg) b_{K\ell-1}(M-1,u)\de u\\
&\;+M\int_0^1\bigg(\E[\omega(u;X)]-\E\left[\omega\Big( \frac{\ell-1}{L};X\Big)\right]  \bigg) b_{K(\ell-1)-1}(M-1,u)\de u\,.
\end{align*}
Since almost surely $|\frac{\partial}{\partial u}\psi(u;X)|\leq C$ for  $ u \in (0,1)$, we realize that  $|\psi(u;X)|$ is bounded by $C$ almost surely, thereby $\omega(u;X)$ is $C$-Lipschitz. Using this along with the triangle inequality we get
\begin{align}
\left|\E\left[\omega\Big(\frac{\ell}{L};X\Big)\right]-\E\left[\omega\Big(\frac{\ell-1}{L};X\Big)\right]-{p_\ell} \right| &\leq MC\int_0^1\Big| \frac{\ell}{L}-u \Big| b_{K\ell-1}(M-1,u)\de u\nonumber \\
&\;+ MC \int_0^1\Big|u-\frac{\ell-1}{L}\Big| b_{K(\ell-1)-1}(M-1,u)\de u \label{eq: tmp-bernstein-df}\,.
\end{align}
We continue by writing the above expressions in terms of a Beta distribution.
Recall that the pdf of Beta distribution $\mathsf{Beta}(\alpha,\beta)$ with shape parameters $(\alpha, \beta)$ is given by $\frac{u^{\alpha-1}(1-u)^{\beta-1}}{B(\alpha,\beta)}$, where  $B(\alpha, \beta):=\int_0^1 u^{\alpha-1}(1-u)^{\beta-1}\de u$. For integer values $a,b$, $B(a,b)$ has a closed form and is given by $B(a,b) = \frac{(a-1)! (b-1)!}{(a+b-1)!}$. Therefore, we have
\begin{align*}
M b_{K\ell-1}(M-1,u) &= M {M-1\choose K\ell-1} u^{K\ell-1}(1-u)^{M-K\ell}\\
& = \frac{M!}{(K\ell-1)!(M-K\ell)!} u^{K\ell-1}(1-u)^{M-K\ell}\\
& = \frac{u^{K\ell-1}(1-u)^{M-K\ell}}{B(K\ell,M-K\ell+1)}\,.
\end{align*}
Using a similar expression for $M b_{K(\ell-1)-1}(M-1,u)$ we write \eqref{eq: tmp-bernstein-df} as follows:
\begin{align}
\frac{1}{C}\left|\E\left[\omega\Big(\frac{\ell}{L};X\Big)\right]-\E\left[\omega\Big(\frac{\ell-1}{L};X\Big)\right]-{p_\ell} \right|
&\leq \E_{u\sim \mathsf{Beta}(K\ell,M-K\ell+1)}\left[\Big| \frac{\ell}{L}-u \Big|\right] \nonumber \\ 
&\;+  \E_{u\sim \mathsf{Beta}(K(\ell-1),M-K(\ell-1)+1)}\left[\Big|u-\frac{\ell-1}{L}\Big|\right] \label{eq: tmp-bernstein2-df}\,.
\end{align}
Since the mean of $\mathsf{Beta}(\alpha,\beta)$ is given by $\alpha/(\alpha+\beta)$, the above Beta distributions have mean values of $\ell/L$ and $(\ell-1)/L$, respectively (recall that $M+1 = KL$). Therefore the terms on the right-hand side of \eqref{eq: tmp-bernstein2-df} are indeed the mean absolute deviation of two Beta distributions. Further, using Jenson's inequality we know that for arbitrary random variable $S$ we have
\[
\E[|S-\E[S]|]\leq \var[S]^{1/2}\,.
\] 
Using this in \eqref{eq: tmp-bernstein2-df} yields
\begin{align}
\frac{1}{C}\left|\E\left[\omega\Big(\frac{\ell}{L};X\Big)\right]-\E\left[\omega\Big(\frac{\ell-1}{L};X\Big)\right]-{p_\ell} \right| &\leq  \var\left[\mathsf{Beta}(K\ell,M-K\ell+1)\right]^{1/2} \nonumber\\ 
&\;+  \var\left[\mathsf{Beta}(K(\ell-1),M-K(\ell-1)+1)\right]^{1/2} \label{eq: tmp-bernstein3-df}\,.
\end{align}
In the next step, note that
\begin{equation}\label{eq: var-beta-df}
\var\left( \mathsf{Beta}(\alpha,\beta)\right)=\frac{\alpha\beta}{(\alpha+\beta)^2(\alpha+\beta+1)}\leq \frac{1}{4(\alpha+\beta)}\,,
\end{equation}
where the last inequality follows from $(\alpha+\beta)^2\geq 4\alpha\beta$. 
Combining \eqref{eq: tmp-bernstein3-df} and \eqref{eq: var-beta-df} yields
\begin{equation}\label{eq: tmp-mean-value-df}
\frac{1}{C}\left|\E\left[\omega\Big(\frac{\ell}{L};X\Big)\right]-\E\left[\omega\Big(\frac{\ell-1}{L};X\Big)\right]-{p_\ell} \right|\leq \frac{2}{2\sqrt{M+1}} = \frac{1}{\sqrt{KL}}\,.
\end{equation}
Now from the mean value theorem, we know that there exists $s_\ell(X) \in \left[\frac{\ell-1}{L}, \frac{\ell}{L}\right]$ such that 
\begin{equation}\label{eq: mean-value-df}
\omega\Big(\frac{\ell}{L};X\Big)-\omega\Big(\frac{\ell-1}{L};X\Big)=\frac{1}{L}\psi(s_\ell(X);X)\,.
\end{equation}
Combining \eqref{eq: mean-value-df} and \eqref{eq: tmp-mean-value-df} yields
\begin{equation}\label{eq: Lp_ell-df}
\lim_{K\to \infty} Lp_\ell=\E\left[\psi(s_\ell(X);X)\right]\,,~ \text{ where } s_\ell(x)\in \left[\frac{\ell-1}{L}, \frac{\ell}{L}\right]\,.
\end{equation}
We are now ready to prove the claim of~\eqref{eq:claim-2-df}. Introduce function $\sigma(u)=\E[\psi(u;X)]$, we have 
\begin{align}\label{eq: f-div-diff-df}
\frac{1}{L}\sum_{\ell=1}^Lf(L p_\ell)-\int_0^1f(\E\left[\psi(u;X)\right])\de u &=\frac{1}{L}\sum_{\ell=1}^L \left[f(L p_\ell)-L\int_{\frac{\ell-1}{L}}^{\frac{\ell}{L}}(f\circ\sigma)(u)\de u\right]\,.
\end{align}
Given that $\psi(u,X)$ is almost surely continuous, it is easy to observe that $\sigma(u)$ is also continuous. In the next step,  by another application of the mean value theorem for the continuous function $f\circ\sigma$,  there exists $t_\ell\in \left[\frac{\ell-1}{L}, \frac{\ell}{L}\right]$ such that 
\begin{equation}\label{eq: t_ell-df}
L\int_{\frac{\ell-1}{L}}^{\frac{\ell}{L}}(f\circ\sigma)(u)\de u=(f\circ \sigma)(t_\ell)\,.
\end{equation}
By combining \eqref{eq: f-div-diff-df} and \eqref{eq: t_ell-df} we get
\begin{align}\label{eq: f-div-diff-2-df}
\frac{1}{L}\sum_{\ell=1}^Lf(L p_\ell)-\int_0^1 f(\E\left[\psi(u;X)\right])\de u &=\frac{1}{L}\sum_{\ell=1}^L \left[f(L p_\ell)-f\circ\sigma(t_\ell) \right]\,,
\end{align}
and using continuity of $f$ along with \eqref{eq: Lp_ell-df} and \eqref{eq: f-div-diff-2-df} we obtain
\begin{align}\label{eq: f-div-diff-3-df} 
\lim_{K\to \infty}\frac{1}{L}\sum_{\ell=1}^Lf(L p_\ell)-\int_0^1f(\E[\psi(u;X)])\de u &=\frac{1}{L}\sum_{\ell=1}^L \left[f(\E[\psi(s_\ell(X);X)]) -f( \E[\psi(t_\ell;X) ] ) \right]\,.
\end{align}

For the rest of the proof, we show that for every $\eps>0$ for sufficiently large $L$ the right hand side of \eqref{eq: f-div-diff-3-df} is smaller than $\eps$. For this end, we start by the fact that since $f$ is a continuous function, therefore on the compact set $[0,C]$ must be uniformly continuous. This implies that there exists $\delta>0$ such that for every $r_1,r_2 \in [0,C] $ with $|r_1-r_2| \leq \delta$ we have $|f(r_1)-f(r_2)|\leq \eps$. In the next step, we define values $r_1^{(\ell)}=\E[\psi(s_\ell(X);X)]$ and $r_2^{(\ell)}=\E[\psi(t_\ell;X)]$.  We then claim that for sufficiently large value of $L$ such that  $L\geq \frac{C}{\delta}$ we have  $|r_1^{(\ell)}-r_2^{(\ell)}|\leq \delta$ and $r_1^{(\ell)},r_2^{(\ell)}\in [0,C]$ (proof of this claim is provided later). An immediate consequence of uniform continuity yields $|f(r_1^{(\ell)})-f(r_2^{(\ell)})|\leq \eps$, by plugging this into \eqref{eq: f-div-diff-3-df} we arrive at

\begin{align*} 
\left|\lim_{K\to \infty}\frac{1}{L}\sum_{\ell=1}^Lf(L p_\ell)-\int_0^1f(\E[\psi(u;X)])\de u \right|&=\frac{1}{L}\sum_{\ell=1}^L \left|f(\E[\psi(s_\ell(X);X)]) -f( \E[\psi(t_\ell;X) ] ) \right| \\
&= \frac{1}{L}\sum_{\ell=1}^L \left|f(r_1^{(\ell)}) -f(r_2^{(\ell)}) \right|\leq \eps\,.
\end{align*}

By letting $L$ go to infinity, $\eps$ can be chosen arbitrarily small and so
\begin{align} 
\lim_{L\to\infty}\lim_{K\to \infty}\frac{1}{L}\sum_{\ell=1}^Lf(L p_\ell)=\int_0^1f(\E[\psi(u;X)])\de u\,.
\end{align}

We are only left to prove our claim that for every $\ell\in [L]$ when $L\geq \frac{B}{\delta}$ we have $r_{1}^{(\ell)},r_2^{(\ell)} \in [0,C]$ and $|r_{1}^{(\ell)}-r_2^{(\ell)}|\leq \delta$. For this purpose,  given that almost surely for every  $ u \in (0,1)$ we have  $\psi(u;X)\leq C$, therefore $r_1^{(\ell)},r_2^{(\ell)} \in [0,C]$. In the next step, by an application of Jenson's inequality we have $|r_1^{(\ell)}-r_2^{(\ell)}| \leq  \E[|\psi(s_\ell(X);X)- \psi(t_\ell;X)|]$. We then use the second assumption stated in Proposition \ref{propo: K_L grows-df}, and get 
$ |r_1^{(\ell)}-r_2^{(\ell)}| \leq B\, \E[ |s_\ell(X)-t_\ell|]$. Finally, given that $s_\ell(X), t_\ell$ both belong to $\left[\frac{\ell-1}{L}, \frac{\ell}{L} \right]$, we arrive at $|r_1^{(\ell)}-r_2^{(\ell)}| \leq \frac{B}{L}$. Using $L \geq \frac{B}{\delta}$ completes the proof.

}

{\color{black} 
\subsection{Proof of Proposition \ref{propo: pval-crt}}\label{proof: propo: pval-crt} 

Proposition \ref{propo: f} states that when the null hypothesis \eqref{eq: null} holds (with $\tau=0$), the random variable $w$ is independent from $x$ and follows a uniform distribution over $[0,1]$.  Because of the symmetry among $\bw,\bw_1,\dots,\bw_M$, it is easy to check that the p-value $p$ takes values in $\{\frac{1}{M+1},\dots, \frac{M}{M+1},1\}$ uniformly at random. Since the p-value $p$ is discrete, for any $t\in [0,1]$, we have $\prob(p\leq t)=\frac{\lfloor (M+1)t\rfloor}{M+1}$. Therefore, $\prob(p\leq t)\leq t$, which implies that under the null hypothesis, the p-value $p$ is superuniform.

}

\subsection{Proof of Proposition \ref{propo: model-x}}\label{proof: propo: model-x}
The proof for the first part, is basically almost similar to the proof of Proposition \ref{propo: discretize-distance}. The minor difference is that the arguments should be followed for the function $\psi(u)$ from Assumption \ref{assumption: cdfs-model-x} instead of conditional functions $\psi(u;x)$ from Assumption \ref{assumption: cdfs}, and also careful treatment of conditional expectations with respect to covariates $x$. We provide the complete proof here for the reader’s convenience.

First, it is easy to observe that the joint distribution of $(V_1,\dots,V_L)$ is a multinomial distribution with probabilities $(p_1,\dots,p_{L})$ where $p_\ell$ denotes the probability of random variables $(X,W)$ admitting label $\ell$ for $\ell\in [L] $. In accordance with Algorithm \ref{algorithm: balls-bins-model-x}, each sample $(X,W)$ has $M$  counterfeits $(\tX_1,\tW_1),\dots,(\tX_M,\tW_M)$ where $\tW_i$ are iid random variables distributed as $\unif[0,1]$, and $\tX_i$ are iid from $\cP_X$. To lighten the notation, we use the shorthands $T=T(X,W)$, and $T_j=T(\tX_j,\tW_j)$ for $j\in[M]$. In this case, the rank value is given by
\[
R={1+\sum_{j=1}^M \ind\big(T\geq T_j\big)}\,.
\]
 We have
\begin{align*}
\prob( (X,W) \text{ admits label } \ell)&=\sum_{j=K(\ell-1)}^{K\ell-1}\prob(R=j+1)\\
&=\sum_{j=K(\ell-1)}^{K\ell-1}\prob(T \text{ is exactly larger than } j \text{ of } T_{j\in [M]})  \,.
\end{align*} 
We know that random variables $T$, $T_{j\in [M]}$ are independent, which gives us the following:
\begin{align}
\prob( (X,W) \text{ admits label } \ell)&=\sum_{j=K(\ell-1)}^{K\ell-1}\int\prob( \text{ exactly } j \text{ of } T_{j\in [M]} \text{ are smaller than } t ) \de P_{T}(t)  \nonumber \\
&=\sum_{j=K(\ell-1)}^{K\ell-1}\int \binom{M}{j} F_c(t)^{j}(1-F_c(t))^{M-j}  f_o(t) \de t  \label{eq: tmp-prob-model-x} \,.
\end{align} 
 The last relation follows from the iid property of random variables $T_{j\in [M]}$ along with the definitions of $F_c(t), F_o(t)$ given in Assumption \ref{assumption: cdfs-model-x}.  By  using the dummy variable $u=F_c(t)$ in the inner integral of \eqref{eq: tmp-prob-model-x} we get
 \[
 \prob( (X,W) \text{ admits label } \ell) =\sum_{j=K(\ell-1)}^{K\ell-1}\int_{0}^1 \binom{M}{j} u^{j}(1-u)^{M-j} \frac{\partial F_o(F_c^{-1}(u))}{\partial u}  \de u  \,.
 \]

Next recall the function $\psi(u)$ as density function of $\omega(u)=F_0(F_c^{-1}(u))$ defined in Assumption \ref{assumption: cdfs-model-x}. We write the above probability as
\begin{equation}\label{eq: p_ell-model-x}
p_\ell=\sum_{j=K(\ell-1)}^{K\ell-1} \binom{M}{j} \int_{0}^{1}u^{j}(1-u)^{M-j}   \psi(u) \de u \,. \quad \forall \ell\in [L]\,.
\end{equation}
For $\ell \in [L]$, we define the polynomial $\beta_\ell:[0,1]\to\reals^+$ as follows:
\[
\beta_\ell(u)=\sum_{j=K(\ell-1)}^{K\ell-1} \binom{M}{j} u^{j}(1-u)^{M-j}\,.
\]
Using the definition of $\beta_\ell(u)$ in \eqref{eq: p_ell-model-x} we write
\begin{equation}\label{eq: p_ell psi}
p_\ell= \int_{0}^{1}\beta_\ell(u)   \psi(u) \de u \,. \quad \forall \ell\in [L]\,,
\end{equation}
and so
\begin{equation}\label{eq: D_f V_ell psi-model-x}
\frac{1}{L}\sum_{\ell=1}^L f(Lp_\ell) =  \frac{1}{L}\sum_{\ell=1}^{L}f\left(L \int_0^1 \beta_\ell(u)\psi(u) \de u\right)\,.
\end{equation}
We continue by upper bounding the right-hand side in the above equation.
Consider the following probability density function over $[0,1]$:
\begin{equation}\label{eq: phi-model-x}
\vphi_\ell(u)=\frac{\beta_\ell(u)}{\int_{0}^1\beta_\ell(u)\de u}\,,\quad \forall u\in[0,1]\,.
\end{equation}
By an application of Jensen's inequality and using the convexity of $f$ we have
\begin{equation}\label{eq: jensen-phi-model-x}
f\left(\E_{u\sim \vphi_\ell}[\psi(u)]\Big]\right)\leq  \E_{u\sim \vphi_\ell}\Big[f (\psi(u))\Big]\,.
\end{equation}
Equivalently this can be rewritten as
\begin{equation}\label{eq: jenson-f-model-x}
f\left(\int_0^1 \vphi_\ell(u) \psi(u) \de u\right) \leq  \int_0^1 \vphi_\ell(u)  f(\psi(u))\de u\,.
\end{equation}
Plugging \eqref{eq: phi-model-x} into \eqref{eq: jenson-f-model-x} yields
\begin{equation}\label{eq: jenson-beta-model-x}
f\left( \frac{\int_0^1 \beta_\ell(u)\psi(u)\de u}{\int_{0}^1\beta_\ell(u)\de u} \right) \leq 
\frac{\int_0^1\beta_\ell(u) f(\psi(u))\de u} {\int_{0}^1\beta_\ell(u)\de u}   \,.
\end{equation}
To use the above inequality in~\eqref{eq: D_f V_ell psi-model-x}, we establish two properties of function $\beta_\ell(u)$.

First from the definition of polynomial $\beta_\ell(u)$ we have
\begin{align}
\int_0^1 \beta_\ell(u)\de u&= \sum_{j=K(\ell-1)}^{K\ell-1} \binom{M}{j}\int_0^1 u^{j}(1-u)^{M-j}\de u\nonumber\\
&=\sum_{j=K(\ell-1)}^{K\ell-1} \binom{M}{j} B(j+1,M-j+1)\nonumber\\
&=\sum_{j=K(\ell-1)}^{K\ell-1} \binom{M}{j} \frac{j!(M-j)!}{(M+1)!}=\frac{1}{L} \,,\label{eq: sum-bernstein-model-x} 
\end{align}
where $B(a,b)$ is the beta function. Second, we have
\begin{align}
\sum_{\ell=1}^L \beta_\ell(u)&=\sum_{\ell=1}^{L}\sum_{j=K(\ell-1)}^{K\ell-1}\binom{M}{j}u^j(1-u)^{M-j}\nonumber\\
&=\sum_{j=0}^{M}\binom{M}{j}u^j(1-u)^{M-j}=(u+1-u)^M=1\,.\label{eq:sum_beta}
\end{align} 

Using \eqref{eq: sum-bernstein-model-x} in \eqref{eq: jenson-beta-model-x} we have
\begin{equation} \label{eq: f-main-inequality-model-x}
f\left(L\int_0^1 \beta_\ell(u)\psi(u)\de u \right) \leq 
L\int_0^1\beta_\ell(u) f(\psi(u))\de u\,, 
\end{equation}
which together with \eqref{eq: D_f V_ell psi-model-x} gives
\begin{align}
\frac{1}{L}\sum_{\ell=1}^L f(Lp_\ell) &\leq  \sum_{\ell=1}^{L} \int_0^1 \beta_\ell(u)f(\psi(u)) \de u= \int_0^1 f(\psi(u))\de u \label{eq: discrete-f-upper-bound-model-x} \,.
\end{align}
The last step above follows from \eqref{eq:sum_beta}.

We next use the definition of $\psi(u)$ in the right hand side of \eqref{eq: discrete-f-upper-bound-model-x} to obtain
\begin{align}
\int_0^1 f(\psi(u))\de u&= \int_0^1  f\Big( \frac{f_o(F_c^{-1}(u))}{f_c(F_c^{-1}(u))}  \Big) \de u \nonumber\\
&=  \int_{-\infty}^{\infty} f\Big( \frac{f_o(t)}{f_c(t)}  \Big) f_c(t) \de t \nonumber\\
&=D_f\left(\cL(T(X,W))\|\cL(T(\tX,\tW))\right)\label{eq: tmp-upper-f-model-x}\,,
\end{align}
where the last relation comes from the definition of $f-$divergence and density functions $f_o(t)$ and $f_c(t)$. 

We next proceed to the proof of the second part. By virtue of characterization in~\eqref{eq: tmp-upper-f-model-x}, we need to show that
\begin{align}\label{eq:claim-2}
 \lim_{L\to \infty}\lim_{K\to \infty}  \frac{1}{L}\sum_{\ell=1}^L f(Lp_\ell) = \int_0^1 f(\psi(u))\de u\,.
\end{align}
Let $b_j(M,u)$ denote the $j$-th Bernstein polynomial of degree $M$, which is given by
\[
b_j(M,u)=\binom{M}{j}u^{j}(1-u)^{M-j} \,.
\]
Then the probability $p_\ell$ given in \eqref{eq: p_ell-model-x} can be written as
\begin{align*}
p_\ell=\sum_{j=K(\ell-1)}^{K\ell-1}  \int_0^1 b_{j}(M,u)\psi(u) \de u \,. \quad \forall \ell\in [L]\,
\end{align*}
Next, partial integration yields 
\begin{equation}\label{eq: p_ell_bernstein}
p_\ell=-\sum_{j=K(\ell-1)}^{K\ell-1}  \int_0^1 \frac{\partial b_{j}(M,u)}{\partial u}\omega(u) \de u \,. \quad \forall \ell\in [L]\,.
\end{equation}
On the other hand, by simple algebraic calculation, it is easy to get the following identity for Bernstein polynomials:
\begin{equation}\label{eq: bernstein-identity}
\frac{\partial b_j(M,u)}{\partial u}=M(b_{j-1}(M-1,u)-b_{j}(M-1,u))\,,
\end{equation}
where we set the convention $\binom{n}{k}=0$ for $k>n$ and $k<0$.  
Using \eqref{eq: bernstein-identity} in \eqref{eq: p_ell_bernstein} gives us
\[
p_\ell=M\sum_{j=K(\ell-1)}^{K\ell-1}\int_{0}^1 \omega(u)(b_j(M-1,u)-b_{j-1}(M-1,u)) \de u \,. \quad \forall \ell\in [L]\,
\]
Moving the summation inside the integration we obtain

\begin{equation}\label{eq: p_ell-bernstein-simple}
p_\ell=M\int_0^1 \omega(u)(b_{K\ell-1}(M-1,u)-b_{K(\ell-1)-1}(M-1,u)) \de u \,. \quad \forall \ell\in [L]\,
\end{equation}
On the other hand, it is easy to check that
\begin{align}\label{eq: bernstein-identity2}
\int_0^1 b_{K\ell-1}(M-1,u)\de u&=\int_0^1 b_{K(\ell-1)-1}(M-1,u)\de u=\frac{1}{M}\,.
\end{align}
Using the identities \eqref{eq: bernstein-identity2} and \eqref{eq: p_ell-bernstein-simple}, we write
\begin{align*}
\omega\Big(\frac{\ell}{L}\Big)-\omega\Big(\frac{\ell-1}{L}\Big)-{p_\ell} &= M\int_0^1\bigg(\omega\Big( \frac{\ell}{L}\Big)- \omega(u) \bigg) b_{K\ell-1}(M-1,u)\de u\\
&\;+M\int_0^1\bigg(\omega(u)-\omega\Big( \frac{\ell-1}{L}\Big)  \bigg) b_{K(\ell-1)-1}(M-1,u)\de u\,.
\end{align*}
Since $\psi(.)$ is continuous on $[0,1]$, therefore it is $B$-bounded. This means that $\omega(.)$ is $B$-Lipschitz. Using this along with the triangle inequality we get
\begin{align}
\left|\omega\Big(\frac{\ell}{L}\Big)-\omega\Big(\frac{\ell-1}{L}\Big)-{p_\ell} \right| &\leq MB\int_0^1\Big| \frac{\ell}{L}-u \Big| b_{K\ell-1}(M-1,u)\de u\nonumber \\
&\;+ MB \int_0^1\Big|u-\frac{\ell-1}{L}\Big| b_{K(\ell-1)-1}(M-1,u)\de u \label{eq: tmp-bernstein}\,.
\end{align}
We continue by writing the above expressions in terms of a Beta distribution.
Recall that the pdf of Beta distribution $\mathsf{Beta}(\alpha,\beta)$ with shape parameters $(\alpha, \beta)$ is given by $\frac{u^{\alpha-1}(1-u)^{\beta-1}}{B(\alpha,\beta)}$, where  $B(\alpha, \beta):=\int_0^1 u^{\alpha-1}(1-u)^{\beta-1}\de u$. For integer values $a,b$, $B(a,b)$ has a closed form and is given by $B(a,b) = \frac{(a-1)! (b-1)!}{(a+b-1)!}$. Therefore, we have
\begin{align*}
M b_{K\ell-1}(M-1,u) &= M {M-1\choose K\ell-1} u^{K\ell-1}(1-u)^{M-K\ell}\\
& = \frac{M!}{(K\ell-1)!(M-K\ell)!} u^{K\ell-1}(1-u)^{M-K\ell}\\
& = \frac{u^{K\ell-1}(1-u)^{M-K\ell}}{B(K\ell,M-K\ell+1)}\,.
\end{align*}
Using a similar expression for $M b_{K(\ell-1)-1}(M-1,u)$ we write \eqref{eq: tmp-bernstein} as follows:
\begin{align}
\frac{1}{B}\left|\omega\Big(\frac{\ell}{L}\Big)-\omega\Big(\frac{\ell-1}{L}\Big)-{p_\ell} \right|
&\leq \E_{u\sim \mathsf{Beta}(K\ell,M-K\ell+1)}\left[\Big| \frac{\ell}{L}-u \Big|\right] \nonumber \\ 
&\;+  \E_{u\sim \mathsf{Beta}(K(\ell-1),M-K(\ell-1)+1)}\left[\Big|u-\frac{\ell-1}{L}\Big|\right] \label{eq: tmp-bernstein2}\,.
\end{align}
Since the mean of $\mathsf{Beta}(\alpha,\beta)$ is given by $\alpha/(\alpha+\beta)$, the above Beta distributions have mean values of $\ell/L$ and $(\ell-1)/L$, respectively (recall that $M+1 = KL$). Therefore the terms on the right-hand side of \eqref{eq: tmp-bernstein2} are indeed the mean absolute deviation of two Beta distributions. Further, using Jenson's inequality we know that for arbitrary random variable $S$ we have
\[
\E[|S-\E[S]|]\leq \var[S]^{1/2}\,.
\] 
Using this in \eqref{eq: tmp-bernstein2} yields
\begin{align}
\frac{1}{B}\left|\omega\Big(\frac{\ell}{L}\Big)-\omega\Big(\frac{\ell-1}{L}\Big)-{p_\ell} \right|
&\leq  \var\left[\mathsf{Beta}(K\ell,M-K\ell+1)\right]^{1/2} \nonumber\\ 
&\;+  \var\left[\mathsf{Beta}(K(\ell-1),M-K(\ell-1)+1)\right]^{1/2} \label{eq: tmp-bernstein3}\,.
\end{align}
In the next step, note that
\begin{equation}\label{eq: var-beta}
\var\left( \mathsf{Beta}(\alpha,\beta)\right)=\frac{\alpha\beta}{(\alpha+\beta)^2(\alpha+\beta+1)}\leq \frac{1}{4(\alpha+\beta)}\,,
\end{equation}
where the last inequality follows from $(\alpha+\beta)^2\geq 4\alpha\beta$. 
Combining \eqref{eq: tmp-bernstein3} and \eqref{eq: var-beta} yields
\begin{equation}\label{eq: tmp-mean-value}
\frac{1}{B}\left|\omega\Big(\frac{\ell}{L}\Big)-\omega\Big(\frac{\ell-1}{L}\Big)-{p_\ell} \right|\leq \frac{2}{2\sqrt{M+1}} = \frac{1}{\sqrt{KL}}\,.
\end{equation}
Now from the mean value theorem, we know that there exists $s_\ell \in \left[\frac{\ell-1}{L}, \frac{\ell}{L}\right]$ such that 
\begin{equation}\label{eq: mean-value}
\omega\Big(\frac{\ell}{L}\Big)-\omega\Big(\frac{\ell-1}{L}\Big)=\frac{1}{L}\psi(s_\ell)\,.
\end{equation}
Combining \eqref{eq: mean-value} and \eqref{eq: tmp-mean-value} yields
\begin{equation}\label{eq: Lp_ell}
\lim_{K\to \infty} Lp_\ell=\psi(s_\ell)\,,~ \text{ for some } s_\ell\in \left[\frac{\ell-1}{L}, \frac{\ell}{L}\right]\,.
\end{equation}
We are now ready to prove the claim of~\eqref{eq:claim-2}. Write
\begin{align}\label{eq: f-div-diff}
\frac{1}{L}\sum_{\ell=1}^Lf(L p_\ell)-\int_0^1f(\psi(u))\de u &=\frac{1}{L}\sum_{\ell=1}^L \left[f(L p_\ell)-L\int_{\frac{\ell-1}{L}}^{\frac{\ell}{L}}(f\circ\psi)(u)\de u\right]\,.
\end{align}
By another application of the mean value theorem for the continuous function $f\circ\psi$,  there exists $t_\ell\in \left[\frac{\ell-1}{L}, \frac{\ell}{L}\right]$ such that 
\begin{equation}\label{eq: t_ell}
L\int_{\frac{\ell-1}{L}}^{\frac{\ell}{L}}(f\circ\psi)(u)\de u=(f\circ \psi)(t_\ell)\,.
\end{equation}
By combining \eqref{eq: f-div-diff} and \eqref{eq: t_ell} we get
\begin{align}\label{eq: f-div-diff-2}
\frac{1}{L}\sum_{\ell=1}^Lf(L p_\ell)-\int_0^1f(\psi(u))\de u &=\frac{1}{L}\sum_{\ell=1}^L \left[f(L p_\ell)-f\circ\psi(t_\ell) \right]\,,
\end{align}
and using continuity of $f$ along with \eqref{eq: Lp_ell} and \eqref{eq: f-div-diff-2} we obtain
\begin{align}\label{eq: f-div-diff-3} 
\lim_{K\to \infty}\frac{1}{L}\sum_{\ell=1}^Lf(L p_\ell)-\int_0^1f(\psi(u))\de u &=\frac{1}{L}\sum_{\ell=1}^L \left[f\circ\psi(s_\ell)-f\circ\psi(t_\ell) \right]\,.
\end{align}
Next since $f\circ \psi$ is continuous on the compact set $[0,1]$, it is uniformly continuous. This implies that for every arbitrary $\eps>0$, there exists $\delta>0$ such that if $|u_1-u_2|\leq \delta$, then we have $|f\circ\psi(u_1)-f\circ\psi(u_2)|\leq \eps$. Choose $L$ sufficiently large such that $1/L <\delta$. Since both $t_\ell,s_\ell$ belong to the interval $[(\ell-1)/L,\ell/L]$ we get that $|s_\ell-t_\ell| <\delta$, and therefore $|f\circ\psi(s_\ell)-f\circ\psi(t_\ell)|\leq \eps$. Using this observation in \eqref{eq: f-div-diff-3} we see that for every arbitrary small $\eps>0$ and sufficiently large $L$ we have
\begin{align*}
\left|\lim_{K\to \infty}\frac{1}{L}\sum_{\ell=1}^Lf(L p_\ell)-\int_0^1f(\psi(u))\de u \right|&\leq \frac{1}{L}\sum_{\ell=1}^L \left|f\circ\psi(s_\ell)-f\circ\psi(t_\ell) \right| \leq \eps\,.
\end{align*}
By letting $L$ go to infinity, $\eps$ can be chosen arbitrarily small and so
\begin{align}\label{eq: f-div-diff-4} 
\lim_{L\to\infty}\lim_{K\to \infty}\frac{1}{L}\sum_{\ell=1}^Lf(L p_\ell)=\int_0^1f(\psi(u))\de u\,.\end{align}
This completes the proof of the claim~\eqref{eq:claim-2}, and so the proof of the second part of Proposition~\ref{propo: model-x}. 
\subsection{Proof of Theorem \ref{thm: discretize-distance:model-x}}\label{proofs: thm: discretize-distance:model-x}
As we showed in Proposition~\ref{propo: model-x} (first part), we have:
\begin{align*}
 \frac{1}{L}\sum_{\ell=1}^L f(Lp_\ell) \leq D_f(\cL(T(X,W))\|\cL(T(\tX,\tW)))\,.
 \end{align*}
 Also, by an application of the data processing inequality we have
  \[
  D_f\left(\cL(T(X,W))\|\cL(T(\tX,\tW))\right)\leq D_f\left(\cL(X,W)\|\cL(\tX,\tW)\right)\,.
  \] 
The result then follows from Proposition \ref{propo: f-model-x}. 

\subsection{Proof of Lemma \ref{lemma: variational}}\label{proof: variational}
Using the definition of $f$-divergence and conjugate function we have
\begin{align*}
D_f(p\|q)&=\int qf\left(\frac{p}{q}\right)\de \mu\\
&\geq \int q \sup_{\vphi\in \cG}\left\{ \frac{\vphi p}{q}-f^*\left(\vphi \right) \right\}\de \mu \\
&\geq \sup_{\vphi\in \cG}  \int \left( \vphi p-f^*(\vphi)q\right) \de \mu \,.
\end{align*}
This completes the proof of the lower bound. Next we show that this bound becomes tight if $\sigma \in \partial f(p/q)$, for any $x\in \mathcal{X}$ (Here, $p,q$ and $\sigma$ are evaluated at any $x\in\mathcal{X}$). The definition of subdifferential implies that for every measurable function $h:\cX\to \reals$ we have
\[
f(h)-f\left(\frac{p}{q}\right)\geq \sigma h- \frac{\sigma p}{q}\,.
\]  
This gives us
\[
f\left(\frac{p}{q}\right)\leq  \frac{\sigma p}{q} - (\sigma h -f(h))\,.
\]
Since this holds for every measurable function $h$ we get 
\[
f\left(\frac{p}{q}\right)\leq  \frac{\sigma p}{q} - \sup_{h}(\sigma h -f(h))\,.
\]
In the next step, by using the definition of the conjugate dual function we get 
\[
qf\left(\frac{p}{q}\right)\leq  {\sigma p} - qf^*(\sigma)\,.
\]
This completes the proof.

\subsection{Proof of Proposition \ref{propo: achievablity}}\label{proof: propo: achievablity}
Since $X_p\sim p$ and $X_q\sim q$, we get $D_f(p\|q)=D_f(\cL(X_p)\|\cL(X_q))$ and by an application of the data processing inequality we obtain
 \begin{align*}
 D_f(p\|q) = D_f(\cL(X_p)\|\cL(X_q))\geq D_f(\cL(\sigma(X_p))\|\cL(\sigma(X_q)))\,.
 \end{align*}
 We next show the other direction, i.e.,
\begin{equation}\label{eq: achieve-tmp3}
D_f(\cL(\sigma(X_p))\|\cL(\sigma(X_q)))\geq D_f(p\|q)\,.
\end{equation}
Let $\tp$ and $\tq$ denote the density functions of $\sigma(X_p)$ and $\sigma(X_q)$ respectively. By using Lemma \ref{lemma: variational} for density functions $\tp,\tq$ and a class of measurable functions $\cG=\{g:\reals\to \reals\}$ we get
\begin{equation}\label{eq: achieve-tmp1}
D_f(\cL(\sigma(X_p))\|\cL(\sigma(X_q)))\geq \sup_{g\in \cG}  \int \left( g \tp-f^*(g)\tq\right) \de \mu\,.
\end{equation}
Next, by rewriting \eqref{eq: achieve-tmp1} in terms of expectations we arrive at
\[
D_f(\cL(\sigma(X_p))\|\cL(\sigma(X_q)))\geq
\sup_{g\in \cG}\left\{\E_{t\sim \tp}[g(t)]-\E_{t\sim \tq}[f^*(g(t))]\right\}\,.
\]
We then change the expectation measures to get
\begin{equation}\label{eq: achieve-tmp2}
D_f(\cL(\sigma(X_p))\|\cL(\sigma(X_q)))\geq
\sup_{g\in \cG}\left\{\E_{}[g(\sigma(X_p))]-\E_{}[f^*(g(\sigma(X_q)))]\right\}\,.
\end{equation}
We consider the identity function $g(t) = t$ defined over the real line, with the convention that $g(\infty) = \infty$. 
Evaluating the right-hand side of~\eqref{eq: achieve-tmp2} at $g$ we get
\begin{align}
D_f(\cL(\sigma(X_p))\|\cL(\sigma(X_q))) &\geq \E_{}[\tg(\sigma(X_p))]-\E_{}[f^*(\tg(\sigma(X_q)))\nonumber\\
&=\E_{}[\sigma(X_p)]-\E_{}[f^*(\sigma(X_q))]\label{eq: achieve-tmp4}\,.
\end{align}
We also know from Lemma \ref{lemma: variational} that for $\sigma\in \partial f(p/q)$ we have
\begin{equation}\label{eq: achieve-tmp5}
\E_{}[\sigma(X_p)]-\E_{}[f^*(\sigma(X_q)))=D_f(p\|q)\,.
\end{equation}
Combining \eqref{eq: achieve-tmp4} and \eqref{eq: achieve-tmp5} yields \eqref{eq: achieve-tmp3} and completes the proof. 


{\color{black}
\section{Proof of Proposition \ref{propo: linear-f}}
The Lagrangian of $\min\limits_{q\in \cU_\tau}^{} q^\sT x$  is given by
\[
L(q;\eta,\lambda)= q^\sT x +\lambda\left( \frac{1}{L}\sum\limits_{\ell=1}^{L} f(Lq_\ell)- \tau\right) +\eta\left(\sum\limits_{\ell=1}^{L}q_\ell-1\right)\,.
\]
In this case, the dual objective function is 
\[
D(\lambda, \eta)=\min_{q\geq 0} \left[ q^\sT x +\frac{\lambda}{L}\sum\limits_{\ell=1}^{L} f(Lq_\ell)-{\lambda \tau} +\eta(\sum\limits_{\ell=1}^{L}q_\ell-1)
  \right]\,.
\]
The constraints are decoupled and minimization can be moved inside, this yields
\[
D(\lambda, \eta)=-\eta-\lambda\tau+\sum\limits_{\ell=1}^L  \min\limits_{q_\ell\geq 0}^{}\left[q_\ell x_\ell +\frac{\lambda}{L}f(Lq_\ell) +\eta q_\ell\right]\,.
\]
Using the conjugate dual definition $f(s)=\sup\limits_{t\geq 0}^{} [st-f(t)]$ yields 
\[
D(\lambda, \eta)=-\eta-\lambda\tau-\frac{\lambda}{L}\sum\limits_{\ell=1}^L f^*\left( -\frac{x_\ell+\eta}{L\lambda}\right)\,.
\]
In this formulation the optimal $q_\ell$ is such that 
\begin{equation}\label{eq: subgradient} 
-\frac{x_\ell+\eta}{L\lambda} \in \partial f(q_\ell)\,.
\end{equation}
Given that in this problem the strong duality holds (Slater's condition for $\tau>0$), using $\lambda^*,\eta^*$ as solutions of $\arg\min_{\lambda\geq 0, \eta} -D(\lambda, \eta)$ in \eqref{eq: subgradient} completes the proof.

}

\end{document}